\documentclass[preprint,preprintnumbers,amsmath,amssymb,floatfix,nofootinbib]
{revtex4}
\usepackage{graphicx}
\usepackage{url}

\begin{document}
\preprint{BNL-HET-07/5}
\preprint{hep-ph/0703299}
\title{\boldmath{Chiral-logarithmic Corrections to the $S$ and $T$ Parameters 
                 in Higgsless Models}}
\author{S.~Dawson}
\email{dawson@quark.phy.bnl.gov}
\author{C.~B.~Jackson}
\email{cbjackson@bnl.gov}
\affiliation{Physics Department, Brookhaven National Laboratory,
Upton, NY 11973-5000, USA}

\date{\today}

\begin{abstract}
Recently, Higgsless models have proven to be 
viable alternatives to the Standard Model (SM) and supersymmetric models in 
describing the breaking
of the electroweak symmetry.  Whether extra-dimensional in nature or 
their deconstructed counterparts, the physical spectrum of these models
typically consists of  ``towers'' of massive vector gauge bosons which 
carry the same quantum numbers as the SM $W$ and $Z$.  In
this paper, we calculate the one-loop, chiral-logarithmic corrections to
the $S$ and $T$ parameters from the lightest (i.e. SM) and the
next-to-lightest gauge bosons using a novel application of the Pinch
Technique.  We perform our calculation using generic Feynman rules with
generic couplings such that our results can be applied to various models.
To demonstrate how to use our results, we calculate the leading 
chiral-logarithmic corrections to the $S$ and $T$ parameters in the 
deconstructed three site Higgsless model.  As we point out, however, our
results are not exclusive to Higgsless models and may, in fact, be used
to calculate the one-loop corrections
from additional gauge bosons in models with fundamental (or composite) 
Higgs bosons. 
\end{abstract}

\maketitle

\section{Introduction}
\label{sec:intro}

The source of electroweak symmetry breaking (EWSB), i.e. the generation
of the $W^\pm$ and $Z^0$ masses, remains as one of the unanswered
questions in particle physics.  If the Standard Model (SM) or one of
its supersymmetric (SUSY) extensions are correct, then one (or more)
 $SU(2)$ scalar doublets are responsible for EWSB and at least 
one physical Higgs boson should be discovered at 
the Large Hadron Collider (LHC).

Unfortunately, the Higgs mechanism as implemented in the SM has several
theoretical shortcomings.  The most troublesome of these is the fact
that the Higgs boson mass is unstable against radiative corrections, 
a situation known as the {\it large hierarchy problem}.  
In other words,
for the Higgs boson to be light (as indicated by electroweak
precision measurements), its bare mass must be highly 
fine-tuned to cancel large loop effects from high-scale physics.  
In SUSY extensions, this 
fine-tuning is avoided due to additional particles which cancel the
quadratic contributions to the Higgs boson mass from SM particles.

In the past several years, an interesting alternative to SUSY models
has emerged in the form of extra-dimensional models 
\cite{Arkani-Hamed:1998rs,Randall:1999ee}.  In most of these
scenarios, the size and shape of the extra dimension(s) are responsible
for solving the large hierarchy problem.  In addition, variations of 
these models can also provide viable
alternatives to the Higgs mechanism.  For example, in models
where the SM gauge fields propagate in a fifth dimension, masses for 
the $W^\pm$ and $Z^0$ bosons can be generated via non-trivial boundary
conditions placed on the five-dimensional wavefunctions 
\cite{Csaki:2003dt,Cacciapaglia:2004rb,Nomura:2003du,Csaki:2003zu}.  Since the 
need for scalar doublets is eliminated in such scenarios, these models
have been aptly dubbed {\it Higgsless models}.  The result of allowing 
the SM gauge fields to propagate in the bulk, however, is towers of 
physical, massive vector gauge bosons (VGBs), the lightest of which are
identified with the SM $W^\pm$ and $Z^0$ bosons.  The heavier 
Kaluza-Klein (KK) modes, which have the $SU(2)\times U(1)$
quantum numbers of the SM  $W^\pm$ and $Z^0$,
play an important role in longitudinal VGB scattering.  In the SM
without a Higgs boson, the scattering amplitudes for these processes
typically violate unitarity around $\sim$ 1.5 TeV \cite{Lee:1977eg}.  
The exchange of light Higgs 
bosons, however, cancels the  unitarity-violating terms and ensures 
perturbativity of the theory up to high scales.  In 
extra-dimensional Higgsless 
models, the exchange of the heavier KK gauge bosons plays the role of the Higgs
boson and cancels the dominant  unitarity-violating terms 
\cite{Csaki:2003dt}.  As a result, the scale of unitarity 
violation can be pushed to the $\sim$ 5-10 TeV range.

The main drawback of extra-dimensional models is that they are 
non-renormalizable and, thus, must be viewed as effective theories up 
to some cutoff scale $\Lambda$ above which new physics must take over.
An extremely efficient and convenient way of studying the phenomenology 
of five-dimensional effective theories in the context of
four-dimensional gauge theories is that of deconstruction 
\cite{Arkani-Hamed:2001ca,Hill:2000mu}.
Deconstructed models possess extended gauge symmetries which 
approximate the fifth dimension, but can be studied in the simplified
language of coupled non-linear sigma models (nl$\sigma$m) 
\cite{Appelquist:1980vg,Longhitano:1980iz,Longhitano:1980tm}.  In fact,
this method allows one to effectively separate the perturbatively calculable
contributions to low-energy observables from the strongly-coupled
contributions due to physics above $\Lambda$.  The former arise
from the new weakly-coupled gauge states, while the latter can be parameterized
by adding higher-dimension operators \cite{Appelquist:1980vg,Longhitano:1980iz,Longhitano:1980tm,Bagger:1992vu,
Perelstein:2004sc,Chivukula:2007ic}.

The phenomenology of deconstructed Higgsless models has been well-studied
\cite{Foadi:2003xa,Hirn:2004ze,Casalbuoni:2004id,Chivukula:2004pk,Perelstein:2004sc,
Georgi:2004iy,SekharChivukula:2004mu}.  Recently, however, the simplest
version of these types of models, which involves only three ``sites''
\cite{Foadi:2003xa,Perelstein:2004sc,SekharChivukula:2006cg}, has
received much attention and been shown to be capable of 
approximating much of the interesting phenomenology associated with 
extra-dimensional models and more complicated deconstructed Higgsless models
\cite{Cacciapaglia:2004rb,Cacciapaglia:2005pa,Foadi:2004ps,Foadi:2005hz,
Chivukula:2005bn,Casalbuoni:2005rs,SekharChivukula:2005xm}.  The gauge 
structure of the three site model is identical to that of the so-called
BESS (Breaking Electroweak Symmetry Strongly) which was first analyzed over 
twenty years ago \cite{Casalbuoni:1985kq,Casalbuoni:1986vq}.
Once EWSB occurs in
the three site model, the gauge sector consists of a massless photon, two 
relatively
light massive VGBs which are identified with the SM $W$ and $Z$ gauge bosons,
as well as two new heavy VGBs which we denote as $W^\prime$ and $Z^\prime$.
The exchange of these heavier states in longitudinal VGB scattering can
delay unitarity violation up to higher scales \cite{Foadi:2003xa}.

Given the prominent role that the heavier VGBs play in the extra-dimensional
and deconstructed Higgsless scenarios, it is important to assess 
their effects on electroweak precision
observables, namely the oblique parameters ($S, T$ and $U$) 
\cite{Peskin:1991sw}.  These 
parameters are defined in terms of the SM gauge boson 
self-energies, $\Pi^{\mu\nu}_{ij}(q^2)$, where $(ij)$ = $(WW), (ZZ), 
(\gamma\gamma)$ and $(Z\gamma)$, and $q$ is the momentum carried by the
external gauge bosons.  Generically, the one-loop contributions to 
the $\Pi_{ij}$ can be split into four separate classes depending on the 
particles circulating in the loops: namely, those involving ({\it i})
only fermions, ({\it ii}) only scalars, ({\it iii}) a mixture of 
scalars and gauge bosons and ({\it iv}) only gauge bosons.  Due to gauge
invariance, class ({\it i}) and the sum of classes ({\it ii}) and 
({\it iii}) are independent of the $R_{\xi}$ gauge used in the calculation.
However, class ({\it iv}), i.e. contributions to the two-point functions
from loops of gauge bosons, are $R_\xi$ gauge-dependent.
This was shown explicitly for the case of one-loop contributions from SM 
gauge bosons in Ref.~\cite{Degrassi:1992ue}.  In that paper, the authors
showed that in a general $R_\xi$ gauge the 
gauge boson self-energies depend non-trivially on the gauge parameter(s)
$\xi_{i}$ ($i = W,Z,\gamma$).  These dependences carry over into the 
calculation of the oblique parameters resulting in gauge-dependent 
expressions for $S, T$ and $U$ \cite{Degrassi:1993kn}.  However, in a 
series of subsequent papers,
it was shown that by isolating the gauge-dependent terms from other 
one-loop corrections (i.e. vertex and box corrections) and combining these
with the self-energy expressions derived from the two-point functions, it 
is possible to define
gauge-invariant forms of the self-energies and, thus, obtain gauge-invariant 
expressions for the oblique parameters \cite{Degrassi:1992ff,Degrassi:1992ue,
Papavassiliou:1994fp,Degrassi:1993kn}.  This method of extracting 
gauge-invariant Green's functions from scattering amplitudes is known as the 
{\it Pinch Technique} (PT) 
\cite{Cornwall:1981ru,Cornwall:1981zr,Cornwall:1989gv,Papavassiliou:1989zd}.

In this paper, we generalize the results of Refs.
\cite{Degrassi:1992ff,Degrassi:1992ue,Papavassiliou:1994fp,Degrassi:1993kn}
to calculate the one-loop, chiral logarithmic corrections to the oblique
parameters in extra-dimensional and deconstructed Higgsless models.  
In our calculation of the PT 
self-energies, we employ the unitary gauge
($\xi \to \infty$) to define the massive VGB propagators.  The attractive
feature of this choice is that unphysical states (i.e. Goldstone 
bosons, ghosts, etc.) decouple and thus the number of diagrams
is drastically reduced.  Green's functions 
calculated in unitary gauge are individually non-renormalizable
in the sense that they contain divergences proportional to higher powers 
of $q^2$ which cannot be removed by the usual counterterms.  However, 
when the PT is applied, these non-renormalizable terms cancel in the 
same manner as the gauge-dependent terms mentioned above 
\cite{Papavassiliou:1994fp}.

The rest of this paper is organized in the following way.  In 
Section~\ref{sec:the-model}, we discuss the generic Feynman rules 
used in our calculation.  We also describe in some detail the three-site
Higgsless model to which we will apply our results in the following sections.
Section~\ref{sec:Ugauge-and-PT} contains a general 
discussion on the Pinch Technique and its use within the unitary gauge.
In Sections~\ref{sec:loops-neutral} and \ref{sec:loops-charged}, we calculate
the one-loop corrections needed to construct the PT self-energies in terms 
of generic couplings.  These corrections are then assembled in 
Section~\ref{sec:PT-SEs} where we explicitly show how to construct the 
PT gauge boson self-energies. Using these expressions,
we calculate the leading chiral-logarithmic corrections to the $S$ and 
$T$ parameters in the three-site model in Section~\ref{sec:ST-higgsless}.
The one-loop corrections to the $S$ and $T$ parameters in the three site
model were first calculated in Refs.~\cite{Chivukula:2007ic,Matsuzaki:2006wn}
to which we compare our results and find excellent agreement.
Finally, in Section~\ref{sec:conclusions}, we conclude.

\section{The Model(s)}
\label{sec:the-model}

Our results apply to a wide class of Higgsless models in 
extra-dimensional and deconstructed theories.  We begin this 
section by outlining the types of models for which our calculation 
is valid.  After defining the generic Feynman rules used in our 
calculation, we discuss the three site Higgsless model in detail 
and show how it fits within the framework described below.

First, assume that the model has an extended gauge symmetry of the 
form:
\begin{equation}
SU(2) \times SU(2)^N \times \times U(1)\,,
\label{eq:gauge-structure}
\end{equation}  
where the $U(1)$ is gauged as the $T_3$ component of a global $SU(2)$ and
the effective four-dimensional Lagrangian for the gauge kinetic terms is:
\begin{eqnarray}
{\cal{L}}_{G} = -\frac{1}{4}B_{\mu\nu}B^{\mu\nu} - 
  \frac{1}{4}\sum_{i=1}^{N+1}W_{i,\mu\nu}^a W_i^{a,\mu\nu}\,.
\label{eq:L-gauge}
\end{eqnarray}
This gauge structure has been implemented in both extra-dimensional models
(where $N=1$) \cite{Agashe:2003zs,Csaki:2003zu,Nomura:2003du,Barbieri:2003pr,Csaki:2003sh,Davoudiasl:2003me,Burdman:2003ya,Cacciapaglia:2004jz,Davoudiasl:2004pwWY,Barbieri:2004qk,Agashe:2004ay,Hewett:2004dv,Agashe:2004cp,Cacciapaglia:2004rb,Lillie:2005pt,Agashe:2006wa,Carena:2006bn,Cacciapaglia:2006gp,Djouadi:2006rk,Cacciapaglia:2006mz,Contino:2006nn,Carena:2007ua,Lillie:2007yh,Agashe:2007zd} as well as deconstructed versions ($N=1,\dots,\infty$)
\cite{Chivukula:2003wj,Foadi:2003xa,Chivukula:2004pk,Georgi:2004iy,Perelstein:2004sc,Chivukula:2004af,SekharChivukula:2004mu,Chivukula:2005bn,SekharChivukula:2005xm,Chivukula:2005ji,SekharChivukula:2005cc,SekharChivukula:2006cg,Matsuzaki:2006wn,SekharChivukula:2006we,Chivukula:2007ic}.  
Once EWSB occurs, mixing in both the
charged and neutral sectors results in a physical spectrum consisting of
a massless photon and ``towers'' of charged and neutral VGBs.  In terms of 
the mass eigenstates, the gauge fields can be written:
\begin{eqnarray}
\label{eq:Wpm-gen}
W_i^{\pm,\mu} &=& \sum_{n=1}^{N+1} a_{in} W_{(n)}^{\pm,\mu}\,, \\
\nonumber\\
\label{eq:B-gen}
B^\mu &=& b_{00}\gamma^\mu + \sum_{n=1}^{N+1} b_{0n} Z_{(n)}^\mu\,, \\
\nonumber\\
\label{eq:W3-gen}
W_{3,i}^\mu &=&  b_{i0}\gamma^\mu + \sum_{n=1}^{N+1} b_{in} Z_{(n)}^\mu\,, 
\end{eqnarray}
where $W_{(n)}^\pm$ and $Z_{(n)}$ represent the mass eigenstates, the 
lightest of which are identified with the SM $W$ and $Z$.  In 
extra-dimensional models, the above expansions would realistically involve
infinite towers of massive states; however, in writing Eqs.~(\ref{eq:Wpm-gen})-
(\ref{eq:W3-gen}), we have assumed that only the lightest (i.e., SM-like)
and next-to-lightest gauge bosons are important for the phenomenology 
attainable at present and near-future collider experiments 
\cite{Birkedal:2004au}.
In general, the mixing angles $a_{ij}$ and $b_{ij}$ can be written in terms 
of the gauge couplings and the mass eigenvalues and are model-dependent.  
Inserting Eqs.~(\ref{eq:Wpm-gen})-(\ref{eq:W3-gen}) into the kinetic energy 
terms for the $SU(2)$ gauge fields in Eq.~(\ref{eq:L-gauge})
generates 3-point and 4-point interactions between the mass eigenstates.
The overall couplings for these interactions are functions of the 
$SU(2)$ gauge couplings and the mixing angles $a_{ij}$ and $b_{ij}$.

Next, we consider the couplings of the fermions to the gauge fields.  Assuming
the $SU(2$) gauge fields couple only to left-handed fermions while the $U(1)$ 
couples to both left- and right-handed fermions, we take as 
the effective Lagrangian:
\begin{eqnarray}
{\cal{L}}^{eff}_f&=& -\sum_{n=1}^{N+1} \sum_{i,j} 
  {g_{ijW^\pm_n}\over 2 \sqrt{2}}
{\overline \psi_i}\gamma^\mu
(1-\gamma_5)\psi_j W^\pm_{n,\mu} +{\hbox{h.c.}}
\nonumber \\
&&
-\sum_{n=1}^{N+1} \sum_i g_{iiV^0_j} {\overline \psi_i}\gamma^\mu
\biggl[g_{V_i}^{(V^0_n)} + g_{A_i}^{(V^0_n)} \gamma_5 \biggr]
\psi_i V^0_{n,\mu}\,,
\label{eq:fermcoups}
\end{eqnarray}
where $W^\pm_n$ and $V^0_n$ represent the mass eigenstates.
Again, the overall couplings $g_{ijW^\pm_n}$ and $g_{iiV^0_n}$, as well
as the coefficients $g_{V_i}^{(V^0_n)}$ and $g_{A_i}^{(V^0_n)}$, are functions
of the gauge couplings as well as the mixing angles $a_{ij}$ and $b_{ij}$.
Note that electromagnetic gauge invariance requires:
\begin{eqnarray}
g_{ff\gamma}\,g_{V_f}^{(\gamma)} &=& e\,Q_f\,, \nonumber\\
g_{ff\gamma}\,g_{A_f}^{(\gamma)} &=& 0\,,
\label{eq:Photon-coups}
\end{eqnarray}
where $Q_f$ is the fermion's charge in units of the electron charge $e$.

In the following sections, we present our results in terms of 
generic 3- and 4-point gauge boson couplings, as well as generic 
fermion-gauge boson couplings.  By taking this approach, our results 
are applicable to any model which fits within the framework outlined
above.
\begin{figure}[t]
\begin{center}
\includegraphics[bb=62 215 611 766]{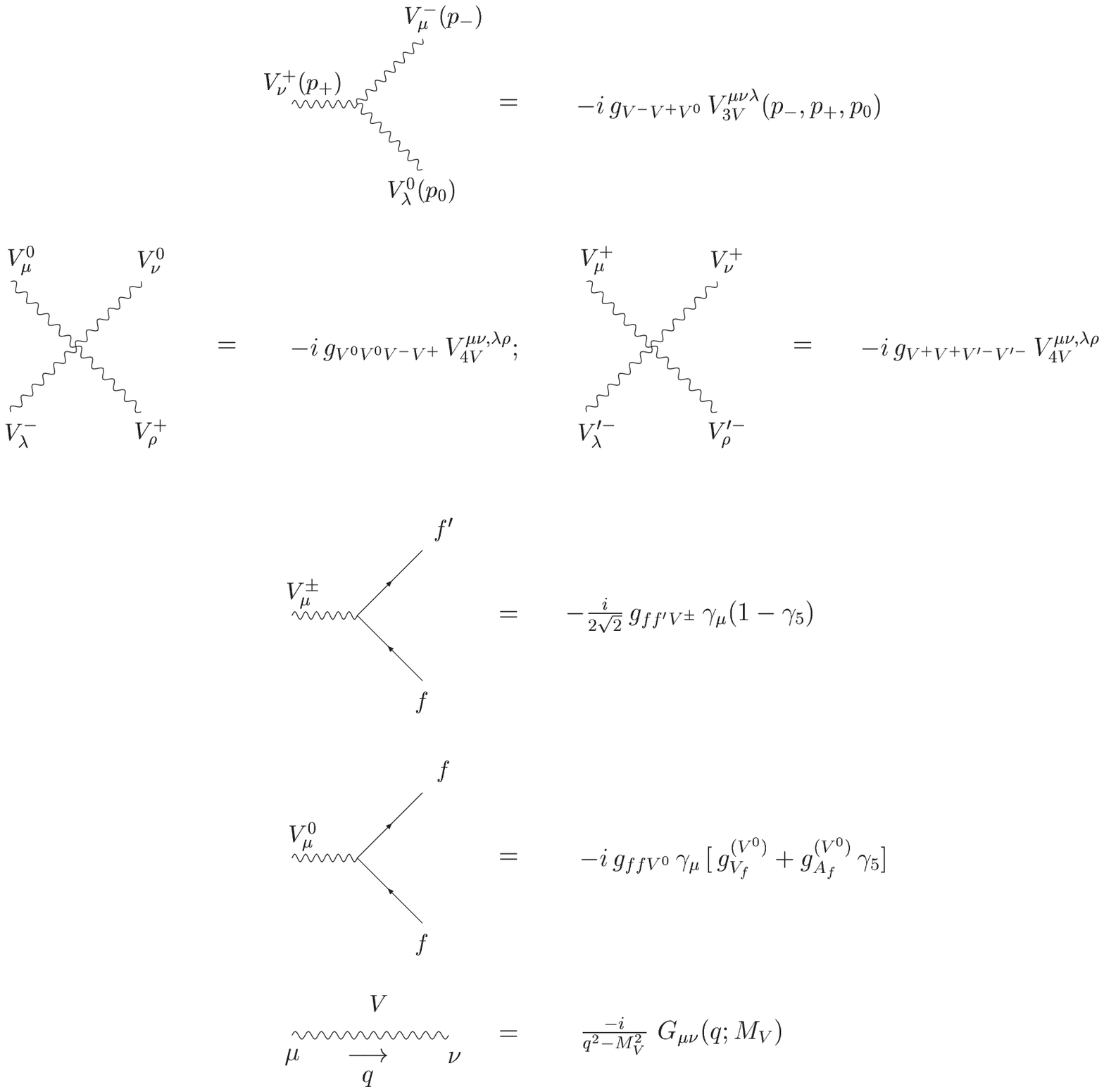} 
\end{center}
\caption[]{Generic Feynman rules used in our calculation.  All kinematic
           functions and coupling constants are defined in the text.}
\label{fg:feyn-rules}
\end{figure}
The Feynman rules used in our calculation are shown in 
Fig.~\ref{fg:feyn-rules}.  In these figures, the momenta of the gauge bosons 
are always defined to be incoming such that the kinematic structures 
$V_{3V}^{\mu\nu\lambda}$ and $V_{4V}^{\mu\nu,\lambda\rho}$ take the forms:
\begin{eqnarray}
V_{3V}^{\mu\nu\lambda}(p_- ,p_+ ,p_0 ) &=& (p_- - p_+ )^{\lambda}\,g^{\mu\nu} +
                                     (p_+ - p_0 )^{\mu}\,g^{\nu\lambda} +
                                     (p_0 - p_- )^{\nu}\,g^{\mu\lambda}\,, \\
\nonumber\\
V_{4V}^{\mu\nu,\lambda\rho} &=& 2g^{\mu\nu}g^{\lambda\rho} - 
                               g^{\mu\lambda}g^{\nu\rho} -
                               g^{\mu\rho}g^{\nu\lambda}\,.
\label{eq:Vertices}
\end{eqnarray}
%
%
Lastly, the massive gauge boson propagator is defined in terms of the 
kinematic structure $G_{\mu\nu}(q;M_V)$ which, in unitary gauge, is 
given by:
\begin{equation}
G_{\mu\nu}(q;M_V) = g_{\mu\nu} - \frac{q_\mu q_\nu}{M_V^2}\,,
\label{eq:Gmunu}
\end{equation}
where $M_V$ is the mass of the propagating gauge boson.

We turn now to the three site Higgsless model which is a prototypical 
example of the models outlined above.

\subsection{The Three Site Higgsless Model}
\label{subsec:3site-model}

\begin{figure}[t]
\begin{center}
\includegraphics[scale=0.8]{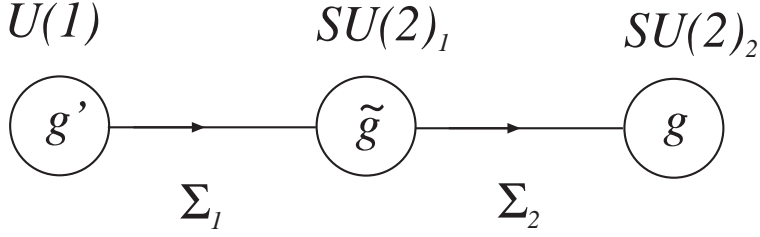} 
\end{center}
\caption[]{Moose diagram for the three site model (from 
Ref.~\cite{Foadi:2003xa}).  The local gauge symmetry
$SU(2)_1 \times SU(2)_2 \times U(1)$ is gauged as the subgroup
of a global $SU(2)^3$ symmetry.}
\label{fg:moose-3site}
\end{figure}

The three site Higgsless model \cite{Foadi:2003xa,Perelstein:2004sc,
SekharChivukula:2006cg} is a nl$\sigma$m based on the global
$SU(2)^3 \to SU(2)$ symmetry breaking pattern, where the remaining 
$SU(2)$ plays the role of the custodial symmetry.  The gauged sub-group
is $SU(2)_1 \times SU(2)_2 \times U(1)$ and the symmetry breaking to 
the SM $SU(2)_L \times U(1)_Y$ is 
achieved by two bifundamental $\Sigma$ fields as depicted in the ``moose''
diagram shown in Fig.~\ref{fg:moose-3site} \footnote{The extended gauge
structure of this model is identical to that of the BESS model
\cite{Casalbuoni:1985kq,Casalbuoni:1986vq}}.

The nl$\sigma$m fields $\Sigma_{1,2}$ consist of two $SU(2)$ triplets
$\pi^a_i$ ($i = 1,2$):
\begin{eqnarray}
\Sigma_1 = \exp\biggl[\frac{2i\pi_1^aT^a}{f_1}\biggr]
\,\,\,\,\,,\,\,\,\,\,
\Sigma_2 = \exp\biggl[\frac{2i\pi_2^aT^a}{f_2}\biggr]\,,
\end{eqnarray}
which are coupled to the gauge fields through the covariant derivatives:
\begin{eqnarray}
D_\mu \Sigma_1 &=& \partial_\mu \Sigma_1 - ig^\prime T^3 B_\mu \Sigma_1
  + i \tilde{g} \Sigma_1 T^a W^a_{1,\mu}\,, \\
D_\mu \Sigma_2 &=& \partial_\mu \Sigma_2 - i\tilde{g} T^a W^a_{1,\mu} \Sigma_2
  + ig \Sigma_2 T^a W^a_{2,\mu}\,,
\end{eqnarray}
where $g^\prime$ is the gauge coupling of the $U(1)$, while $\tilde{g}$
and $g$ are the gauge couplings of $SU(2)_1$ and $SU(2)_2$, respectively.  

The effective Lagrangian for the three site model can be written as an 
expansion in derivatives (or momenta).  At lowest order (dimension-2), 
the relevant terms which obey the custodial symmetry are:
\begin{eqnarray}
{\cal{L}}_2 = \sum_{i=1}^2 \frac{f_i^2}{4} \mbox{Tr}[D^\mu \Sigma_i 
  (D_\mu \Sigma_i)^\dagger] - \frac{1}{4} \sum_{i=1}^2
W_i^{a,\mu\nu}W_{i,\mu\nu}^a - \frac{1}{4}B^{\mu\nu}B_{\mu\nu}\,.
\label{eq:L2}
\end{eqnarray}
In addition to these terms, there is one additional 
dimension-2 operator which violates the custodial symmetry:
\begin{eqnarray}
{\cal{L}}_2^\prime = \beta_{(2)}\,f_1^2\, \mbox{Tr}[\Sigma_1^\dagger
 (D_\mu \Sigma_1) T^3]\,\mbox{Tr}[\Sigma_1^\dagger(D^\mu \Sigma_1) T^3]\,,
\label{eq:L2-CT}
\end{eqnarray}
as well as dimension-4 operators that respect the symmetries of the
theory \cite{Chivukula:2007ic}:
\begin{eqnarray}
{\cal{L}}_4 &=& \alpha_{(1)1} \mbox{Tr}[W_{2,\mu\nu}\Sigma_2 W_1^{\mu\nu}
  \Sigma_2^\dagger] + \alpha_{(2)1} \mbox{Tr}[W_{1,\mu\nu}\Sigma_1 T^3 B^{\mu\nu}
  \Sigma_1^\dagger] \nonumber\\
\nonumber\\
&&- 2i\alpha_{(1)2} \mbox{Tr}[(D_\mu \Sigma_2)^\dagger (D_\nu \Sigma_2) W_1^{\mu\nu}]
- 2i\alpha_{(2)2} \mbox{Tr}[(D_\mu \Sigma_1)^\dagger (D_\nu \Sigma_1)T^3 B^{\mu\nu}]
\nonumber\\
\nonumber\\
&& + \sum_{i=1}^2 \biggl[
  -2i\alpha_{(i)3} \mbox{Tr}[W_i^{\mu\nu}(D_\mu \Sigma_i)(D_\nu \Sigma_i)^\dagger] +
  \alpha_{(i)4} \mbox{Tr}[(D_\mu \Sigma_i)(D_\nu \Sigma_i)^\dagger]
  \mbox{Tr}[(D^\mu \Sigma_i)(D^\nu \Sigma_i)^\dagger] \nonumber\\
\nonumber\\
&& \,\,\,\,\,\,\,\,\,\,\,\,\,\,\,\,\,\,
  + \alpha_{(i)5} \mbox{Tr}[(D_\mu \Sigma_i)(D^\mu \Sigma_i)^\dagger]
  \mbox{Tr}[(D_\nu \Sigma_i)(D^\nu \Sigma_i)^\dagger] \biggr]\,.
\label{eq:L4}
\end{eqnarray}
The coefficients of these terms act as counterterms for the divergences
which appear at one-loop order and serve to parameterize the effects of
unknown high-scale physics \cite{Appelquist:1980vg,Longhitano:1980iz,Longhitano:1980tm,Bagger:1992vu,
Perelstein:2004sc,Chivukula:2007ic}.
As we will discuss later, the $\beta_{(2)}$ 
coefficient contributes to the $T$ parameter while the $\alpha_{(i)1}$ 
coefficients are relevant to the $S$ parameter.

In unitary gauge ($\Sigma_{1,2} \to 1$), the kinetic energy terms for 
the $\Sigma$ fields in Eq.~(\ref{eq:L2}) only serve to give mass to the
various gauge fields.  Diagonalizing the resulting charged- and neutral-sector 
mass matrices, one finds that the spectrum consists of a (massless) photon, 
relatively light charged and neutral gauge bosons ($W$ and $Z$), as well as
heavy charged and neutral gauge bosons ($W^\prime$ and $Z^\prime$).
At this point, there are five free parameters in the model: 
$g,g^\prime, \tilde{g}, f_1$ and $f_2$.  For the purposes of our calculation, 
we find it useful to follow
Ref.~\cite{Foadi:2003xa} and exchange these parameters for the masses of the
light VGBs ($M_W$ and $M_Z$), the masses of the heavy VGBs ($M_{W^\prime}$
and $M_{Z^\prime}$) and the electromagnetic charge $e$.  The latter of which
is defined in this model to be:
\begin{eqnarray}
\frac{1}{e^2} = \frac{1}{g^2} + \frac{1}{\tilde{g}^2} + \frac{1}{g^{\prime 2}}
\,.
\label{eq:e-def}
\end{eqnarray}

The gauge fields can be expanded in terms of the mass eigenstates.
The charged fields can be written as:
\begin{eqnarray}
\label{eq:W1pm}
W_1^\pm &=& a_{11} W^{\prime \pm} + a_{12} W^\pm\,, \\
\label{eq:W2pm}
W_2^\pm &=& a_{21} W^{\prime \pm} + a_{22} W^\pm\,,
\end{eqnarray}
while the neutral fields are given by:
\begin{eqnarray}
\label{eq:B}
B &=& b_{00}\gamma + b_{01}Z^\prime + b_{02}Z\,, \\
\label{eq:W13}
W_1^3 &=& b_{10}\gamma + b_{11}Z^\prime + b_{12}Z\,, \\
\label{eq:W23}
W_2^3 &=& b_{20}\gamma + b_{21}Z^\prime + b_{22}Z\,.
\end{eqnarray}
Precise formulae for the gauge couplings, the decay constants ($f_1$ and 
$f_2$) and the mixing angles ($a_{ij}$ and $b_{ij}$) in terms of the masses 
of the gauge bosons can be found in Appendix \ref{app:3site-params}.

We can now make connection with the generic Feynman rules
for the 3- and 4-gauge boson interactions shown in Fig.~\ref{fg:feyn-rules}.  
Inserting Eqs.~(\ref{eq:W1pm})-(\ref{eq:W23}) into the 
gauge kinetic terms in Eq.~(\ref{eq:L2}), we find that the 3- and 4-point
couplings relevant to the calculation of the $S$ and $T$ parameters are
given by:
\begin{eqnarray}
\label{eq:coupWWV}
g_{W^- W^+ V^0_i} &=& g\,a_{22}^2\,b_{2i} + \tilde{g}\,a_{12}^2\,b_{1i}\,,\\
\label{eq:coupWWpV}
g_{W^- W^{\prime +} V^0_i} &=& g\,a_{21}\,a_{22}\,b_{2i} + 
  \tilde{g}\,a_{11}\,a_{12}\,b_{1i}\,,\\
\label{eq:coupWpWpV}
g_{W^{\prime -} W^{\prime +} V^0_i} &=& g\,a_{21}^2\,b_{2i} + 
  \tilde{g}\,a_{11}^2\,b_{1i}\,,
\end{eqnarray}
and:
\begin{eqnarray}
\label{eq:coupVVWW}
g_{V^0_i V^0_j W^- W^+} &=& g^2\,a_{22}^2\,b_{2i}\,b_{2j} +
  \tilde{g}^2\,a_{12}^2\,b_{1i}\,b_{2j}\,, \\
\label{eq:coupVVWpWp}
g_{V^0_i V^0_j W^{\prime -} W^{\prime +}} &=& g^2\,a_{21}^2\,b_{2i}\,b_{2j} +
  \tilde{g}^2\,a_{11}^2\,b_{1i}\,b_{2j}\,, \\
\label{eq:coupWWpWWp}
g_{W^+ W^{\prime +} W^- W^{\prime -}} &=& g^2\,a_{21}^2\,a_{22}^2 + 
  \tilde{g}^2\,a_{11}^2\,a_{12}^2\,, \\
\label{eq:coupWWWW}
g_{W^+ W^+ W^- W^-} &=& g^2\,a_{22}^4 + \tilde{g}^2\,a_{12}^4\,,
\end{eqnarray}
where ($V^0_0,V^0_1,V^0_2$) = ($\gamma,Z^\prime,Z$). 

Next, we consider the couplings of the fermions to the gauge fields.
In the simplest version of the three site model, the left-handed fermions
only couple directly to $SU(2)_2$, while the left- and right-handed
fermions couple directly to the $U(1)$ with charges $Y_L$ and $Y_R$, 
respectively \cite{Foadi:2003xa}.  In the language of deconstruction,
the fermions are {\it localized} on the first and third sites.
However, it has been shown that this setup leads to an unacceptably large
tree-level contribution to the $S$ parameter \cite{Foadi:2003xa,
Perelstein:2004sc,SekharChivukula:2006cg}.  A solution to this problem
is obtained by allowing the fermions to have a small (but non-zero)
coupling to the ``middle'' $SU(2)$ of Fig.~\ref{fg:moose-3site}
\cite{SekharChivukula:2006cg,Matsuzaki:2006wn}.  By appropriately tuning
the amount of ``delocalization'', one can reduce (or even cancel
altogether) the large contribution to the $S$ parameter from the tree
level.

The effective Lagrangian describing the coupling of fermions to gauge bosons
in the delocalized scenario is then given by:
\begin{eqnarray}
{\cal{L}}_f = g^\prime\,\bar{\psi}\,\gamma_\mu (Y_L P_L + Y_R P_R)B^\mu \psi
  + g\,(1-x_1)\,\bar{\psi}\,\gamma_\mu\,T^a W_2^{a,\mu} P_L \psi 
  + \tilde{g}\,x_1\,\bar{\psi}\,\gamma_\mu\,T^a W_1^{a,\mu} P_L \psi \,,
\label{eq:Lf}
\end{eqnarray}
where $P_{L,R}$ are projection operators:
\begin{equation}
P_{L,R} = \frac{1}{2}(1 \mp \gamma_5)\,,
\label{eq:PLR}
\end{equation}
and the parameter $x_1$ is a measure of the amount of fermion delocalization
$(0 \leq x_1 \ll 1)$ \cite{SekharChivukula:2006cg,Matsuzaki:2006wn}.
In principle, the value of $x_1$ for a given fermion species depends 
indirectly on the mass of the fermion.  This implies that, in general,
one should define a different $x_1$ for each fermion species.  However,
since we are only interested in light fermions (i.e., all SM fermions
except the top quark), we can safely neglect these differences and assume
that the amount of delocalization for all light fermions is the same
\cite{Matsuzaki:2006wn}.

Expressing the gauge fields in terms of the mass eigenstates using Eqs.
(\ref{eq:B}) and (\ref{eq:W23}), we can identify the couplings and 
coefficients used in our generic Feynman rules.  For example, the 
couplings for the charged-current interactions are given by:
\begin{eqnarray}
\label{eq:coupffWp}
g_{ff^\prime W^{\prime \pm}} &=& g\,(1-x_1)\,a_{21} + 
    \tilde{g}\,x_1\,a_{11}\,, \\
\label{eq:coupffW}
g_{ff^\prime W^{\pm}} &=& g\,(1-x_1)\,a_{22} +
    \tilde{g}\,x_1\,a_{12}\,. 
\end{eqnarray}
Next, the expressions for the neutral-current couplings and coefficients 
can be simplified by making the identification $Y_R = Y_L + T_f^3 = Q_f$.
In fact, we find:
\begin{eqnarray}
\label{eq:coupffV0}
g_{ff V_i^0} &=& g\,(1-x_1)\,b_{2i} + \tilde{g}\,x_1\,b_{1i} - 
   g^\prime\,b_{0i}\,,
\end{eqnarray}
and:
\begin{eqnarray}
\label{eq:gvV0}
g_{V_f}^{(V^0_i)} &=& \frac{1}{2}T_f^3 + \frac{g^\prime b_{0i}}
  {(g\,(1-x_1)\,b_{2i} + \tilde{g}\,x_1\,b_{1i} - g^\prime\,b_{0i})}\,Q_f \,,\\
\label{eq:gaV0}
g_{A_f}^{(V^0_i)} &=& -\frac{1}{2}T_f^3\,.
\end{eqnarray}

Having now specified the types of models we are interested in, let us
discuss the pinch technique in more detail as well as 
its application to models with extra dimensions and/or 
extended gauge symmetries. 

\section{Unitary Gauge and the PT}
\label{sec:Ugauge-and-PT}

As mentioned in the Introduction, vector-bosonic loop corrections to the VGB
self-energies suffer from two troublesome issues:
({\it i}) the final expressions are non-trivially dependent on the 
particular $R_\xi$ gauge used and ({\it ii}) use of the unitary gauge
($R_\xi \to \infty$) results in non-renormalizable terms.  The first 
issue has been studied in detail in Refs.~\cite{Degrassi:1992ff,Degrassi:1992ue,Papavassiliou:1994fp,Degrassi:1993kn}.  In this paper, we  
employ the unitary gauge in order to reduce the number of diagrams.  
Therefore, let us discuss the second issue and its resolution in more detail.

While the unitary gauge is known to result in renormalizable $S$-matrix
elements, Green's functions calculated in this gauge are 
{\it individually non-renormalizable}.  These terms are 
non-renormalizable in the sense
that they cannot be removed by the usual mass- and field-renormalization
counterterms.  To see how this arises, consider the form of the massive 
VGB propagator in unitary gauge:  
\begin{equation}
\label{eq:Dmunu}
D_{i}^{\mu\nu} = \frac{-i}{q^2 - M_i^2}\biggl[ g^{\mu\nu} - 
\frac{q^\mu q^\nu}{M_i^2} \biggr]\,\,.
\end{equation}
The problem arises in the limit $q^2 \to \infty$ where $D_{i} \sim 1$.
In this limit, one-loop amplitudes containing one or more propagators
of the form in Eq.~(\ref{eq:Dmunu}) become highly divergent.  In particular,
if dimensional regularization is applied, this divergent behavior 
manifests itself in poles proportional to higher powers of the external
momentum-squared ($q^2$) \cite{Papavassiliou:1994fp}.  For example, two-point 
functions calculated in unitary gauge contain poles proportional to $q^4$ 
and $q^6$.

The Pinch Technique supplies a solution to both the gauge-dependence
and the appearance of the $q^4$ and $q^6$ terms  
via a systematic algorithm which leads to the rearrangement of 
one-loop Feynman graphs contributing to a gauge-invariant and renormalizable
amplitude 
\cite{Cornwall:1981ru,Cornwall:1981zr,Cornwall:1989gv,Papavassiliou:1989zd}.  
The end results of the rearrangement are individually
gauge-independent propagator-, vertex- and box-like structures which are 
void of any higher powers of $q^2$.  In other words, propagator-like
or ``pinch'' terms coming from vertex and box corrections are isolated in 
a systematic manner and added to the self-energies.  These 
pinch pieces carry the exact 
gauge-dependent and non-renormalizable terms needed to cancel those of the 
two-point functions.  Finally, to construct gauge-invariant 
expressions for the oblique parameters at one-loop, one needs only replace
the various $\Pi_{ij}$ calculated from two-point diagrams with their PT
counterparts, $\Pi_{ij}^{\mbox{\tiny PT}}$ \cite{Degrassi:1993kn}.  In the 
following sections, we will 
demonstrate how to construct the PT self-energies in models with additional,
massive gauge bosons.

\begin{figure}[t]
\includegraphics[bb=70 642 507 752]{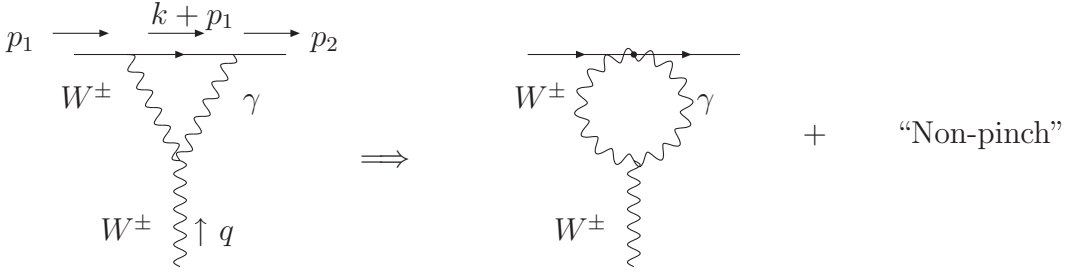}
\caption[]{Schematic example of the extraction of pinch pieces from
vertex corrections.}
\label{fg:pinchex}
\end{figure}

Before moving on to our results, though, let us first give a simple example 
of how
the pinch terms are isolated.  Consider the vertex diagram shown on the 
left side of Fig.~\ref{fg:pinchex} where the external and internal fermions 
are considered to be massless.  
When the $W$ propagator (with loop 
momentum $k$) is contracted with the $Wff^\prime$ vertex, a term arises of 
the form:
\begin{eqnarray}
{\cal{A}}_{V}^{\mu} &\sim& \int \frac{d^n k}{(2\pi)^n}
\bar{u}(p_2)\biggl\{ \cdots (\not{k} + \not{p_1})\not{k} \cdots
 \biggr\} u(p_1) \frac{1}{(k^2 - M_W^2)(k+p_1)^2(k-q)^2}\nonumber\\
\nonumber\\
&=& \int \frac{d^n k}{(2\pi)^n}
 \bar{u}(p_2)\biggl\{ \cdots (\not{k} + \not{p_1})
 ((\not{k}+\not{p_1})-\not{p_1}) \cdots
 \biggr\} u(p_1) \frac{1}{(k^2 - M_W^2)(k+p_1)^2(k-q)^2}\nonumber\\
\nonumber\\
&=& \int \frac{d^n k}{(2\pi)^n}
\bar{u}(p_2)\biggl\{ \cdots (k + p_1)^2 \cdots
 \biggr\} u(p_1) \frac{1}{(k^2 - M_W^2)(k+p_1)^2(k-q)^2} +
 \cdots\,\,.
\end{eqnarray}
In the second line, we have written the second factor of ${\not k}$ in 
terms of adjacent, inverse fermion propagators.
Canceling the factor of $(k+p_1)^2$ in the numerator and denominator, 
we see that the first term in the third line resembles
a correction to the $W$ propagator, i.e. it is {\it propagator-like}, 
and
can be represented schematically as shown on the right side of 
Fig.~\ref{fg:pinchex}.  This {\it pinch} term 
(along with others coming from other vertex and box corrections) is then 
combined 
with the loop-corrected two-point function to construct the self-energy
for the $W$ boson.

\section{One-loop Corrections to the Neutral Currents}
\label{sec:loops-neutral}

In this section, we outline the calculation of the one-loop corrections
needed to construct the self-energies for the neutral
gauge bosons using the Pinch Technique \cite{Degrassi:1992ff,Degrassi:1992ue}.
We write all 
amplitudes in terms of the generic couplings defined in 
Fig.~\ref{fg:feyn-rules} and reduce all tensor integrals to the 
usual Passarino-Veltman (P-V) tensor integral coefficients 
\cite{Passarino:1978jh} and scalar integrals defined in 
Appendix ~\ref{app:P-V}.

\begin{figure}[t]
\begin{center}
\includegraphics[bb=44 635 526 732]{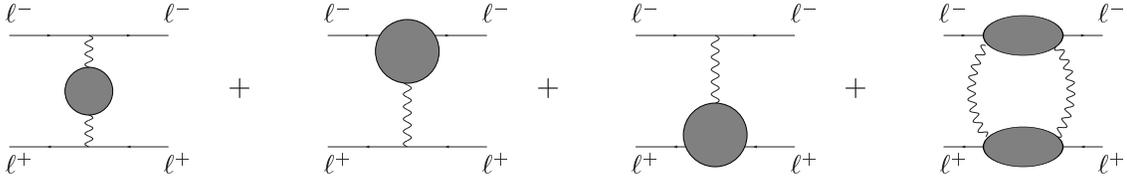} 
\end{center}
\caption[]{One-loop corrections to $t$-channel $\ell^-\ell^+$ scattering.
  From left to right, the corrections consist of one-loop corrections
  to the gauge boson propagator, corrections to the $V^0 \ell\ell$
  vertices and box corrections.}
\label{fg:blobs-neutral}
\end{figure}

The PT self-energies for the neutral VGBs are calculated in the context of 
four-fermion scattering, in particular $\ell^- \ell^+ \to \ell^- \ell^+$,
with all external (and internal) fermions considered
to be massless \footnote{It is straightforward to show that the results
given below are independent of the particular choice of four-fermion
scattering process}.  The one-loop corrections are shown schematically in 
Fig.~\ref{fg:blobs-neutral}.  In the following, we calculate the 
corrections to the gauge boson propagators, as well as the pinch pieces 
from both vertex and box corrections.    

\subsection{Corrections to the Gauge Boson Propagators}
\label{subsec:neutral-two-pt}

\begin{figure}[t]
\begin{center}
\includegraphics[bb=46 649 379 725]{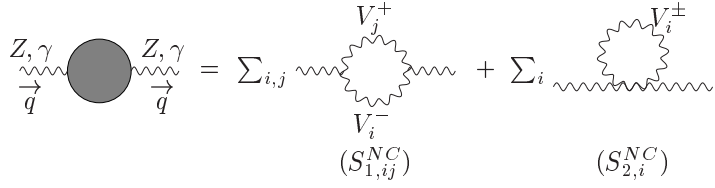} 
\end{center}
\caption[]{General corrections to the two-point functions of the 
           neutral gauge bosons.}
\label{fg:two-pt-neutral}
\end{figure}

The one-loop corrections to the neutral gauge boson propagators 
are shown in Fig.~\ref{fg:two-pt-neutral}.  In terms of the amplitudes
of these diagrams, the transverse two-point functions for the SM neutral 
gauge bosons can be constructed as:
\begin{eqnarray}
i\Pi_{mn}(q^2)\,g^{\mu\nu} = i \biggl[ \sum_{i,j}S_{1,ij}^{NC,\mu\nu} +
  \sum_{i}S_{2,i}^{NC,\mu\nu} \biggr]\,,
\label{eq:gen-Pi-neutral}
\end{eqnarray}
where $(mn) = (\gamma\gamma), (ZZ)$ or $(Z\gamma)$.  The structures of the 
individual amplitudes take the forms:
\begin{eqnarray}
\label{eq:S1nc}
S_{1,ij}^{NC,\mu\nu} &=& g_{V^-_i V^+_j \gamma(Z)}^2 \biggl[
  C_{S}^{(0)} + C_{S}^{(2)}\,q^2 + C_{S}^{(4)}\,q^4 \biggr]
  g^{\mu\nu}\,, \\
\nonumber\\
\label{eq:S2nc}
S_2^{NC,\mu\nu} &=& g_{\gamma\gamma(ZZ)V^-_i V^+_i} \biggl[
  -\frac{9}{2} + \frac{15}{4}\epsilon \biggr]\,A_0(M_i)\,g^{\mu\nu}\,,
\end{eqnarray}
where:
\begin{eqnarray}
\label{eq:C_S0}
C_{S}^{(0)} &=& \biggl(10 + \frac{M_i^2}{M_j^2} + \frac{M_j^2}{M_i^2}
                            - 8\epsilon\biggr)\,B_{22}(q^2;M_i,M_j) + 
                  \biggl(M_i^2 + M_j^2 - \frac{M_i^4}{M_j^2} - 
                         \frac{M_j^4}{M_i^2} \biggr)B_0(q^2;M_i,M_j)
\nonumber\\
&& \,\,\,\,\,\,\,\,+
                  \biggl(\frac{1}{4} - \frac{M_j^2}{M_i^2} +
                         \frac{\epsilon}{8} \biggr)A_0(M_i) +
                  \biggl(\frac{1}{4} - \frac{M_i^2}{M_j^2} +
                         \frac{\epsilon}{8} \biggr)A_0(M_j)\,,
\\
\nonumber\\
\label{eq:C_S2}
C_{S}^{(2)} &=& -2\biggl(\frac{1}{M_i^2} + \frac{1}{M_j^2}\biggr)
                     B_{22}(q^2;M_i,M_j) + 
                   2\biggl(2 + \frac{M_i^2}{M_j^2} + 
                     \frac{M_j^2}{M_i^2}\biggr)B_0(q^2;M_i,M_j)
\nonumber\\
&& \,\,\,\,\,\,\,\,+
	          \frac{1}{M_i^2}A_0(M_i) + \frac{1}{M_j^2}A_0(M_j)\,,
\\
\nonumber\\
\label{eq:C_S4}
C_{S}^{(4)} &=& \frac{1}{M_i^2 M_j^2} \biggl[
                   B_{22}(q^2;M_i,M_j) - (M_i^2 + M_j^2)B_0(q^2;M_i,M_j)
                   \biggr]\,.
\end{eqnarray}
In the above and the following, $A_0$ and $B_0$ represent the one- and 
two-point scalar integrals, respectively, while the $B_{ij}$'s represent
the P-V tensor integral coefficients \cite{Passarino:1978jh} (see
Appendix~\ref{app:P-V}).

\subsection{Pinch Contributions from Vertex Corrections}
\label{subsec:neutral-vertex}

\begin{figure}[t]
\begin{center}
\includegraphics[bb=37 517 544 729]{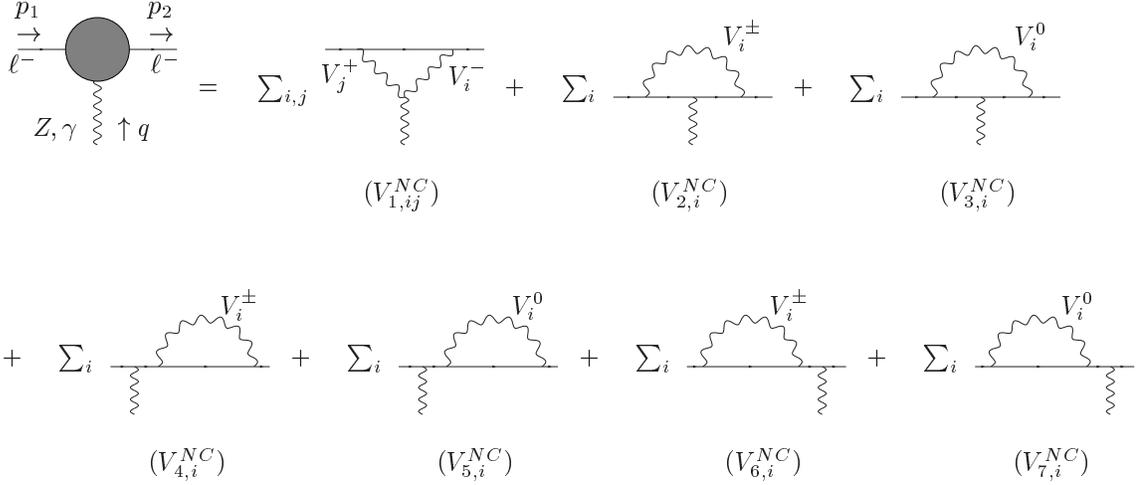} 
\end{center}
\caption[]{General one-loop corrections to the neutral gauge-boson-fermion 
           vertices and the external legs which give rise to pinch
           contributions.}
\label{fg:vertex-neutral}
\end{figure}

The one-loop vertex corrections for the neutral current are shown in 
Fig.~\ref{fg:vertex-neutral}.  Note that we have included 
the external leg corrections in addition to the traditional vertex 
corrections.  In Appendix~\ref{app:light-heavy-mixing}, we discuss 
additional corrections which can arise due to mixing between the light
and heavy gauge bosons at the one-loop level.  When the mass of the
heavy gauge boson is much larger than $q^2$, the corrections from these
mixings become ``pinch-like'' and should be combined with the corrections
from vertices and boxes.  The pinch contributions to the total amplitude 
from vertex corrections take the form:
\begin{eqnarray}
\Delta{\cal{A}}_{V,\gamma(Z)}^\mu |_{pinch} &=& \{V_{\gamma(Z)}\}\,
  [\bar{u}(p_i)\,\gamma^\mu(1-\gamma_5)\,u(p_i)] \nonumber\\
&\equiv&
\{V_{\gamma(Z)}\}\,\Gamma_W^\mu\,,
\label{eq:delta-Av}
\end{eqnarray}
where $\{V_{\gamma(Z)}\}$ represents the sum of the pinch contributions
calculated from the diagrams shown in Fig.~\ref{fg:vertex-neutral}
and $\Gamma_W$ is the current associated with the SM-like $W$.  
In applying the PT to the neutral currents, we find it useful to
rewrite $\Gamma_W$ in terms of the
currents associated with the SM-like $Z$ and photon.  For example, 
using Eqs.~(\ref{eq:coupffV0})-(\ref{eq:gaV0}), this structure 
in the three site model can be rewritten as:
\begin{eqnarray}
\Gamma_W^\mu \equiv \bar{u}(p_j)\gamma^\mu(1-\gamma_5)u(p_i) &=&
  \frac{2}{T^3_f}\,\bar{u}(p_j)\,\gamma^\mu\biggl[ \biggl( \frac{1}{2}
  T^3_f + \frac{g^\prime\,b_{02}}{(g\,(1-x_1)\,b_{22} 
  + \tilde{g}\,x_1\,b_{12} - g^\prime\,b_{02})}\,Q_f
 \biggr) \nonumber\\
\nonumber\\
&& \,\,\,\,\,\,\,\,\,\,  
  -  \frac{g^\prime\,b_{02}}{(g\,(1-x_1)\,b_{22} 
  + \tilde{g}\,x_1\,b_{12} - g^\prime\,b_{02})}\,Q_f
  - \frac{1}{2}T^3_f\,\gamma_5 \biggr]\,u(p_i) \nonumber\\
\nonumber\\
&=& \frac{2}{T^3_f}\,\bar{u}(p_j)\,\gamma^\mu \biggl[
  (g_{V_f}^{(Z)} + g_{A_f}^{(Z)}\,\gamma_5) \nonumber\\
\nonumber\\
&& \,\,\,\,\,\,\,\,\,\,  
+ 
  \frac{g^\prime\,b_{02}}{(g\,(1-x_1)\,b_{22} 
  + \tilde{g}\,x_1\,b_{12} - g^\prime\,b_{02})}\,Q_f
  \biggl]\,u(p_i) \nonumber\\
\nonumber\\
\label{eq:ZPhocurrent}
&\equiv& \frac{2}{T^3_f} \biggl[ \Gamma_Z^\mu + 
  \frac{g^\prime\,b_{02}}{(g\,(1-x_1)\,b_{22} 
  + \tilde{g}\,x_1\,b_{12} - g^\prime\,b_{02})}\,Q_f\,
  \Gamma_\gamma^\mu \biggr]\,,
\end{eqnarray}
where we have defined the currents associated with the SM $Z$ and photon
respectively as:
\begin{eqnarray}
\label{eq:Gamma_Z}
\Gamma_Z^\mu &=& \bar{u}(p_j)\,\gamma^\mu (g_{V_f}^{(Z)} + 
  g_{A_f}^{(Z)}\gamma_5)\,u(p_i)\,,\\
\nonumber\\
\label{eq:Gamma_pho}
\Gamma_\gamma^\mu &=& \bar{u}(p_j)\,\gamma^\mu\,u(p_i)\,.
\end{eqnarray}

Using the Feynman rules defined in Fig.~\ref{fg:feyn-rules} and reducing all
amplitudes to P-V tensor coefficients and scalar integrals, we find that the
pinch pieces from the individual diagrams shown in Fig.~\ref{fg:vertex-neutral}
are given by:
\begin{eqnarray}
\label{eq:V1nc}
\{V_{1,ij}^{NC}\} &=& -g_{\ell\nu V_i^\pm}\,g_{\ell\nu V_j^\pm}\,
    g_{\gamma(Z)V^+_j V^-_i}\,\,
    [\,C_{V}^{(0)} + C_{V}^{(2)}\,q^2\,]\\
\nonumber\\
\label{eq:V2nc}
\{V_{2,i}^{NC}\} &=& -g^2_{\ell\nu V_i^\pm}\,g_{\nu\nu \gamma(Z)}\,\,
    [g_{V_\nu}^{(\gamma,Z)} - g_{A_\nu}^{(\gamma,Z)}]\,\,
    \frac{A_0(M_i)}{M_i^2}\\
\nonumber\\
\label{eq:V3nc}
\{V_{3,i}^{NC}\}^\mu &=& g^2_{\ell\ell V_i^0}\,g_{\ell\ell \gamma(Z)}\,\,
    \frac{A_0(M_i)}{M_i^2}\,\bar{u}(p_2)\,\gamma^\mu
    \biggl[ (2g_{V_\ell}^{(V^0)}g_{A_\ell}^{(V^0)}g_{A_\ell}^{(\gamma,Z)}
    + (g_{V_\ell}^{(V^0)})^2 g_{V_\ell}^{(\gamma,Z)} 
    + (g_{A_\ell}^{(V^0)})^2 g_{V_\ell}^{(\gamma,Z)}) \nonumber\\
&&+
    \gamma_5(2g_{V_\ell}^{(V^0)}g_{A_\ell}^{(V^0)}g_{V_\ell}^{(\gamma,Z)}
    + (g_{V_\ell}^{(V^0)})^2 g_{A_\ell}^{(\gamma,Z)} 
    + (g_{A_\ell}^{(V^0)})^2 g_{A_\ell}^{(\gamma,Z)})\biggr]u(p_1)\\
\nonumber\\
\label{eq:V4nc}
\{V_{4,i}^{NC}\} &=& \frac{g_{\ell\nu V_i^\pm}^2\,g_{\ell\ell \gamma(Z)}}{2}\,
         (g_{V_\ell}^{(\gamma,Z)} - g_{A_\ell}^{(\gamma,Z)})
         \frac{A_0(M_i)}{M_i^2}\\
\nonumber\\
\label{eq:V5nc}
\{V_{5,i}^{NC}\}^\mu &=& -\frac{1}{2}\{V_{3,i}^{NC}\}^\mu \\
\nonumber\\
\label{eq:V6nc}
\{V_{6,i}^{NC}\} &=& \{V_{4,i}^{NC}\} \\
\nonumber\\
\label{eq:V7nc}
\{V_{7,i}^{NC}\}^\mu &=& \{V_{5,i}^{NC}\}^\mu = 
     -\frac{1}{2}\{V_{3,i}^{NC,\mu}\}\,,
\end{eqnarray}
where:
\begin{eqnarray}
\label{eq:C_V0}
C_{V}^{(0)} &=& \biggl(-3 + 2\epsilon\biggr)
    \biggl(\frac{1}{M_i^2} + \frac{1}{M_j^2}\biggr)B_{22}(q^2;M_i,M_j)
    + 2B_0(q^2,M_i,M_j) \\
\nonumber\\
\label{eq:C_V2}
C_{V}^{(2)} &=& \biggl[\frac{1}{M_j^2}B_{0}(q^2;M_i,M_j) 
  -\frac{2}{M_i^2}B_{11}(q^2;M_i,M_j) - \biggl(\frac{1}{M_i^2} + 
   \frac{1}{M_j^2}\biggr)B_{21}(q^2;M_i,M_j) \nonumber\\
&&\,\,\,\,\,\,\,\, - 
   \frac{1}{M_i^2 M_j^2}B_{22}(q^2;M_i,M_j) \biggr]\,.
\end{eqnarray}
Thus, we immediately see that the pinch pieces from the vertex corrections 
and external leg corrections containing a virtual, neutral gauge 
boson ($V^0$) cancel amongst themselves, i.e.:
\begin{equation}
\{V_{3,i}^{NC}\}^\mu + \{V_{5,i}^{NC}\}^\mu + \{V_{7,i}^{NC}\}^\mu = 0\,.
\label{eq:sumV3V5V7}
\end{equation}
Note that this is true regardless of the exact form of the couplings.
Finally, in terms of the above amplitudes, the total pinch contribution
from the vertex corrections is given by:
\begin{eqnarray}
\label{eq:Vtotal-nc}
\{V_{\gamma(Z)}\} = \sum_{i,j}\{V_{1,ij}^{NC}\} + \sum_i \biggl[
  \{V_{2,i}^{NC}\} + \{V_{4,i}^{NC}\} + \{V_{6,i}^{NC}\} \biggr]\,.
\end{eqnarray}

\subsection{Pinch Contributions from Box Corrections}
\label{subsec:neutral-box}

\begin{figure}[t]
\begin{center}
\includegraphics[bb=34 636 302 726]{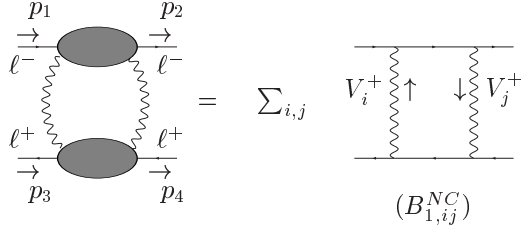} 
\end{center}
\caption[]{General one-loop box corrections from charged VGBs 
           which contain pinch contributions.}
\label{fg:box-neutral}
\end{figure}

The one-loop box diagrams which give rise to pinch contributions are depicted
in Fig.~\ref{fg:box-neutral}.  The total pinch amplitude arising from
these corrections can be written as:
\begin{eqnarray}
\Delta{\cal{A}}_{B,\gamma(Z)}^\mu |_{pinch} = \{B_{\gamma(Z)}\}\,
  \Gamma_W^\mu \, \Gamma_{W,\mu}\,,
\label{eq:delta-Ab}
\end{eqnarray}   
where $\{B_{\gamma(Z)}\}$ represents the sum of the pinch contributions
from the box diagrams.  
Individually, the amplitudes for these diagrams take the compact
form:
\begin{eqnarray}
\label{eq:B1nc}
\{B_{1,ij}^{NC}\} &=& \frac{g_{\ell\nu V_i^\pm}^2 g_{\ell\nu V_j^\pm}^2}
  {M_i^2 M_j^2}
  \biggl[B_{22}(q^2;M_i,M_j) - (M_i^2 + M_j^2)B_{0}(q^2;M_i,M_j)\biggr]\,.
\label{eq:Box1}
\end{eqnarray}
such that the total pinch contributions from box corrections 
$\{B_{1,ij}^{NC}\}$ is given by:
\begin{eqnarray}
\label{eq:Btotal-nc}
\{B_{\gamma(Z)}\} = \sum_{i,j} \{B_{1,ij}^{NC}\}\,.
\end{eqnarray}

\section{One-Loop Corrections to the Charged Current}
\label{sec:loops-charged}

\begin{figure}[t]
\begin{center}
\includegraphics[bb=50 634 523 728]{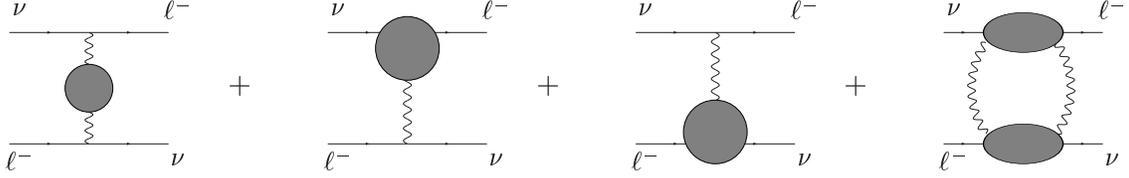} 
\end{center}
\caption[]{Schematic depiction of the one-loop corrections to $t$-channel
           $\nu\ell^-$ scattering.}
\label{fg:blobs-charged}
\end{figure}

In this section, we calculate the one-loop corrections needed to 
construct the $W$ boson self-energy using the PT \cite{Papavassiliou:1994fp}.  
The loop-corrected amplitudes are again calculated in the context of 
four-fermion scattering. In particular, we consider the one-loop
corrections to $\nu \ell^- \to \nu \ell^-$ which are schematically depicted in 
Fig.~\ref{fg:blobs-charged}.

\subsection{Corrections to the $W$ Boson Propagator}
\label{subsec:charged-two-pt}

\begin{figure}[t]
\begin{center}
\includegraphics[bb=62 625 550 720]{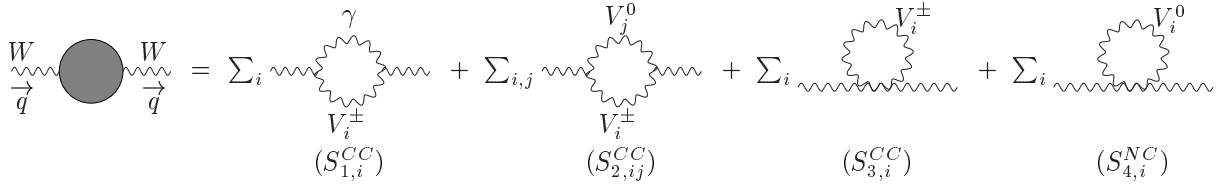} 
\end{center}
\caption[]{General corrections to the two-point functions of the 
           $W$ gauge boson.}
\label{fg:two-pt-charged}
\end{figure}

The one-loop corrections to the $W$ boson propagator are shown in
Fig.~\ref{fg:two-pt-charged}.  In terms of these diagrams, the transverse
two-point function of the $W$ boson is:
\begin{eqnarray}
i\Pi_{WW}(q^2)g^{\mu\nu} = i \biggl[ \sum_i S_{1,i}^{CC,\mu\nu} +
  \sum_{i,j} S_{2,ij}^{CC,\mu\nu} + \sum_i S_{3,i}^{CC,\mu\nu} +
  \sum_i S_{4,i}^{CC,\mu\nu}\biggr]\,. 
\label{eq:gen-Pi-charged}
\end{eqnarray}
Note that we have distinguished the photon
from the other neutral gauge bosons in Fig.~\ref{fg:two-pt-charged}.  
Since the photon is massless, the 
kinematic structures of these diagrams are slightly different than those 
for massive, neutral gauge bosons.  The amplitudes for all of these diagrams 
take compact forms:
\begin{eqnarray}
\label{eq:S1cc}
S_{1,i}^{CC,\mu\nu} &=& g_{W^- V_i^+ \gamma}^2 \biggl[
  K_S^{(0)} + K_S^{(2)}\,q^2 + K_S^{(4)}\,q^4 \biggr]\,g^{\mu\nu}\,, \\
\nonumber\\
\label{eq:S2cc}
S_{2,ij}^{CC,\mu\nu} &=& g_{W^- V_i^+ V_j^0}^2 \biggl[
  C_S^{(0)} + C_S^{(2)}\,q^2 + C_S^{(4)}\,q^4 \biggr]\,g^{\mu\nu}\,, \\
\nonumber\\
\label{eq:S3cc}
S_{3,i}^{CC,\mu\nu} &=& g_{W^+V_i^+ W^- V_i^-}\biggl[ 
  -\frac{9}{4} + \frac{15}{8}\epsilon \biggr]\,A_0(M_i)\,g^{\mu\nu}\,, \\
\nonumber\\
\label{eq:S4cc}
S_{4,i}^{CC,\mu\nu} &=& g_{V_i^0 V_i^0 W^- W^+}\biggl[ 
  -\frac{9}{4} + \frac{15}{8}\epsilon \biggr]\,A_0(M_i)\,g^{\mu\nu}\,,
\end{eqnarray}
where the $C_S^{(i)}$ coefficients are the same as those in 
Eqs.~(\ref{eq:C_S0})-(\ref{eq:C_S4}) and the $K_S^{(i)}$ coefficients
are given by:
\begin{eqnarray}
\label{eq:K_S0}
K_S^{(0)} &=& (10 - 8\epsilon)B_{22}(q^2;M_i,0) + M_i^2 B_0(q^2;M_i,0) +
  \biggl(\frac{1}{4} + \frac{\epsilon}{8}\biggr)A_0(M_i)\,,\\
\nonumber\\
\label{eq:K_S2}
K_S^{(2)} &=& -\frac{2}{M_i^2}B_{22}(q^2;M_i,0) + 4B_0(q^2;M_i,0) +
  \frac{A_0(M_i)}{M_i^2}\,,\\
\nonumber\\
\label{eq:K_S4}
K_S^{(4)} &=& -\frac{1}{M_i^2}B_0(q^2;M_i,0)\,.
\end{eqnarray}

\subsection{Pinch Contributions from Vertex Corrections}
\label{subsec:charged-vertex}

\begin{figure}[t]
\begin{center}
\includegraphics[bb=61 490 524 782]{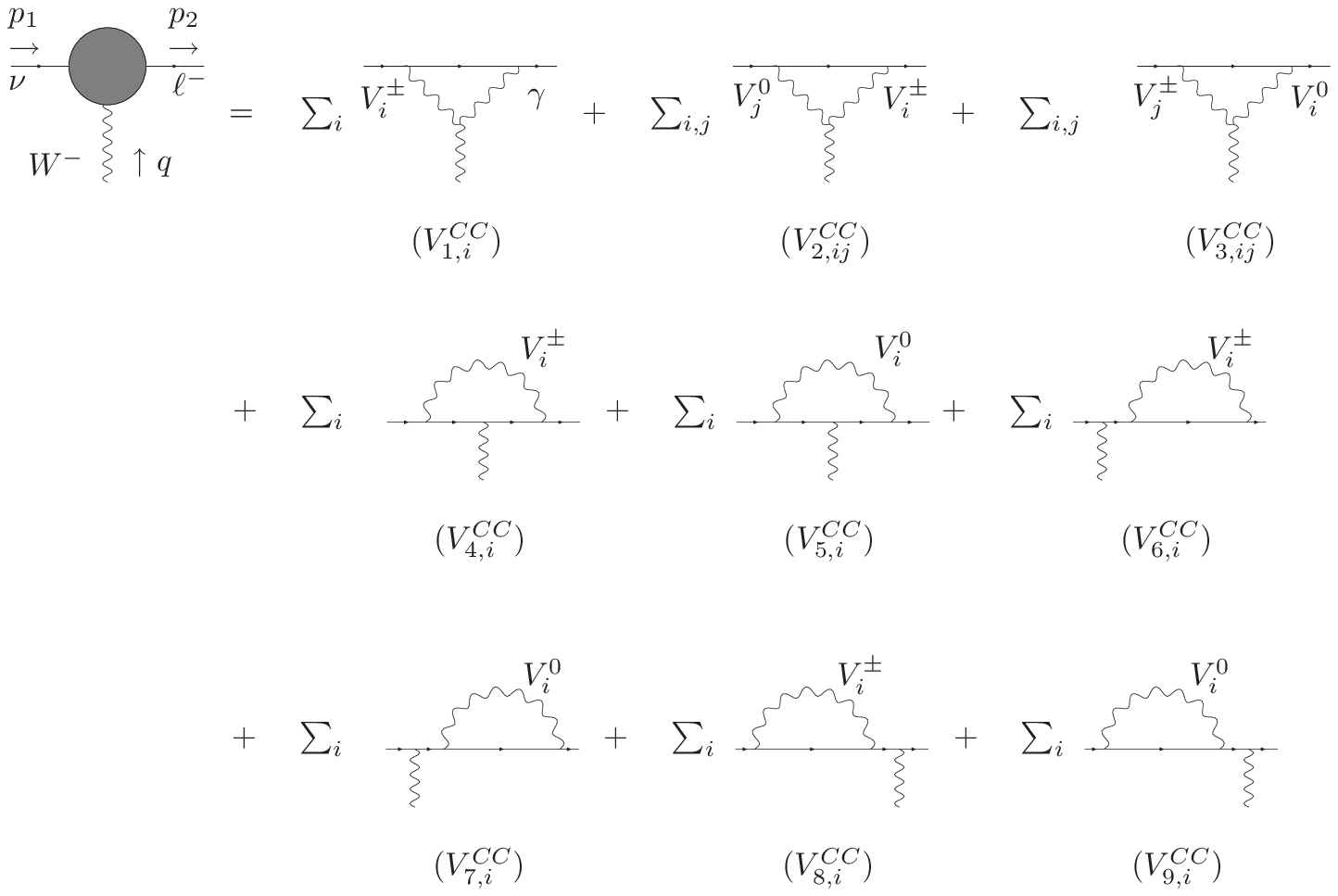} 
\end{center}
\caption[]{General one-loop corrections to the gauge-boson-fermion 
           vertices and external fermion legs which give rise to
           pinch contributions.}
\label{fg:vertex-charged}
\end{figure}

The one-loop vertex corrections which give rise to pinch contributions 
are shown in Fig.~\ref{fg:vertex-charged} \footnote{As mentioned earlier, 
corrections
which mix the light and heavy gauge bosons also give rise to pinch-like 
contributions as discussed in Appendix~\ref{app:light-heavy-mixing}.}.  
The amplitude structure of 
the vertex diagrams is very similar to the neutral
current amplitudes with the exception of diagrams $(V_{1,i}^{CC})$ 
and $(V_{2,i}^{CC})$.  The pinch contributions from the vertex corrections 
take the form:
\begin{eqnarray}
\Delta{\cal{A}}_{V,W}^\mu |_{pinch} = \{V_{W}\}\,
  \Gamma_W^\mu\,,
\label{eq:delta-Avw}
\end{eqnarray}
where $\{V_{W}\}$ is the sum of the pinch contributions from the diagrams
in Fig.~\ref{fg:vertex-charged} and $\Gamma_W$ is defined in 
Eq.~(\ref{eq:delta-Av}).  The individual amplitudes which contribute
to $\{V_{W}\}$ can be written as:
\begin{eqnarray}
\label{eq:V1cc}
\{V_{1,i}^{CC}\} &=& \frac{g_{\ell\nu V_i^\pm}\,g_{\ell\ell\gamma}\,
    g_{W^- V_i^+ \gamma}}
    {2\sqrt{2}}\,\,(g_{V_\ell}^{(\gamma)} - g_{A_\ell}^{(\gamma)})\,
    [\,K_{V}^{(0)} + K_{V}^{(2)}\,q^2\,]\,,\\
\nonumber\\
\label{eq:V2cc}
\{V_{2,ij}^{CC}\} &=& -\frac{g_{\ell\nu V_i^\pm}\,g_{\nu\nu V_j^0}\,
    g_{W^- V_i^+ V_j^0}}
     {2\sqrt{2}}\,\,
    (g_{V_\nu}^{(V^0)} - g_{A_\nu}^{(V^0)})\,
    [\,C_{V}^{(0)} + C_{V}^{(2)}\,q^2\,]\,,\\
\nonumber\\
\label{eq:V3cc}
\{V_{3,ij}^{CC}\} &=& \frac{g_{\ell\nu V_j^\pm}\,g_{\ell\ell V_i^0}\,
    g_{W^- V_j^+ V_i^0}}
     {2\sqrt{2}}\,\,
    (g_{V_\ell}^{(V^0)} - g_{A_\ell}^{(V^0)})\,
    [\,C_{V}^{(0)} + C_{V}^{(2)}\,q^2\,]\,,\\
\nonumber\\
\label{eq:V4cc}
\{V_{4,i}^{CC}\} &=& \frac{g_{\ell\nu V_i^\pm}^2 g_{\ell\nu W^\pm}}{2\sqrt{2}}
  \biggl(\frac{1}{2}\biggr)\,\frac{A_0(M_i)}{M_i^2}\,,\\
\nonumber\\
\label{eq:V5cc}
\{V_{5,i}^{CC}\} &=& \frac{g_{\nu\nu V_i^0}g_{\ell\ell V_i^0} 
  g_{\ell\nu W^\pm}}{2\sqrt{2}}
  (g_{V_\ell}^{(V^0)} - g_{A_\ell}^{(V^0)})
  (g_{V_\nu}^{(V^0)} - g_{A_\nu}^{(V^0)})
  \,\frac{A_0(M_i)}{M_i^2}\,,\\
\nonumber\\
\label{eq:V6cc}
\{V_{6,i}^{CC}\} &=& - \{V_{4,i}^{CC}\} \,,\\
\nonumber\\
\label{eq:V7cc}
\{V_{7,i}^{CC}\} &=& -\frac{g_{\ell\ell V^0}^2 g_{\ell\nu W^\pm}}{2\sqrt{2}}
  \biggl(\frac{1}{2}\biggr)(g_{V_\ell}^{(V^0)} - g_{A_\ell}^{(V^0)})^2\,
  \frac{A_0(M_i)}{M_i^2}\,,\\
\nonumber\\
\label{eq:V8cc}
\{V_{8,i}^{CC}\} &=& \{V_{6,i}^{CC}\} = - \{V_{4,i}^{CC}\}\,, \\
\nonumber\\
\label{eq:V9cc}
\{V_{9,i}^{CC}\} &=& -\frac{g_{\nu\nu V^0}^2 g_{\ell\nu W^\pm}}{2\sqrt{2}}
  \biggl(\frac{1}{2}\biggr)(g_{V_\nu}^{(V^0)} - g_{A_\nu}^{(V^0)})^2\,
  \frac{A_0(M_i)}{M_i^2}\,,
\end{eqnarray}
where the $C_V^{(i)}$ coefficients are given by Eqs.~(\ref{eq:C_V0}) and
(\ref{eq:C_V2}) and the $K_V^{(i)}$ coefficients are given by:
\begin{eqnarray}
\label{eq:K_V0}
K_V^{(0)} &=& \frac{1}{M_i^2}(-3+2\epsilon)B_{22}(q^2;M_i,0) + 
  2B_0(q^2;M_i,0)\,, \\
\nonumber\\
\label{eq:K_V2}
K_V^{(2)} &=& \frac{1}{M_i^2}\biggl[B_0(q^2;0,M_i) - B_{21}(q^2;0,M_i)
  \biggr]\,.
\end{eqnarray}

Finally, the total pinch contribution from the vertex corrections can then be
calculated by summing the above amplitudes:
\begin{eqnarray}
\{V_{W}\} &=& \sum_i \biggl[
  \{V_{1,i}^{CC}\} + \{V_{4,i}^{CC}\} + \{V_{5,i}^{CC}\} +
  \{V_{6,i}^{CC}\} + \{V_{7,i}^{CC}\}
+ \{V_{8,i}^{CC}\} +
  \{V_{9,i}^{CC}\} \biggr]\nonumber\\
\nonumber\\
&& +
\sum_{i,j}\biggl[\{V_{2,ij}^{CC}\} + \{V_{3,ij}^{CC}\}\biggr] \,.
\end{eqnarray}

\subsection{Pinch Contributions from Box Corrections}
\label{subsec:charged-box}

\begin{figure}[t]
\begin{center}
\includegraphics[bb=63 580 520 774]{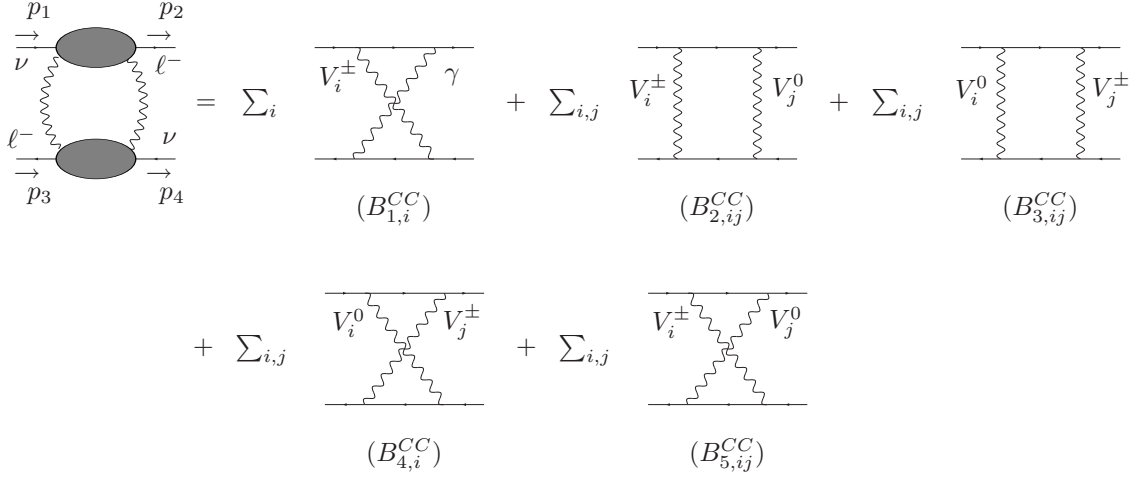} 
\end{center}
\caption[]{General one-loop box corrections to the charged current
           process.}
\label{fg:box-charged}
\end{figure}

The one-loop box corrections which contribute to the $W$ boson
PT self-energy are shown in Fig~\ref{fg:box-charged}.  Extracting
the pinch contributions, the amplitude from box corrections takes
the form:
\begin{eqnarray}
\Delta {\cal{A}}_{B,W}^\mu |_{pinch} = \{B_{W}\}\,\Gamma_W^\mu\,
  \Gamma_{W,\mu}\,,
\label{eq:Abw}
\end{eqnarray}
where $\{B_{W}\}$ represents the pinch piece of the total box amplitude.
Since the photon only couples to 
charged fermions, there is only one diagram involving a photon 
which gives a non-zero contribution to the total pinch amplitude:
\begin{eqnarray}
\{B_{1,i}^{CC}\} &=& -\frac{g_{\ell\ell\gamma}^2 g_{\ell\nu V_i^\pm}^2}
  {(2\sqrt{2})^2}
  \frac{(g_{V_\ell}^{(\gamma)} - g_{A_\ell}^{(\gamma)})^2}{M_i^2}\,
  B_0(q^2;M_i,0)\,.
\end{eqnarray}
The other four diagrams, those which contain a massive neutral gauge boson as 
well as a massive charged gauge boson, have kinematic structures identical
to $\{B_{1,i}^{NC}\}$ (Eq.~(\ref{eq:B1nc})).  In fact, we find:
\begin{eqnarray}
\label{eq:B2cc}
\{B_{2,ij}^{CC}\} &=& -\frac{g_{\ell\ell V_j^0} g_{\nu\nu V_j^0} 
  g_{\ell\nu V_i^\pm ff}^2}{(2\sqrt{2})^2}
  \frac{(g_{V_\ell}^{(V^0)}-g_{A_\ell}^{(V^0)})
        (g_{V_\nu}^{(V^0)}-g_{A_\nu}^{(V^0)})}{M_i^2 M_j^2} \nonumber\\
\nonumber\\
&& \,\,\,\,\,\,\,\,\,\,\,\times
  \biggl[B_{22}(q^2;M_i,M_j) - (M_i^2 + M_j^2)B_{0}(q^2;M_i,M_j)\biggr]\,,\\
\nonumber\\
\label{eq:B3cc}
\{B_{3,ij}^{CC}\} &=& \{B_{2,ij}^{NC}\} 
  \,\,\,\,\,\,\,\,\mbox{(with $i \leftrightarrow j$)}\,, \\
\nonumber\\
\label{eq:B4cc}
\{B_{4,ij}^{CC}\} &=& 
  -\frac{g_{\nu\nu V_i^0}^2 g_{\ell\nu V_j^\pm}^2}{(2\sqrt{2})^2}
  \frac{(g_{V_\nu}^{(V^0)}-g_{A_\nu}^{(V^0)})^2}{M_i^2 M_j^2} \nonumber\\
\nonumber\\
&& \,\,\,\,\,\,\,\,\,\,\,\times
  \biggl[B_{22}(q^2;M_i,M_j) - (M_i^2 + M_j^2)B_{0}(q^2;M_i,M_j)\biggr]\,,\\
\nonumber\\
\label{eq:B5cc}
\{B_{5,ij}^{CC}\} &=& 
  -\frac{g_{\ell\ell V_i^0}^2 g_{\ell\nu V_j^\pm}^2}{(2\sqrt{2})^2}
  \frac{(g_{V_\ell}^{(V^0)}-g_{A_\ell}^{(V^0)})^2}{M_i^2 M_j^2} \nonumber\\
\nonumber\\
&& \,\,\,\,\,\,\,\,\,\,\,\times
  \biggl[B_{22}(q^2;M_i,M_j) - (M_i^2 + M_j^2)B_{0}(q^2;M_i,M_j)\biggr]\,.
\end{eqnarray}
Finally, in terms of these amplitudes, the total pinch contribution from 
box corrections is given by:
\begin{eqnarray}
\{B_{W}\} &=& \sum_{i}\,\{B_{1,ij}^{CC}\} + \sum_{i,j} \biggl[\{B_{2,ij}^{CC}\}
 + \{B_{3,ij}^{CC}\} + \{B_{4,ij}^{CC}\} + \{B_{5,ij}^{CC}\} \biggr]\,.
\end{eqnarray}

\section{The Gauge Boson Self-energies in the PT}
\label{sec:PT-SEs}

In this section, we demonstrate how to construct the self-energies
for the SM-like gauge bosons using the various pieces calculated in the
previous sections.  We will do this first for a general model and then,
in the next section, apply our results to the three site model.
As stated earlier, we consider the process
$\ell^-(p_1) + \ell^+(p_3) \to \ell^-(p_2) + \ell^+(p_4)$ for the neutral
currents and the process
$\nu(p_1) + \ell^-(p_3) \to \ell^-(p_2) + \nu(p_4)$ for the charged
current where both the neutral and charged gauge bosons are 
exchanged in the $t$-channel as depicted in Figs.~\ref{fg:blobs-neutral} 
and \ref{fg:blobs-charged} respectively.  The results given below, however,
are independent of the particular process \cite{Degrassi:1992ue}.

\subsection{The Neutral Gauge Boson Self-energies}
\label{subsec:photonPT}

Let us begin by constructing the the PT self-energy for the photon. 
The tree-level amplitude for the $t$-channel exchange of a photon is given by:
\begin{eqnarray}
A^0_\gamma \equiv -\frac{i\,e^2}{q^2}\,\Gamma_\gamma^\mu \, 
   \Gamma_{\gamma,\mu}\,.
\label{eq:A0pho}
\end{eqnarray}
where we have made use of Eq.~(\ref{eq:Photon-coups}).
The amplitude from the loop-corrected photon propagator diagrams takes the 
form :
\begin{eqnarray}
A^S_\gamma = \biggl(-\frac{i\,e^2}{q^2}\,
  \Gamma_\gamma^\mu \, 
   \Gamma_{\gamma,\mu}\biggr)\,\frac{\Pi_{\gamma\gamma}}{q^2} 
\equiv A^0_\gamma \, \frac{\Pi_{\gamma\gamma}}{q^2}\,,
\label{eq:ASpho}
\end{eqnarray}
where $\Pi_{\gamma\gamma}$ represents the sum of the diagrams
contributing to the photon's two-point function 
as given by Eq.~(\ref{eq:gen-Pi-neutral}).  

Next, we consider the pinch pieces coming from the $\gamma\ell\ell$ 
vertex corrections.  In this
case, we sum the two middle diagrams of Fig.~\ref{fg:blobs-neutral} to 
find:
\begin{eqnarray}
A^V_\gamma = 2\,\biggl(-\frac{i\,e^2}{q^2}\,
  \Gamma_\gamma^\mu \, 
   \Gamma_{\gamma,\mu}\biggr)\,\{V_\gamma\} 
\equiv 2\,A^0_\gamma \, \{V_\gamma\}\,,
\label{eq:AVpho}
\end{eqnarray}
where $\{V_\gamma\}$ represents the pinch contributions from the diagrams
in Fig.~\ref{fg:vertex-neutral} plus any contributions from mixing between
the light and heavy gauge bosons.  The factor of two accounts
for the contribution from both $\gamma \ell\ell$ vertices.

Finally, for the box corrections, we have the amplitude:
\begin{eqnarray}
A^B_\gamma = \biggl(-\frac{i\,e^2}{q^2}\,
  \Gamma_\gamma^\mu \, 
   \Gamma_{\gamma,\mu}\biggr)\,\, q^2\{B_\gamma\} 
\equiv A^0_\gamma \, q^2\{B_\gamma\}\,,
\label{eq:ABpho}
\end{eqnarray}
where $\{B_\gamma\}$ represents the pinch contributions coming from the 
diagrams shown in Fig.~\ref{fg:box-neutral}.

Now, we can construct the photon's self-energy using
the PT.  Summing Eqs.~(\ref{eq:ASpho}), (\ref{eq:AVpho}) and 
(\ref{eq:ABpho}), we find for the PT loop-corrected amplitude
\cite{Degrassi:1992ff,Degrassi:1992ue}:
\begin{eqnarray}
{\cal{A}}^{one-loop}_{\gamma} &=& 
  \frac{{\cal{A}}^0_\gamma}{q^2}\biggl[\, \Pi_{\gamma\gamma} + 
   2\,q^2\,\{V_\gamma\} +
   q^4\,\{B_\gamma\} \,\biggr] \nonumber\\
\nonumber\\
&\equiv& \frac{{\cal{A}}^0_\gamma}{q^2}\,\, 
   \Pi^{\mbox{\tiny{PT}}}_{\gamma\gamma}\,.
\label{eq:PiPhoPho-PT}
\end{eqnarray}

The calculation of the PT self-energy for the $Z$ follows along the same
lines as that of the photon.  Tree-level exchange of a $Z$ boson in the
$t$-channel results in the amplitude:
\begin{eqnarray}
{\cal{A}}^0_Z \equiv \frac{i\,g_{\ell\ell Z}^2}{q^2-M_Z^2}\,\Gamma_Z^\mu \, 
   \Gamma_{Z,\mu}\,.
\label{eq:A0Z}
\end{eqnarray}
Then, in terms of Eq.~(\ref{eq:A0Z}), the amplitudes for the loop-corrected
$Z$ boson propagator, vertex and box diagrams are given respectively by:
\begin{eqnarray}
\label{eq:ASz}
{\cal{A}}^S_Z &=& \biggl(\frac{i\,g_{\ell\ell Z}^2}{q^2-M_Z^2}\,\Gamma_Z^\mu
  \, \Gamma_{Z,\mu}\biggr)\,\frac{\Pi_{ZZ}}
   {q^2-M_Z^2} 
\equiv {\cal{A}}^0_Z \, \frac{\Pi_{ZZ}}{q^2-M_Z^2}\,,\\
\nonumber\\
\label{eq:AVz}
{\cal{A}}^V_Z &=& 2\,\biggl(\frac{i\,g_{\ell\ell Z}^2}{q^2-M_Z^2}\,\Gamma_Z^\mu
  \,\Gamma_{Z,\mu}\biggr)\,\{V_Z\} 
\equiv 2\,{\cal{A}}^0_Z \, \{V_Z\}\,, \\
\nonumber\\
\label{eq:ABz}
{\cal{A}}^B_Z &=& \biggl(\frac{i\,g_{\ell\ell Z}^2}{q^2-M_Z^2}\,\Gamma_Z^\mu 
  \,\Gamma_{Z,\mu}\biggr)\,\, (q^2-M_Z^2)\{B_Z\} 
\equiv {\cal{A}}^0_Z\,(q^2-M_Z^2)\{B_Z\}\,.
\end{eqnarray}
where the quantities  $\Pi_{ZZ}$, $\{V_Z\}$ and $\{B_Z\}$ 
can be calculated using the results from Section~\ref{sec:loops-neutral}
and the factor of two in Eq.~(\ref{eq:AVz}) accounts for both
of the $Z\ell\ell$ vertices.
Summing Eqs.~(\ref{eq:ASz})-(\ref{eq:ABz}), the PT one-loop corrected amplitude
takes the form:
\begin{eqnarray}
{\cal{A}}^{one-loop}_{Z} &=& \frac{{\cal{A}}^0_Z}{q^2-M_Z^2}\,\, 
   \Pi^{\mbox{\tiny{PT}}}_{ZZ}\,,
\end{eqnarray}
where the $Z$ PT self-energy is given by 
\cite{Degrassi:1992ff,Degrassi:1992ue}:
\begin{eqnarray}
\Pi^{\mbox{\tiny{PT}}}_{ZZ} = \Pi_{ZZ} + 2\,(q^2-M_Z^2)\,\{V_Z\} + 
  (q^2-M_Z^2)^2\,\{B_Z\}\,.
\label{eq:PiZZ-PT}
\end{eqnarray}

The calculation of the PT $Z-\gamma$ mixing self-energy follows in complete
analogy to the cases of the photon and $Z$ self-energies with the exception
that there are no tree-level exchange diagrams.  The one-loop diagrams which 
mix the photon and $Z$ propagators give rise to the amplitude:
\begin{eqnarray}
A^S_{Z\gamma} = \biggl(\frac{i\,e\,g_{\ell\ell Z}}
  {q^2(q^2-M_Z^2)} \Gamma_Z^\mu \, \Gamma_{\gamma,\mu} \biggr)\,\Pi_{Z\gamma}
  \,.
\label{eq:ASzpho}
\end{eqnarray}
The pinch contributions from vertex corrections are found by summing the 
second and third diagrams in Fig.~\ref{fg:blobs-neutral}:
\begin{eqnarray}
A^V_{Z\gamma} = \biggl(\frac{i\,e\,g_{\ell\ell Z}}
  {q^2(q^2-M_Z^2)} \Gamma_Z^\mu \,\Gamma_{\gamma,\mu}\biggr)\,
 \biggl[ (q^2-M_Z^2)\{V_{Z\gamma}^{(1)}\} + q^2\{V_{Z\gamma}^{(2)}\} \biggr]
  \,,
\label{eq:AVzpho}
\end{eqnarray}
where $\{V_{Z\gamma}^{(1)}\}$ comes from the $\Gamma_Z^\mu$ 
pieces of the $\gamma \ell \ell$ vertex corrections and $\{V_{Z\gamma}^{(2)}\}$
comes from the $\Gamma_{\gamma,\mu}$ pieces of the $Z \ell \ell$
corrections (see Eqs.~(\ref{eq:delta-Av})-(\ref{eq:Gamma_pho})).
Lastly, the pinch contributions to the $Z-\gamma$ mixing arising from box 
corrections is: 
\begin{eqnarray}
A^B_{Z\gamma} = \biggl(\frac{i\,e\,g_{\ell\ell Z}}
  {q^2(q^2-M_Z^2)} \Gamma_Z^\mu \,
  \Gamma_{\gamma,\mu} \biggr)\,q^2(q^2-M_Z^2)\,\{B_{Z\gamma}\}\,.
\label{eq:ABzpho}
\end{eqnarray}

Thus, summing Eqs.~(\ref{eq:ASzpho}), (\ref{eq:AVzpho}) and 
(\ref{eq:ABzpho}), the $Z-\gamma$ mixing PT self-energy can be extracted
and we find \cite{Degrassi:1992ff,Degrassi:1992ue}:
\begin{eqnarray}
\Pi^{\mbox{\tiny{PT}}}_{Z\gamma} &=& 
 \Pi_{Z\gamma} + (q^2-M_Z^2)\{V_{Z\gamma}^{(1)}\} + q^2\{V_{Z\gamma}^{(2)}\} 
  + q^2(q^2-M_Z^2)\,\{B_{Z\gamma}\}\,.
\label{eq:PiZPho-PT}
\end{eqnarray}

\subsection{The $W$ Boson Self-energy}
\label{subsec:WPT}

We now consider the PT self-energy for the $W$ boson.  The amplitude
for tree-level $W$-exchange in t-channel $\nu\ell^-$ scattering is given
by:
\begin{eqnarray}
{\cal{A}}_W^0 \equiv \frac{i}{q^2-M_W^2}\,
  \biggl(\frac{\,g_{\ell\nu W^\pm}}{2\sqrt{2}}\biggr)^2\,
  \Gamma_W^\mu \, \Gamma_{W,\mu} \,.
\label{eq:A0W}
\end{eqnarray}

As in the neutral current cases, the one-loop corrections to the $W$
boson propagator, as well as the pinch contributions from the vertex
and box corrections, are proportional to the tree-level amplitude 
${\cal{A}}_W^0$:
\begin{eqnarray}
\label{eq:ASw}
{\cal{A}}^S_W &=& \biggl[\frac{i}{q^2-M_W^2}\,
  \biggl(\frac{g_{\ell\nu W^\pm}}{2\sqrt{2}}\biggr)^2\,
   \Gamma_W^\mu \, \Gamma_{W,\mu}\biggr]\,\frac{\Pi_{WW}}
   {q^2-M_W^2} 
\equiv {\cal{A}}^0_W \, \frac{\Pi_{WW}}{q^2-M_W^2}\,,\\
\nonumber\\
\label{eq:AVw}
{\cal{A}}^V_W &=& 2\,\biggl[\frac{i}{q^2-M_W^2}\,
  \biggl(\frac{g_{\ell\nu W^\pm}}{2\sqrt{2}}\biggr)^2\,
   \Gamma_W^\mu \, \Gamma_{W,\mu}\biggr]\,\{V_W\} 
\equiv 2\,{\cal{A}}^0_W \, \{V_W\}\,,\\
\nonumber\\
\label{eq:ABw}
{\cal{A}}^B_W &=& \biggl[\frac{i}{q^2-M_W^2}\,
  \biggl(\frac{g_{\ell\nu W^\pm}}{2\sqrt{2}}\biggr)^2\,
   \Gamma_W^\mu \, \Gamma_{W,\mu}\biggr]\,\{B_W\} 
\equiv {\cal{A}}^0_W \, \{B_W\}\,.
\end{eqnarray}
where the factor of two in ${\cal{A}}^V_W$ accounts for both 
loop-corrected $W\nu\ell$ vertices. 
Then, summing 
Eqs.~(\ref{eq:ASw})-(\ref{eq:ABw}), the PT one-loop corrected amplitude
is given by:
\begin{eqnarray}
{\cal{A}}^{one-loop}_{W} &=& \frac{{\cal{A}}^0_W}{q^2-M_W^2}\,\, 
   \Pi^{\mbox{\tiny{PT}}}_{WW}\,.
\end{eqnarray}
where the $W$ PT self-energy is defined to be
\cite{Degrassi:1992ff,Degrassi:1992ue,Papavassiliou:1994fp}:
\begin{eqnarray}
\Pi^{\mbox{\tiny{PT}}}_{WW} \equiv \Pi_{WW} + 2\,(q^2-M_W^2)\,\{V_W\} + 
  (q^2-M_W^2)^2\,\{B_W\}\,.
\label{eq:PiWW-PT}
\end{eqnarray}
%

\subsection{The $S$ and $T$ Parameters in the PT}
\label{subsec:S-and-T-PT}

Finally, having constructed the PT expressions for the 
self-energies, we can calculate the one-loop
corrections to the oblique parameters \cite{Peskin:1991sw}.  Since most 
experimental analyses require $U=0$ \cite{Yao:2006px}, we will focus on 
the calculation of the $S$ and $T$ parameters.

In the PT framework, gauge-invariant expressions for the oblique parameters
are constructed by replacing the self-energies calculated from two-point 
functions alone by their PT counterparts \cite{Degrassi:1993kn}.  In other
words, using the standard definitions of the $S$ and $T$ parameters from
Ref.~\cite{Peskin:1991sw}, the PT versions of the $S$ and $T$ parameters
are:
\begin{equation}
\frac{\alpha S}{4 s_w^2 c_w^2} =
  \Pi_{ZZ}^{\mbox{\tiny{PT}}\prime}(0) - 
  \Pi_{\gamma\gamma}^{\mbox{\tiny{PT}}\prime}(0) - 
  \frac{c_w^2 - s_w^2}{s_w c_w}
  \Pi_{Z\gamma}^{\mbox{\tiny{PT}}\prime}(0)\,,
\label{eq:gen-S1loop}
\end{equation}  
and:
\begin{equation}
\alpha T = \frac{\Pi_{WW}^{\mbox{\tiny{PT}}}(0)}{M_W^2} - 
           \frac{\Pi_{ZZ}^{\mbox{\tiny{PT}}}(0)}{M_Z^2}\,,
\label{Tparam}
\end{equation}
where primes indicate the derivative with respect to $q^2$ and the PT
self-energies $\Pi_{\gamma\gamma}^{\mbox{\tiny{PT}}}(q^2), 
\Pi_{ZZ}^{\mbox{\tiny{PT}}}(q^2),\Pi_{Z\gamma}^{\mbox{\tiny{PT}}}(q^2)$ 
and $\Pi_{WW}^{\mbox{\tiny{PT}}}(q^2)$ are given by 
Eqs.~(\ref{eq:PiPhoPho-PT}), (\ref{eq:PiZZ-PT}), (\ref{eq:PiZPho-PT}) 
and (\ref{eq:PiWW-PT}), respectively. 

In the following numerical analysis, we define $c_w$ and $s_w$ to take 
their on-shell values, i.e.:
\begin{eqnarray}
\label{eq:cw}
c_w^2 &=& \frac{M_W^2}{M_Z^2} \\
\label{eq:sw}
s_w^2 &=& 1 - \frac{M_W^2}{M_Z^2}\,,
\end{eqnarray}
while we take the other parameters to be \cite{Yao:2006px}:
\begin{eqnarray}
\alpha^{-1}(M_Z) &=& 127.904 \\
M_W &=& 80.450 \,\,\,\mbox{GeV}\\
M_Z &=& 91.1874 \,\,\,\mbox{GeV}\,. 
\end{eqnarray}

\section{Results for the Three Site Higgsless Model}
\label{sec:ST-higgsless}

\begin{figure}[t]
\begin{center}
\includegraphics[scale=0.5]{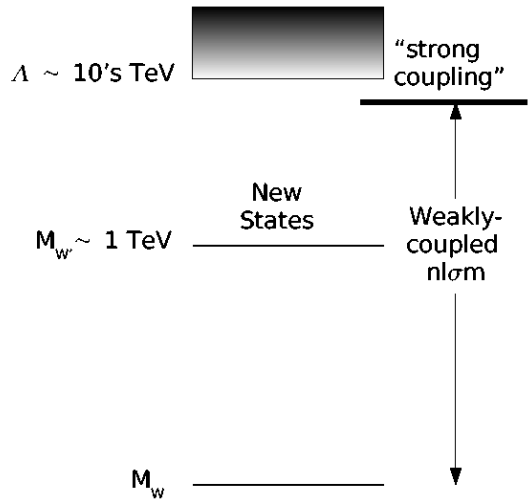} 
\end{center}
\caption[]{Fundamental scales of the three site model which are relevant
           to the calculation of the chiral-logarithmic corrections to
           the $S$ and $T$ parameters.}
\label{fg:scales-3site}
\end{figure}

In this section, we calculate the one-loop, chiral-logarithmic corrections
to the $S$ and $T$ parameters in the three site Higgsless model.  
To first approximation, the three site model contains three {\it fundamental} 
scales as depicted in Fig.~\ref{fg:scales-3site}: the mass of the
SM-like $W$, the mass of the heavy charged gauge boson $W^\prime$
\footnote{We assume that the mass splitting between the $W^\prime$
and the $Z^\prime$ is small compared to the differences between the 
scales depicted in Fig.~\ref{fg:scales-3site}.} 
and the cutoff scale of the effective theory $\Lambda$.  In order to 
estimate the size of the one-loop contributions in this model, we assume 
that the hierarchy is such that $M_W^2 \ll M_{W^\prime}^2 \ll \Lambda^2$.
In this scenario, contributions to the one-loop corrected $S$ and $T$ 
parameters are then dominated by the leading chiral logarithms and any constant
terms may safely be neglected \cite{Li:1971vr,Ling-Fong:1972hu,
Langacker:1973hh}.

To extract the leading chiral logarithms, we apply the following algorithm.  
First, all tensor integral coefficients are written in terms of scalar 
integrals as given by Eqs.~(\ref{eq:B11-coeff})-(\ref{eq:B22-coeff}) 
in Appendix~\ref{app:P-V} \cite{Passarino:1978jh}.  Then, using 
Eqs.~(\ref{eq:A0-chiral}) and (\ref{eq:B0-chiral}), the poles in $\epsilon$
are identified with the appropriate chiral logarithms.  In particular,
chiral logarithms coming from diagrams which contain only light, SM-like
particles are scaled from the cutoff $\Lambda$ down to $M_W$, while
poles originating from diagrams which contain at least one heavy VGB
(either $W^\prime$ or $Z^\prime$) are identified with the logarithm
$\log(\Lambda^2/M_{W^\prime}^2)$.
%
%

Finally, in the limit $M_W^2 \ll M_{W^\prime}^2$, the couplings of the 
$U(1)$ and the $SU(2)_2$ gauge groups reduce to the corresponding 
SM values (up to corrections of ${\cal{O}}(M_W^2/M_{W^\prime}^2)$)
\cite{Foadi:2003xa,SekharChivukula:2006cg}:
\begin{eqnarray}
g^\prime \simeq \frac{e}{c_w}\,\,\,\,\,,\,\,\,\,\,g \simeq \frac{e}{s_w}\,,
\end{eqnarray}
where we have used the tree-level definitions for $c_w$ and $s_w$ given by 
Eqs.~(\ref{eq:cw}) and (\ref{eq:sw}).  The chiral-logarithmic corrections to
the $S$ and $T$ parameters for the three site model with delocalized 
fermions in the limit $M_W \ll M_{W^\prime}^2$ have been previously 
calculated in Feynman gauge ($\xi = 1$) 
\cite{Matsuzaki:2006wn} and Landau gauge ($\xi=0$) \cite{Chivukula:2007ic}
with identical results.  In the present work, we use the exact expressions
for the gauge couplings and mixing angles as given in 
Appendix~\ref{app:3site-params}.  By using the exact expressions
for these parameters, our one-loop results retain
sub-leading terms in $M_W^2/M_{W^\prime}^2$ which can be important for 
smaller values of $M_{W^\prime}$ (and $M_{Z^\prime}$).  We have checked that
our (unitary gauge) results in the limit $M_W^2 \ll M_{W^\prime}^2$
agree with those of Refs.~\cite{Matsuzaki:2006wn} and \cite{Chivukula:2007ic}, 
thus proving the gauge-independence of our calculation\footnote{However,
this agreement is only achieved for the particular choice 
$M_{Z^\prime}^2 = M_{W^\prime}^2 + (M_Z^2 - M_W^2)$.}.

\subsection{The $S$ Parameter}
\label{subsec:S-higgsless}

In the three site model with localized fermions, the $S$ parameter receives
large corrections at tree level \cite{Foadi:2003xa,Perelstein:2004sc}.  
As alluded to earlier, however, this 
problem can be alleviated by allowing the light fermions to have a small
coupling to the middle $SU(2)$ of Fig.~\ref{fg:moose-3site}.  In this 
situation, the tree-level contribution to the $S$ parameter is given
by \cite{Chivukula:2005bn,Anichini:1994xx}:
\begin{equation}
\alpha S_{tree} = \frac{4 s_w^2 M_W^2}{M_{W^\prime}^2}
  \biggl(1 - \frac{x_1\,M_{W^\prime}^2}{2\,M_W^2}\biggr)
  + {\cal{O}}\biggl(\frac{M_W^4}{M_{W^\prime}^4}\biggr)\,.
\label{eq:Stree}
\end{equation}
For localized fermions $(x_1 = 0)$, where the fermions only directly couple
to the end gauge groups of the moose diagram, $S_{tree}$ can only be made to 
agree
with constraints from experimental data for very large values of 
$M_{W^\prime}$ ($\sim 2-3$ TeV).  This spoils the restoration of unitarity
in $V_L V_L$ scattering (where $V = W, Z$) which requires 
$M_{W^\prime, Z^\prime} \leq 1.5$ TeV.  However, from Eq.~(\ref{eq:Stree}),
we see that delocalizing the fermions provides a negative contribution
to $S_{tree}$
which reduces the overall value at tree level.  In fact, for 
$x_1 = 2 M_W^2/M_{W^\prime}^2$, the tree-level contribution to $S$
completely vanishes, a situation which is referred to as {\it ideal 
delocalization} \cite{SekharChivukula:2005xm}.  Thus, assessing the 
one-loop contributions to the $S$ parameter in the three site model becomes 
an important issue.  In particular, the one-loop results are useful to 
answer an important question in Higgsless models: is there a unique 
choice for $x_1$ which is ideal for all orders or, in the case that the
one-loop corrections are large, does $x_1$ need to be
tuned order-by-order in perturbation theory in order to keep the value of $S$
within experimental limits.  We will address this issue in the following.

Using the generic results for the PT self-energies from the previous sections
and identifying poles in $\epsilon$ with the appropriate chiral logarithms
as discussed above, the one-loop-corrected $S$ parameter in the 
three site model can be written as:
\begin{eqnarray}
S_{3-site} &=& S_{tree} + A_W^S\,\log\biggl(\frac{\Lambda^2}{M_W^2}
 \biggr) + 
  A_{W^\prime}^S\,\log\biggl(\frac{\Lambda^2}{M_{W^\prime}^2}
 \biggr) + S_0 \nonumber\\
\nonumber\\
&=& S_{tree} + A_W^S\,\log\biggl(\frac{M_{W^\prime}^2}{M_W^2}
 \biggr) + 
  (A_{W^\prime}^S + A_W^S)\,\log\biggl(\frac{\Lambda^2}{M_{W^\prime}^2}
 \biggr) + S_0 
\label{eq:S-3site}
\end{eqnarray}
where, in the second line, the second term represents the contributions 
from the low-energy
region (below $M_{W^\prime}$), the third term comprises the 
high-energy contributions and $S_0$ represents contributions from 
higher-dimension operators.  Specifically, $S_0$ arises from the first
two operators of Eq.~(\ref{eq:L4}).  Inserting the expressions for the
gauge fields in terms of the mass eigenstates (Eqs.(\ref{eq:B})-
(\ref{eq:W23})) into these operators, we can isolate shifts to the 
kinetic energy terms of the mass eigenstates.  The effective
Lagrangian describing these shifts takes the form \cite{Csaki:2005vy}:
\begin{eqnarray}
{\cal{L}}_{S_0} = - \frac{A}{4}\,F_{\mu\nu}F^{\mu\nu} 
  - \frac{C}{4}\,Z_{\mu\nu}Z^{\mu\nu} 
  + \frac{G}{2}\,F_{\mu\nu}Z^{\mu\nu}\,,     
\label{eq:L-deltaS}
\end{eqnarray}
where $F_{\mu\nu}$ and $Z_{\mu\nu}$ are the usual Abelian field strengths
and the coefficients $A, C$ and $G$ in the three site model are found
to be:
\begin{eqnarray}
A &=& -2( \alpha_{(1)1}\,b_{10}\,b_{20} + \alpha_{(2)1}\,b_{00}\,b_{10})\,, \\
C &=& -2( \alpha_{(1)1}\,b_{12}\,b_{22} + \alpha_{(2)1}\,b_{02}\,b_{12})\,, \\
G &=& \alpha_{(1)1}\,(b_{10}\,b_{22} + b_{12}\,b_{20}) + 
      \alpha_{(2)1}\,(b_{00}\,b_{12} + b_{02}\,b_{10})\,.
\label{eq:A-C-G}
\end{eqnarray} 
Finally, in terms of these coefficients, the contribution to $S$ from 
higher-dimension operators is \cite{Csaki:2005vy}:
\begin{eqnarray}
S_0 = \frac{4 s_w^2 c_w^2}{\alpha} \biggl[
  A - C - \frac{c_w^2 - s_w^2}{s_w c_w}\,G \biggr]\,.
\label{eq:S0}
\end{eqnarray}
Thus, as stated earlier, the coefficients $\alpha_{(i)1}$ serve as 
counterterms which absorb the logarithmic divergences of 
Eq.~(\ref{eq:S-3site}), namely the $\log(\Lambda^2/M_{W^\prime}^2)$
terms.  In other words, these coefficients parameterize
the effects of unknown physics above the scale $\Lambda$.  Since we
are mainly interested in studying the behavior of the one-loop results,
we will set $S_0$ to zero in the following analysis.

\begin{figure}[t]
\begin{center}
\includegraphics[scale=0.5,angle=-90]{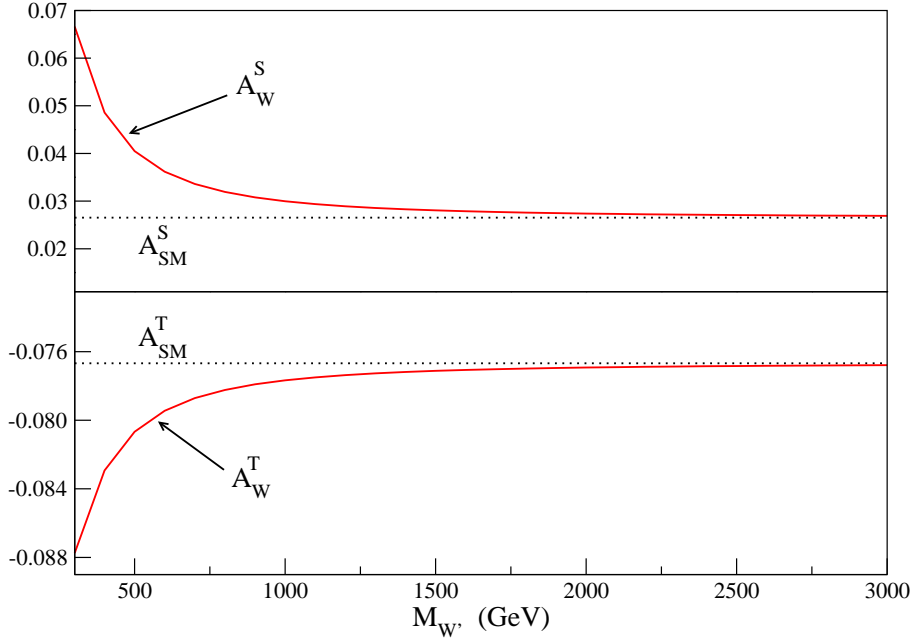} 
\end{center}
\caption[]{Top (Bottom) Panel: Coefficient for the low-energy contribution 
           to $S_{3-site}$ ($T_{3-site}$) as a function of the $W^\prime$ 
           mass.  These plots assume ideal delocalization of the fermions
           and $M_{Z^\prime}^2 = M_{W^\prime}^2 + (M_Z^2 - M_W^2)$.}
\label{fg:AS-and-AT}
\end{figure}

At this point, an important check on our calculation is the numerical value 
of the coefficient ($A_W^S$) of the low-energy contribution.  At energies
well below 
$M_{W^\prime}$, the symmetries of the three site model are the same as 
those of the SM with a heavy Higgs boson \cite{Longhitano:1980iz,Longhitano:1980tm}.  
This implies that the dimension-two interactions in the two models at 
low energy 
are identical which, in turn, requires that the chiral-logarithmic corrections 
calculated from these interactions take the same form in the two theories \cite{Bando:1984ej,Bando:1985rf,Bando:1987ym,Bando:1987br}.  One would therefore
expect that, in the limit that $M_W^2 \ll M_{W^\prime}^2$, the coefficient
of the low-energy contribution in the three site model should reduce to the 
value one would obtain 
in the SM with a heavy Higgs boson, i.e. \cite{Peskin:1991sw,Herrero:1993nc,Herrero:1994iu,Dittmaier:1995ee,Matias:1996hf,Alam:1997nk}:
\begin{equation}
A_{SM}^S = \frac{1}{12\pi}\,.
\label{eq:AS-SM}
\end{equation}
In the top panel of Fig.~\ref{fg:AS-and-AT}, we plot $A_W^S$ as a function
of $M_{W^\prime}$ assuming ideal delocalization of the light fermions
and that the mass of the $Z^\prime$ satisfies the relation $M_{Z^\prime}^2 = 
M_{W^\prime}^2 + (M_Z^2 - M_W^2)$.  Clearly, 
$A_W^S$ saturates at the SM value for masses 
$M_{W^\prime} \simeq$ 1.5 TeV such that we find it useful to rewrite 
$A_W^S$ as:
\begin{equation}
A_W^S = \frac{1}{12\pi} + \kappa_S\,,
\end{equation}
where $\kappa_S$ represents the contributions which decouple in the 
$M_{W^\prime} \to \infty$ limit \footnote{We have checked that this
agreement is independent of the particular choice of $M_{Z^\prime}$
and $x_1$.}.  It is interesting to note, however, that 
the sub-leading terms in $M_W^2/M_{W^\prime}^2$ can have a significant
impact for masses in the 300-700 GeV range leading to differences of a
factor of two or so.

Precision electroweak data can now be used to constrain $S_{3-site}$ and,
consequently, some of the relevant parameters (e.g., $M_{W^\prime}$, 
$\Lambda$ or the $\alpha_{(i)1}$ coefficients).  However, this process
is complicated by the fact that most global analyses are performed in
the context of the SM with a fundamental Higgs boson.  The 
physically-allowed region for $S$ (and $T$) is extracted in these analyses
by performing a $\chi^2$ fit to fourteen precisely measured electroweak 
observables.  For the case of a heavy Higgs boson, however, these analyses 
can be easily converted to a Higgsless scenario \cite{Bagger:1999te,
Chivukula:2000px}.  This is accomplished by first subtracting the leading
chiral-logarithmic contribution from a heavy Higgs boson 
\cite{Peskin:1991sw,Herrero:1993nc,Herrero:1994iu,Dittmaier:1995ee,
Matias:1996hf,Alam:1997nk,Chivukula:1999az}:
\begin{equation}
S_{Higgs} = \frac{1}{12\pi}
   \log\biggl(\frac{M_H^2}{M_W^2}\biggr)\,,
\label{eq:S-Higgs}
\end{equation}
and then adding back in the contribution from Eq.~(\ref{eq:S-3site}).
Thus, the value of the $S$ parameter to be used in the
$\chi^2$ fit is simply given by:
\begin{eqnarray}
S(S_0,M_{W^\prime},\Lambda) &=& S_{ref}(M_H^{ref}) - S_{Higgs} + 
  S_{3-site} \nonumber\\
&\equiv& [S_{ref}(M_H^{ref})-S_{Higgs}] + S_{1-loop} + S_0\,,
\label{eq:S-chi2}
\end{eqnarray}
where $S_{ref}(M_H^{ref})$ is the SM $S$ parameter as a function of the 
reference Higgs boson mass, $M_H^{ref}$.  In principle, for a heavy Higgs
boson, $S_{ref}$ is dominated by the chiral-logarithmic term 
(Eq.(\ref{eq:S-Higgs})) such that any dependence on $M_H^{ref}$ cancels in the 
square-bracketed term in Eq.(\ref{eq:S-chi2}).  The total one-loop 
contribution $S_{1-loop}$ from the three site model is then given by:
\begin{eqnarray}
S_{1-loop} = S_{tree} + \delta S\,
\label{eq:S-1loop}
\end{eqnarray}
where $\delta S$, the contribution from the loop diagrams alone, is: 
\begin{eqnarray}
\delta S = \frac{1}{12\pi}\,\log\biggl(\frac{M_{W^\prime}^2}{M_{W}^2}
  \biggr)
  + \kappa_S\,\log\biggl(\frac{M_{W^\prime}^2}{M_W^2}\biggr) + 
  (A_{W^\prime}^S + A_{W}^S)\,\log\biggl(\frac{\Lambda^2}
  {M_{W^\prime}^2}\biggr)\,.
\label{eq:deltaS}
\end{eqnarray}

Comparing Eq.~(\ref{eq:S-Higgs}) with the first term of Eq.~(\ref{eq:deltaS})
makes it clear that, in some sense, the role of the Higgs boson in the 
three site model is played by the $W^\prime$ \cite{SekharChivukula:2006cg}.
In other words, the Higgs mass, which cuts off the logarithmic divergences
in the SM, is replaced by the mass of the $W^\prime$.
In the following, this observation will allow us to compare our one-loop
results directly with experimental constraints to obtain bounds
on the three site model without carrying out the full analysis outlined above. 
This is accomplished provided we identify the mass of the $W^\prime$ with 
the corresponding Higgs boson mass used in the global analysis.  In 
particular, we will consider two values: $M_{W^\prime}= M_H^{ref} = 340$ 
GeV and 1 TeV, for which the 90\% C.L. limits on $S$ are 
\cite{Yao:2006px}:
\begin{eqnarray}
&&-0.33 \le S \le 0.05 \,\,\,\,\,\,\,\,(M_{W^\prime}=M_H^{ref}=340 \,\,\mbox{GeV})
\\
\nonumber\\
&&-0.45 \le S \le 0.00 \,\,\,\,\,\,\,\,(M_{W^\prime}=M_H^{ref}=1 \,\,\mbox{TeV})
\,.
\label{S-bounds}
\end{eqnarray}

\begin{figure}[t]
\begin{center}
\includegraphics[scale=0.5,angle=-90]{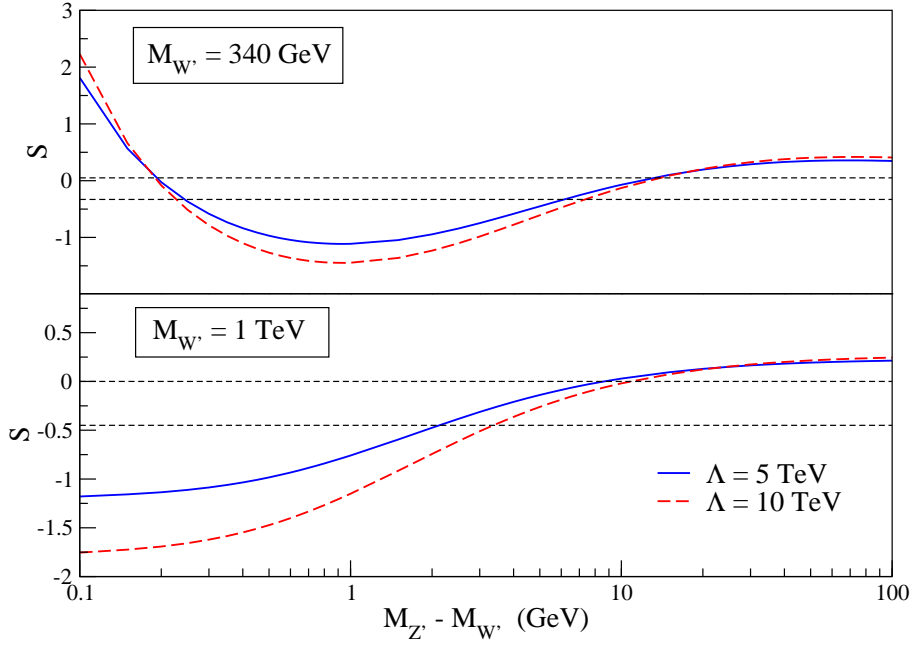}
\end{center}
\caption[]{The $S$ parameter in the three site Higgsless
model at the one-loop level as a function of the mass difference
$M_{Z^\prime}-M_{W^\prime}$ for different values of 
the cutoff of the effective theory, $\Lambda$.  Ideal delocalization
is assumed in these plots such that $S_{tree}$ = 0 (\ref{eq:Stree}).  
The upper (lower)
panel corresponds to $M_{W^\prime}$ = 340 GeV (1 TeV).  The horizontal
dashed lines indicate the 90\% C.L. bounds on the $S$ parameter 
for Higgs boson of the same mass \cite{Yao:2006px}.}
\label{fg:S-massdiff}
\end{figure}

At this point, the one-loop result $\delta S$ is a function of four 
parameters: the masses $M_{W^\prime}, M_{Z^\prime}$, the delocalization
parameter $x_1$ and the cutoff scale $\Lambda$.  In comparing the three
site model to experimental limits, we will identify the mass of the 
$W^\prime$ with the particular Higgs boson mass used in the global fit.
Thus, we are left with only $M_{Z^\prime}$, 
$x_1$ and $\Lambda$ as free parameters.  In Fig.~\ref{fg:S-massdiff}, we plot 
$S_{1-loop}$ as a function of the mass splitting $M_{Z^\prime}-M_{W^\prime}$
for $M_{W^\prime}$ = 340 GeV (top panel) and $M_{W^\prime}$ = 1 TeV
(bottom panel).  We also consider two values of the cutoff $\Lambda$ for 
each mass.  In these plots, we have assumed ideal delocalization for 
the light fermions such that $S_{tree} = 0$.  For both $W^\prime$ masses
considered, we note that the corrections to $S$ in the three site model
can be large ($\sim {\cal{O}}(\pm 1)$) even with $S_{tree} = 0$.  Indeed,
there appear to be only small windows in the mass difference where 
$S_{1-loop}$ can be brought
into approximate agreement with the experimental constraints.  In particular,
for $M_{W^\prime} = 340$ GeV, the allowed regions are $M_{Z^\prime} -
M_{W^\prime} \approx 0.2$ GeV where the $Z^\prime$ and $W^\prime$ are 
nearly degenerate and $M_{Z^\prime}-M_{W^\prime} \approx M_Z - M_W$.
Finally, it is interesting to note that in both cases, 
$S_{1-loop}$ becomes nearly independent of $\Lambda$ above $M_{Z^\prime}-M_{W^\prime} \approx M_Z - M_W$ which (from Eq.~(\ref{eq:deltaS})) implies:
\begin{equation}
A_W^S + A_{W^\prime}^S \simeq 0\,,
\end{equation}
in this range.

\begin{figure}[t]
\begin{center}
\includegraphics[scale=0.5,angle=-90]{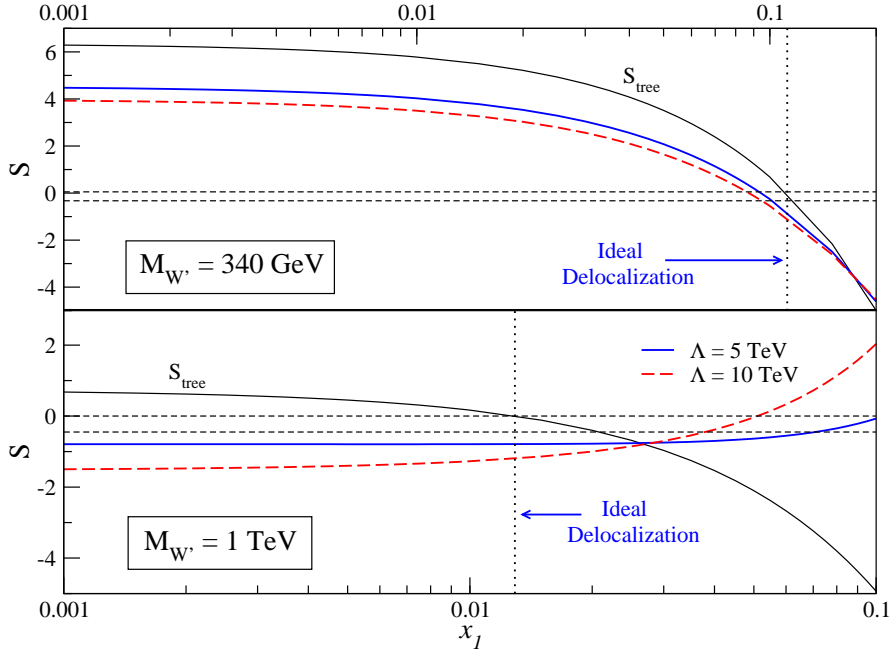}
\end{center}
\caption[]{The $S$ parameter in the three site Higgsless
model at the one-loop level as a function of the delocalization
parameter $x_1$ for $M_{W^\prime}$ = 340 GeV (top) and 1 TeV (bottom).  
The 90\% C.L. limits on $S$ for a Higgs boson of the
same mass are indicated by horizontal dashed lines \cite{Yao:2006px}.}
\label{fg:S-x1dep}
\end{figure}

Finally, we consider the dependence of $S_{1-loop}$ (and $S_{tree}$) on
the delocalization parameter $x_1$ as shown in Fig.~\ref{fg:S-x1dep}.
Again, we consider two values of $M_{W^\prime}$ and we have set
$M_{Z^\prime}^2 = M_{W^\prime}^2 + (M_Z^2 - M_W^2)$.  We denote by a
vertical dotted line the point at which $S_{tree}$ = 0.  The dependence of
the $S$ parameter on the delocalization procedure itself is an important issue
in the three site model and other Higgsless models in general.  The 
outstanding question is whether $x_1$ must be tuned order-by-order in
perturbation theory in such a way to bring $S$ into agreement with precision 
electroweak data or the 
value of $x_1$ which cancels $S_{tree}$ is {\it ideal} at all orders.
As we see from the top panel of Fig.~\ref{fg:S-x1dep}, going from tree
level to one-loop level requires a tuning of $x_1$ at the  $\sim$ 
20-30\% level for smaller values of $M_{W^\prime}$.  This relatively
small tuning is due to the fact that the total contribution to $S$ is
dominated by the tree-level value.  However, as evidenced by the bottom 
panel, the tuning becomes much more severe for heavier masses.  Indeed, 
for $M_{W^\prime} =$ 1 TeV, one must tune $x_1$ by a factor of $\sim$ 5 
in order to reconcile $S_{1-loop}$ with the constraints from 
data.  Of course, this conclusion is highly dependent on the particular 
choice of $M_{Z^\prime}$ (see Fig.~\ref{fg:S-massdiff}), but the relationship
between $M_{Z^\prime}$ and $M_{W^\prime}$ 
we have chosen for these plots is a preferred one in the three site model 
since it results in maximal suppression of unitarity-violating terms in 
$W_L W_L$ scattering (see Fig. 2 of Ref. \cite{Foadi:2003xa}). 

\subsection{The $T$ Parameter}
\label{subsec:T-higgsless}

At tree level in the three site model, the $T$ parameter exactly vanishes
due to the presence of an $SU(2)$ custodial symmetry.  When the fermions
are delocalized to negate large corrections to the $S$ parameter, the 
SM fermions develop heavy partners with the same SM quantum numbers.  At
the one-loop level, these new fermions, especially the partners of the
top and bottom quarks, can make potentially sizable contributions to the 
$T$ parameter \footnote{In general, the corrections to the $S$ parameter
from the heavy fermions are believed to small, so we have neglected their 
contribution in the previous section.} \cite{SekharChivukula:2006cg}.  We 
will not consider these corrections here, but we note that they are typically 
of the same size and opposite sign relative to the gauge sector contributions 
which we discuss below.  The end result is a large cancelation between the 
two contributions such that the $T$ parameter is safe in the three site model 
even at the one-loop level.

The one-loop, chiral-logarithmic corrections to $T$ from the gauge sector
of the three site model naturally separate into low- and high-energy 
contributions:
\begin{eqnarray}
T_{3-site} &=& A_W^T\,\log\biggl(\frac{\Lambda^2}{M_W^2}
 \biggr) + A_{W^\prime}^T\,\log\biggl(\frac{\Lambda^2}{M_{W^\prime}^2}
 \biggr) + T_0 \nonumber\\
\nonumber\\
&=& A_W^T\,\log\biggl(\frac{M_{W^\prime}^2}{M_W^2}
 \biggr) + (A_{W^\prime}^T + A_W^T)\,\log\biggl(\frac{\Lambda^2}{M_{W^\prime}^2}
 \biggr) + T_0\,, 
\label{eq:T-3site}
\end{eqnarray}
where $T_0$ represents the contribution from the dimension-two operator
of Eq.~(\ref{eq:L2}).  The expression for $T_0$ can be extracted by inserting 
the expansions of the gauge fields in terms of the mass eigenstates 
(Eqs.~(\ref{eq:B}) and (\ref{eq:W13})) into Eq.~(\ref{eq:L2}).
Isolating corrections to the SM-like $Z$ boson mass, we find that
${\cal{L}}_2^\prime$ produces a term of the form:
\begin{eqnarray} 
{\cal{L}}_{T^0} = -\frac{z}{2}\,M_Z^2\,Z_\mu Z^\mu\,,
\end{eqnarray}
where $z$ is given by:
\begin{eqnarray}
z = \frac{\beta_{(2)}\,f_1^2}{2 M_Z^2}\,(g^\prime\,b_{02} - 
  \tilde{g}\,b_{12})^2\,.
\label{eq:z}
\end{eqnarray}
In contrast, higher-dimension operators do not contribute
to a shift in the $W$ boson mass and $T_0$ is given by
\cite{Csaki:2005vy}:
\begin{eqnarray}
T_0 = - z = -\frac{\beta_{(2)}\,f_1^2}{2 M_Z^2}\,(g^\prime\,b_{02} - 
  \tilde{g}\,b_{12})^2\,.
\label{eq:T0}
\end{eqnarray}
Combining Eq.~(\ref{eq:T0}) with Eq.~(\ref{eq:T-3site}) makes it clear
that the $\beta_{(2)}$ coefficient acts as a counterterm for the $T$
parameter.  Therefore, as in the previous section, we will set $T_0$
to zero for our analysis.

As a check of our calculation for $T$ in the three site model,
we plot the coefficient of the low-energy contribution, $A_W^T$, in 
the bottom panel of Fig.~(\ref{fg:AS-and-AT}).  In 
the energy region below $M_{W^\prime}$, the operators which generate
corrections to $T$ in the three site model are identical to the operators
of the SM with a heavy Higgs boson \cite{Bando:1984ej,Bando:1985rf,
Bando:1987ym,Bando:1987br}.  Therefore, in the limit 
$M_W^2 \ll M_{W^\prime}^2$, $A_W^T$ reduces to the SM value:
\cite{Peskin:1991sw,Herrero:1993nc,Herrero:1994iu,Dittmaier:1995ee,
Matias:1996hf,Alam:1997nk}:
\begin{equation}
A_{SM}^T = - \frac{3}{16\pi c_w^2}\,,
\end{equation}
as seen in the bottom panel of Fig.~(\ref{fg:AS-and-AT}).
Thus, to simplify our analysis, we find it convenient 
to rewrite the low-energy coefficient as:
\begin{equation}
A_W^T = - \frac{3}{16\pi c_w^2} + \kappa_T\,,
\end{equation}
where $\kappa_T$ parameterizes the piece of the low-energy contributions 
which decouples in the large $M_{W^\prime}$ limit.  In contrast to the 
low-energy coefficient for the $S$ parameter, we see that the sub-leading
terms in $M_W^2/M_{W^\prime}^2$ have a much smaller effect on $A_W^T$ for 
lower values of $M_{W^\prime}$.

In analogy to the previous section, the one-loop prediction for $T$ 
can now be compared to precision electroweak data in order to constrain 
some (or all) of the parameters of the three site model.  The analysis
follows along the same lines as the case of the $S$ parameter.
First, the chiral-logarithmic contribution from a heavy Higgs boson
\cite{Peskin:1991sw,Herrero:1993nc,Herrero:1994iu,Dittmaier:1995ee,
Matias:1996hf,Alam:1997nk,Chivukula:1999az}:
\begin{eqnarray}
T_{Higgs} = - \frac{3}{16\pi c_w^2}\log\biggl( \frac{M_H^2}{M_W^2} \biggr)\,,
\label{eq:T-higgs}
\end{eqnarray}
must be subtracted from the global analysis.  Then, adding back in the 
contribution from Eq.~(\ref{eq:T-3site}), the value of $T$ to be used in the 
$\chi^2$ fit is given by:
\begin{eqnarray}
T(T_0,M_{W^\prime},\Lambda) &=& T_{ref}(M_H^{ref}) - T_{Higgs} + 
  T_{3-site} \\
&\equiv& [T_{ref}(M_H^{ref})-T_{Higgs}] + T_{1-loop} + T_0\,,
\end{eqnarray}
where $T_{ref}(M_H^{ref})$ is the SM $T$ parameter as a function of the 
reference Higgs boson mass, $M_H^{ref}$.  For large enough values of the
Higgs boson mass, $T_{ref}$ is dominated by the chiral-logarithmic 
contribution from the Higgs such that the quantity in square brackets is
independent of $M_H^{ref}$.  Lastly, the one-loop contribution $T_{1-loop}$ 
is found to be:
\begin{eqnarray}
T_{1-loop} = -\frac{3}{16\pi c_w^2}\,
  \log\biggl(\frac{M_{W^\prime}^2}{M_W^2}\biggr)
  + \kappa_T\,\log\biggl(\frac{M_{W^\prime}^2}{M_W^2}\biggr) + 
  (A_{W^\prime}^T + A_W^T)\,
  \log\biggl(\frac{\Lambda^2}{M_{W^\prime}^2}\biggr)\,.
\label{eq:T-1loop}
\end{eqnarray}
Again, comparing Eq.~(\ref{eq:T-higgs}) with the first term of 
Eq.~(\ref{eq:T-1loop}), we see that the cutoff of the logarithmic
divergences from the low-energy sector, which is typically provided 
by the Higgs boson mass, is being played by the $W^\prime$ mass.
Lastly, we mention that, given the potentially
large (and positive definite) corrections coming from the fermionic sector,
we will not exhibit any limits in the following plots.  To be consistent, 
though, we will consider
the same mass values as in the previous section:  $M_{W^\prime} = M_H^{ref}
= 340$ GeV and 1 TeV, for which the 90\% C.L. limits on $T$ are
\cite{Yao:2006px}:
\begin{eqnarray}
&&-0.15 \le T \le 0.27 \,\,\,\,\,\,\,\,(M_{W^\prime}=M_H^{ref}=340 \,\,\mbox{GeV})
\\
\nonumber\\
&&0.02 \le T \le 0.42 \,\,\,\,\,\,\,\,(M_{W^\prime}=M_H^{ref}=1 \,\,\mbox{TeV})
\,.
\label{T-bounds}
\end{eqnarray}

\begin{figure}[t]
\begin{center}
\includegraphics[scale=0.5,angle=-90]{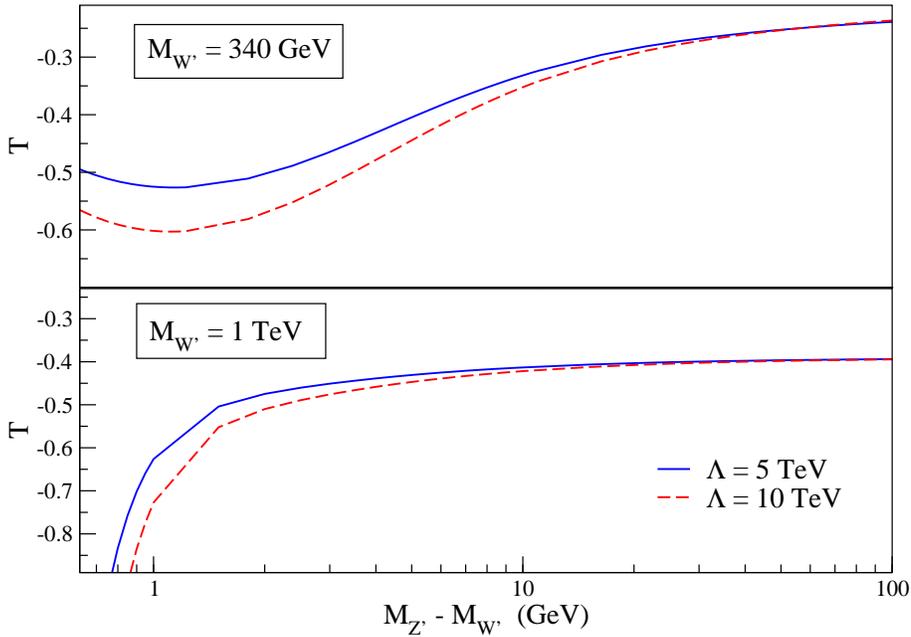}
\end{center}
\caption[]{One-loop, chiral-logarithmic corrections to the $T$ 
parameter in the three site model as a function of the mass difference
$M_{Z^\prime}-M_{W^\prime}$ and $\Lambda$.  The lower (upper) panel 
corresponds to $M_{W^\prime}$ = 340 GeV (1 TeV) and we have assumed
ideally delocalized fermions.}
\label{fg:T-massdiff}
\end{figure}

First, we consider the dependence of $T$ on the mass difference $M_{Z^\prime}
- M_{W^\prime}$ in Fig.~\ref{fg:T-massdiff}.  In
comparison to the results for the $S$ parameter, we note that the overall
size of the corrections to $T$ are much smaller.  We also point out that,
for $M_{Z^\prime}-M_{W^\prime} \ge 1$ GeV, the corrections are of the same
size and opposite sign relative to the fermionic contribution 
\cite{SekharChivukula:2006cg}.  The result is a large cancelation which
brings the full three site model contribution to $T$ within the experimental bounds
given above.  Finally, we again note that, above $M_{Z^\prime}-M_{W^\prime}
= M_Z - M_W$, the one-loop results become nearly independent of the cutoff
scale $\Lambda$, which implies:
\begin{equation}
A_W^T + A_{W^\prime}^T \simeq 0
\end{equation}
in this range.

\begin{figure}[t]
\begin{center}
\includegraphics[scale=0.5,angle=-90]{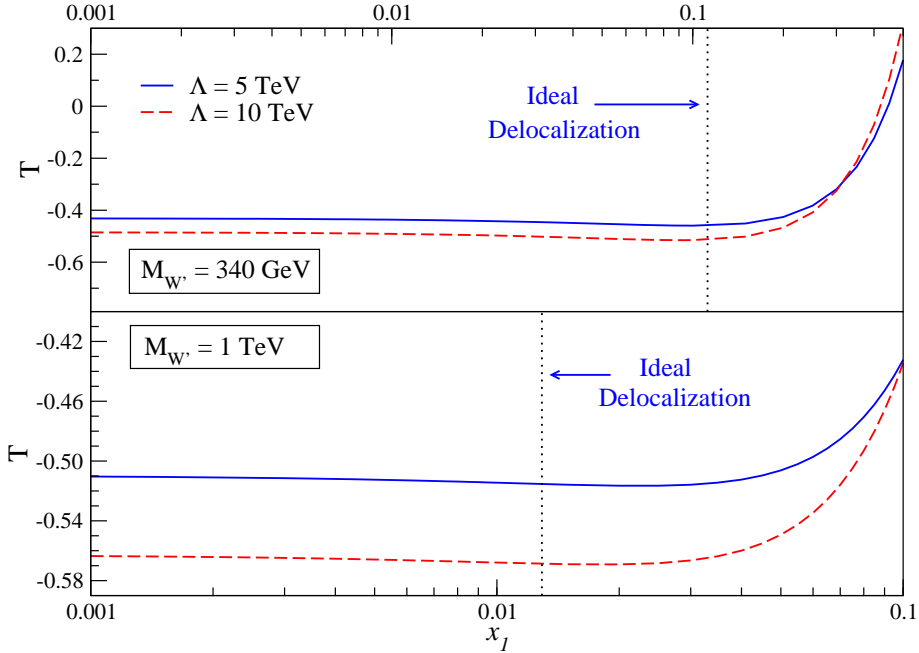}
\end{center}
\caption[]{One-loop, chiral-logarithmic corrections to the $T$ 
parameter in the three site model as a function of the delocalization
parameter $x_1$.  The lower (upper) panel 
corresponds to $M_{W^\prime}$ = 340 GeV (1 TeV).  The point of ideal
fermion delocalization is depicted by a vertical dotted line.}
\label{fg:T-x1dep}
\end{figure}

Finally, we consider the $x_1$ dependence of the one-loop contributions
to $T$.  Since $x_1$ respects the custodial symmetry present in the three 
site model, one would expect this dependence to be negligible 
\cite{SekharChivukula:2006cg}.  In Fig.~\ref{fg:T-x1dep}, we plot our results
as a function of $x_1$ for two separate $M_{W^\prime}$ values.  From these
plots, it is apparent that the one-loop corrections are, in fact, independent of
$x_1$ over the majority of the range considered.  Only at the larger values
of $x_1$ do the values of $T$ start to show some dependence.  However, these
larger values of $x_1$ are typically ruled out by experimental constraints on
the $ZWW$ vertex \cite{SekharChivukula:2006cg}.

\section{Conclusions}
\label{sec:conclusions}

We have calculated the one-loop corrections to the $S$ and $T$ parameters
in models which contain extra vector gauge bosons, but which are devoid of
any fundamental scalars (i.e., Higgs bosons).  We have performed this 
calculation using a novel application of the Pinch Technique (PT) which requires
including certain pieces from vertex and box corrections, along with the 
usual loop-corrected two-point functions, in order to obtain gauge-independent
expressions for the gauge boson self-energies \cite{Degrassi:1992ff,Degrassi:1992ue,Papavassiliou:1994fp,Degrassi:1993kn}.  All of the 
diagrams needed to construct the PT self-energies have been calculated using generic 
couplings for the gauge boson self-interactions as well as the interactions between 
the light fermions and gauge bosons.  This permits our results to be applied to
various models by simply identifying the generic couplings of our expressions
with the fundamental parameters of a specific
model.  To conclude our algorithm, we have demonstrated how to assemble all of
the one-loop diagrams in order to obtain gauge-independent expressions for the 
$S$ and $T$ parameters. 

As an example of how the algorithm presented here may be applied, we have calculated 
the one-loop, chiral-logarithmic corrections to $S$ and $T$ in the highly-deconstructed three
site model \cite{Foadi:2003xa,Perelstein:2004sc,SekharChivukula:2006cg}.  The gauge
sector of this model, which is identical to that of the BESS model 
\cite{Casalbuoni:1985kq,Casalbuoni:1986vq}, consists of a SM-like set of gauge bosons 
(massless photon and light vector gauge bosons, $W^\pm$ and $Z$) plus an extra set 
of heavy vector gauge bosons
($W^{\prime \pm}$ and $Z^\prime$).  At tree level, mixing between the various
gauge eigenstates generates a large contribution to the $S$ parameter.  However, when
the fermions of the model are allowed to derive their couplings from all three
gauge groups, they provide a negative contribution to the $S$ parameter which
can reduce or even negate the large contribution from the gauge sector.  This 
underlines the need for a one-loop calculation of the $S$ parameter in this model
and other Higgsless models where similar delocalization procedures can be employed
to reduce large corrections from the extended gauge sectors.  The $T$ parameter in 
the three site model vanishes at tree level due to the presence of a custodial 
$SU(2)$ symmetry.  Thus, assessing the one-loop corrections to $T$ in this model
also becomes important.

The loop-corrected values of the $S$ and $T$ parameters in the three site model
were previously calculated in both the Feynman \cite{Matsuzaki:2006wn} 
and Landau \cite{Chivukula:2007ic} gauges in the limit $M_W^2 \ll M_{W^\prime}^2$.
In the calculation presented here, however, we have worked with the {\it exact} 
expressions for the free parameters of the model.  In other words, we have
retained sub-leading terms in $M_W^2/M_{W^\prime}^2$ which can be important
for smaller values of $M_{W^\prime}$.  We have compared our results, which
were obtained by employing the unitary gauge, in the limit
$M_W^2 \ll M_{W^\prime}^2$ with those of Refs. \cite{Matsuzaki:2006wn} and
\cite{Chivukula:2007ic} and found excellent agreement.  This was an important
check of our algorithm and, in fact, proves the gauge-independence of our
results.

In our approach, the one-loop expressions for $S$ and $T$ in the three site 
model reduce to functions of only four parameters: the cutoff of the effective theory 
($\Lambda$), the degree of delocalization for the light fermions ($x_1$)
and the masses of the heavy gauge bosons ($M_{W^\prime}$ and $M_{Z^\prime}$).
The first of these only appears in the chiral logarithms which are present due to
the non-renormalizability of the theory and is assumed to be in the range
5 TeV $< \Lambda <$ 10 TeV. We have shown that, for both $S$ and
$T$, the low-energy contribution in the three site model reduces to the usual
SM value with the Higgs boson mass dependence replaced by the $W^\prime$ mass.

In particular, we studied the dependence of the $S$ and $T$ parameters on
the mass difference $M_{Z^\prime} - M_{W^\prime}$ and the delocalization
parameter $x_1$.  While the dependence of $T$ on these quantities is minimal, 
the $S$ parameter exhibits strong dependence on both and, in fact, can only
be reconciled with experimental limits in small ranges of both quantities.
The dependence of $S$ on $x_1$ is of particular interest
in the three site model.  The outstanding issue is whether or not $x_1$ must be 
tuned order-by-order in perturbation theory to bring $S$ into agreement with 
experimental constraints.
In our analysis, we have found that the tuning is minimal for lighter $W^\prime$
masses.  This is mainly due to dominance of the tree-level contribution over
the one-loop contributions for small $M_{W^\prime}$.  However, for larger masses, 
the tuning can be much more severe.  In particular, for $M_{W^\prime}$ = 1 TeV,
we found that $x_1$ must be tuned by a factor of five in going from tree level
to the one-loop level.  

Finally, it should be stated that our calculation is not exclusive to 
Higgsless models.  As mentioned in the Introduction, one-loop corrections 
to the VGB self-energies can always be separated into gauge-invariant 
contributions from fermions, scalars and gauge bosons.  In other words, 
our calculation could be used to  calculate the gauge-bosonic contributions 
to oblique parameters in models which contain fundamental (or composite) 
Higgs bosons.

\begin{acknowledgements}
We are very grateful to Sekhar Chivukula and Shinya Matsuzaki for useful
discussions on the three site model.  We would also like to thank
Hooman Davoudiasl for a careful reading of the manuscript.  This manuscript 
has 
been authored by employees of Brookhaven Science Associates, LLC under 
Contract No. DE-AC02-98CH10886 with the U.S. Department of Energy. The 
publisher by accepting the manuscript for publication acknowledges that the 
United States Government retains a non-exclusive, paid-up, irrevocable, 
world-wide license to publish or reproduce the published form of this 
manuscript, or allow others to do so, for United States Government purposes.
\end{acknowledgements}

\appendix

\section{Scalar Integrals and Tensor Coefficients}
\label{app:P-V}

The scalar integrals that appear in the calculation of the PT self-energies
are the one-point integral $A_0(M)$:
\begin{equation}
A_0(M) \equiv \int \frac{d^nk}{(2\pi)^n} \frac{1}{k^2 - M^2}\,,
\label{eq:A0}
\end{equation}
and the two-point integral $B_0(q^2;M_1,M_2)$:
\begin{equation}
B_0(q^2;M_1,M_2) = \int \frac{d^nk}{(2\pi)^n} \frac{1}{(k^2 - M_1^2)
  ((k+q)^2 - M_2^2)}\,.
\label{eq:B0}
\end{equation}
In order to extract the chiral-logarithmic corrections,
we only need to calculate the poles of the scalar integrals which are then
identified with the appropriate chiral logarithms:
\begin{eqnarray}
\label{eq:A0-chiral}
A_0(M)\bigl|_{pole} &=& \frac{i}{16\pi^2\epsilon} M^2 \to 
   \frac{i}{16\pi^2}\,\log\biggl( \frac{\Lambda^2}{M^2} \biggr) M^2\,,
\end{eqnarray}
and:
\begin{eqnarray}
\label{eq:B0-chiral}
B_0(q^2;M_1,M_2)\bigl|_{pole} &=& \frac{i}{16\pi^2\epsilon} \to
   \frac{i}{16\pi^2}\,\log\biggl( \frac{\Lambda^2}{M^2} \biggr)\,.
\end{eqnarray}

The tensor integrals that arise in our calculation consist of the rank-one
and rank-two two-point integrals:
\begin{eqnarray}
\label{eq:B1-tensor}
B^{\mu}(q^2;M_1,M_2) &=& \int \frac{d^n k}{(2\pi)^n} 
  \frac{k^\mu}{(k^2 - M_1^2)((k+q)^2 - M_2^2)}\,, \\
\nonumber\\
B^{\mu\nu}(q^2;M_1,M_2) &=& \int \frac{d^n k}{(2\pi)^n} 
  \frac{k^\mu k^\nu}{(k^2 - M_1^2)((k+q)^2 - M_2^2)}\,.
\label{eq:B2-tensor}
\end{eqnarray}
Tensor integrals can always be expanded in terms of external momenta and
the metric tensor $g^{\mu\nu}$ \cite{Passarino:1978jh}.  Specifically, the 
integrals in Eqs.~(\ref{eq:B1-tensor}) and (\ref{eq:B2-tensor}) can be 
written as 
\begin{eqnarray}
\label{eq:B1-PV}
B^{\mu}(q^2;M_1,M_2) &=& q^\mu \, B_{11}(q^2;M_1,M_2) \\
\nonumber\\
\label{eq:B2-PV}
B^{\mu\nu}(q^2;M_1,M_2) &=& q^\mu q^\nu \, B_{21}(q^2;M_1,M_2) +
  g^{\mu\nu}\,B_{22}(q^2;M_1,M_2)\,.
\end{eqnarray}
Finally, equating the the tensor integral with its respective expansion
and contracting both sides with external momenta and $g^{\mu\nu}$, one
can solve the system of equations for the coefficients in 
Eqs.~(\ref{eq:B1-PV}) and (\ref{eq:B2-PV})
in terms of the scalar integrals.  Specifically,:
\begin{eqnarray}
\label{eq:B11-coeff}
B_{11}(q^2;M_1,M_2) &=& \frac{1}{2q^2} \biggl[
  A_0(M_1) - A_0(M_2) - (q^2 + M_1^2 - M_2^2)B_0(q^2;M_1,M_2)\biggr] \\
\nonumber\\
\label{eq:B21-coeff}
B_{21}(q^2;M_1,M_2) &=& \frac{1}{3q^2} \biggl[
  A_0(M_2) + 2(M_2^2 - M_1^2 - q^2)\,B_{11}(q^2;M_1,M_2) \nonumber\\
&& \,\,\,\,\,\,\,\,\,\,\,\,\,\,\,\,\,\,\,\,- 
  M_1^2\,B_0(q^2;M_1,M_2) \biggr] \\
\nonumber\\
\label{eq:B22-coeff}
B_{22}(q^2;M_1,M_2) &=& \frac{1}{3} \biggl[ \frac{1}{2}A_0(M_2) + 
  M_1^2\,B_0(q^2;M_1,M_2) \nonumber\\
&& \,\,\,\,\,\,\,\,\,\,\,\,\,\,\,\,\,\,\,\,-\frac{1}{2}
  (M_2^2 - M_1^2 - q^2)B_{11}(q^2;M_1,M_2) \biggr]\,.
\end{eqnarray}

\section{Formulae for the Three Site Model}
\label{app:3site-params}

In this appendix, we summarize the relevant formulae for the three site model
\cite{Foadi:2003xa,Perelstein:2004sc,SekharChivukula:2006cg}.
We begin by finding the mass eigenvalues.  First, in the charged sector, 
the mass matrix is:
\begin{equation}
 M_{CC} = \frac{1}{4} 
  \left( \begin{array}{cc}
\tilde{g}^2(f_1^2 + f_2^2) & \,\,\,\,\,-g\tilde{g}f_2^2  \\
-g\tilde{g}f_2^2  & g^2f_2^2  \end{array} \right)\,,
\label{eq:M-CC}
\end{equation} 
for which we find the eigenvalues:
\begin{eqnarray}
M_{W,W^\prime}^2 = \frac{1}{8} \biggl\{ \tilde{g}^2(f_1^2 + f_2^2) +
  g^2\,f_2^2 \mp \biggl[(\tilde{g}^2(f_1^2 + f_2^2) +
  g^2\,f_2^2)^2 - 4f_1^2 f_2^2 g^2 \tilde{g}^2 \biggr]^{\frac{1}{2}}
  \biggr\}\,,
\label{eq:CC-eigen}
\end{eqnarray}
where the SM-like $W$ is identified with the lighter of the two eigenvalues.

Next, the mass matrix for the neutral sector is:
\begin{equation}
 M_{NC} = \frac{1}{8} 
  \left( \begin{array}{ccc}
 g^{\prime 2} f_1^2 & -g^\prime \tilde{g} f_1^2 & 0   \\
-g^\prime \tilde{g} f_1^2  & \,\,\,\,\tilde{g}^2(f_1^2+f_2^2) & 
\,\,\,\,-g\tilde{g}f_2^2 \\
0 & -g\tilde{g}f_2^2 & g^2 f_2^2  \end{array} \right)\,.
\label{eq:M-NC}
\end{equation} 
Diagonalizing this matrix results in a massless eigenstate, which is identified
with the SM photon, and two massive states with mass eigenvalues:
\begin{eqnarray}
M_{Z,Z^\prime} &=& \frac{1}{16} \biggl\{
  \tilde{g}^2(f_1^2 + f_2^2) + g^{\prime 2}f_1^2 + g^2 f_2^2 \nonumber\\
&& \,\,\,\,\,\,\,\,\mp
  \biggl[ (\tilde{g}^2(f_1^2 + f_2^2) + g^{\prime 2}f_1^2 + g^2 f_2^2)^2 -
  4 f_1^2 f_2^2 (g^2\tilde{g}^2 + g^{\prime 2}(g^2+\tilde{g}^2)) 
  \biggr]^{\frac{1}{2}} \biggr\}\,.
\label{eq:NC-eigen}
\end{eqnarray}  
where the SM-like $Z$ is identified with the lighter of the two states.

In its original form, the three site model contains five free parameters: 
$g,g^\prime,\tilde{g},f_1$ and $f_2$.  For our analysis, we find it 
convenient to 
exchange these parameters for the four masses ($M_W, M_Z, M_{W^\prime}$ and
$M_{Z^\prime}$) defined through Eqs.~(\ref{eq:CC-eigen}) and 
(\ref{eq:NC-eigen}).  As the fifth parameter, we choose the electromagnetic
coupling $e$ which is defined by Eq.~(\ref{eq:e-def}).  Solving these
equations for the original parameters, we find \cite{Foadi:2003xa}:
\begin{eqnarray}
g^{\prime2}&=&{e^2M_Z^2M_{Z^\prime}^2\over M_W^2M_{W^\prime}^2}\ ,\nonumber\\
\nonumber\\
\tilde{g}^{2}&=&g^{\prime2}\left[
{(M_W^2+M_{W^\prime}^2)(M_Z^2+M_{Z^\prime}^2-M_W^2-M_{W^\prime}^2)
+M_W^2M_{W^\prime}^2-M_Z^2M_{Z^\prime}^2\over
(M_Z^2+M_{Z^\prime}^2-M_W^2-M_{W^\prime}^2)^2}\right]\ ,\nonumber\\
\nonumber\\
g^{2}&=&{g^{\prime2}M_W^2M_{W^\prime}^2}\left[
{(M_W^2+M_{W^\prime}^2)(M_Z^2+M_{Z^\prime}^2-M_W^2-M_{W^\prime}^2)
+M_W^2M_{W^\prime}^2-M_Z^2M_{Z^\prime}^2\over
(M_{Z}^2-M_{W}^2)(M_{Z^\prime}^2-M_{W^\prime}^2)
(M_{Z^\prime}^2-M_{W}^2)(M_{W^\prime}^2-M_{Z}^2)}\right]\ ,\nonumber\\
\nonumber\\
f_1^{2}&=&{4\over g^{\prime2}}
(M_Z^2+M_{Z^\prime}^2-M_W^2-M_{W^\prime}^2)\ ,\nonumber\\
\nonumber\\
f_2^{2}&=&{16M_W^2M_{W^\prime}^2\over\tilde{g}^2g^2f_1^2}\ ,
\label{eq:3site-params}
\end{eqnarray}
where we have assumed in the above relations that 
$M_{Z^\prime} > M_{W^\prime}$.

Finally, in order to compute the couplings relevant to the calculation
of the $S$ and $T$ parameters, we need to calculate the mixing angles defined
through Eqs.~(\ref{eq:W1pm})-(\ref{eq:W23}).  First, in the charged sector,
we have $a_{11}=a_{22}$ and $a_{12}=-a_{21}$ where:
\begin{eqnarray}
a_{11} &=& \left[{M_{W^\prime}^2(M_{W^\prime}^2-M_{Z}^2)
(M_{Z^\prime}^2-M_{W^\prime}^2)\over
M_{W^\prime}^2(M_{W^\prime}^2-M_{Z}^2)
(M_{Z^\prime}^2-M_{W^\prime}^2)+M_{W}^2(M_{Z^\prime}^2-M_{W}^2)
(M_{Z}^2-M_{W}^2)}
\right]^{1/2}\ ,\nonumber\\
\nonumber\\
a_{12} &=& \left[{M_{W}^2(M_{Z^\prime}^2-M_{W}^2)
(M_{Z}^2-M_{W}^2)\over
M_{W^\prime}^2(M_{W^\prime}^2-M_{Z}^2)
(M_{Z^\prime}^2-M_{W^\prime}^2)+M_{W}^2(M_{Z^\prime}^2-M_{W}^2)
(M_{Z}^2-M_{W}^2)}
\right]^{1/2}\ .
\label{eq:Wmix}
\end{eqnarray}
The mixing angles in the neutral sector are given by:
\begin{eqnarray}
b_{00} &=& \frac{e}{g^\prime}\,, \\ 
\nonumber\\
b_{10} &=& \frac{e}{\tilde{g}}\,, \nonumber\\
\nonumber\\
b_{20} &=& \frac{e}{g}\,, \nonumber\\
\nonumber\\
b_{01}&=&-\left[{
(M_{Z^\prime}^2-M_{W}^2)(M_{Z^\prime}^2-M_{W^\prime}^2)
\over M_{Z^\prime}^2(M_{Z^\prime}^2-M_{Z}^2)}
\right]^{1/2}\ ,\nonumber\\
b_{11}&=&\left[{
(M_{Z^\prime}^2-M_{W}^2)(M_{Z^\prime}^2-M_{W^\prime}^2)
\over M_{Z^\prime}^2(M_{Z^\prime}^2-M_{Z}^2)
\Bigl[(M_W^2+M_{W^\prime}^2)(M_Z^2+M_{Z^\prime}^2-M_W^2-M_{W^\prime}^2)
+M_W^2M_{W^\prime}^2-M_Z^2M_{Z^\prime}^2\Bigr]
}
\right]^{1/2}\nonumber\\
&&\times\biggl(M_{W^\prime}^2+M_{W}^2-M_Z^2\biggr)\ ,\nonumber\\
b_{21}&=&-\left[{M_W^2M_{W^\prime}^2
(M_{Z}^2-M_{W}^2)(M_{W^\prime}^2-M_{Z}^2)
\over M_{Z^\prime}^2(M_{Z^\prime}^2-M_{Z}^2)
\Bigl[(M_W^2+M_{W^\prime}^2)(M_Z^2+M_{Z^\prime}^2-M_W^2-M_{W^\prime}^2)
+M_W^2M_{W^\prime}^2-M_Z^2M_{Z^\prime}^2\Bigr]
}
\right]^{1/2}\ ,\nonumber\\
b_{02}&=&-\left[{
(M_{Z}^2-M_{W}^2)(M_{W^\prime}^2-M_{Z}^2)
\over M_{Z}^2(M_{Z^\prime}^2-M_{Z}^2)}
\right]^{1/2}\ ,\nonumber\\
b_{12}&=&\left[{
(M_{Z}^2-M_{W}^2)(M_{W^\prime}^2-M_{Z}^2)
\over M_{Z}^2(M_{Z^\prime}^2-M_{Z}^2)
\Bigl[(M_W^2+M_{W^\prime}^2)(M_Z^2+M_{Z^\prime}^2-M_W^2-M_{W^\prime}^2)
+M_W^2M_{W^\prime}^2-M_Z^2M_{Z^\prime}^2\Bigr]
}
\right]^{1/2}\nonumber\\
&&\times\biggl(M_{W^\prime}^2+M_{W}^2-M_{Z^\prime}^2\biggr)\ ,\nonumber\\
b_{22}&=&\left[{M_W^2M_{W^\prime}^2
(M_{Z^\prime}^2-M_{W}^2)(M_{Z^\prime}^2-M_{W^\prime}^2)
\over M_{Z}^2(M_{Z^\prime}^2-M_{Z}^2)
\Bigl[(M_W^2+M_{W^\prime}^2)(M_Z^2+M_{Z^\prime}^2-M_W^2-M_{W^\prime}^2)
+M_W^2M_{W^\prime}^2-M_Z^2M_{Z^\prime}^2\Bigr]
}
\right]^{1/2}. \nonumber
\end{eqnarray}
%

\section{Pinch Contributions from Light-Heavy Gauge Boson Mixing}
\label{app:light-heavy-mixing}

\begin{figure}[t]
\begin{center}
\includegraphics[bb=59 642 293 724,scale=1.25]{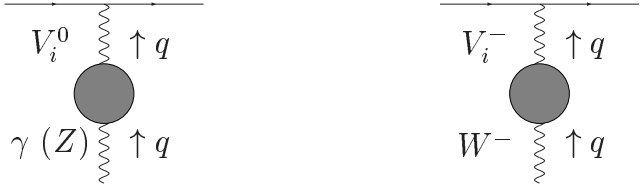} 
\end{center}
\caption[]{One-loop corrections which result in mixing between the 
light and heavy gauge bosons.  In the limit where the mass of the heavy
gauge boson is much larger than the momentum $q$, these diagrams give
rise to pinch-like contributions.}
\label{fg:light-heavy-mix}
\end{figure}

In addition to the pinch contributions arising from vertex corrections
as discussed in Sections \ref{subsec:neutral-vertex} and 
\ref{subsec:charged-vertex}, one-loop diagrams which mix the light and
heavy gauge boson propagators can also give rise to pinch-like contributions.
The diagrams which contribute to these types of corrections are depicted 
in Fig.~\ref{fg:light-heavy-mix} for both the neutral- and charged-current
cases.  The amplitudes for these corrections can easily be calculated by
using the results of Sections ~\ref{subsec:neutral-two-pt} and 
\ref{subsec:charged-two-pt} with one of the external light gauge bosons 
replaced by a heavy gauge boson and subsequently coupling these diagrams to a 
fermion line.  Below, we outline the calculation of these corrections for the 
three site model.

First, in the case of the neutral currents, we must rewrite the current
associated with the $Z^\prime$ ($\Gamma^\mu_{Z^\prime}$) in terms of those
associated with the photon ($\Gamma^\mu_\gamma$) and the SM-like $Z$ 
($\Gamma^\mu_Z$).  Using Eqs.~(\ref{eq:coupffV0})-(\ref{eq:gaV0}), we
find:
\begin{eqnarray}
\Gamma^\mu_{Z^\prime} &\equiv& \bar{u}(p_2)\,\gamma^\mu\,
  (g_{V_f}^{(Z^\prime)} + g_{A_f}^{(Z^\prime)}\,\gamma_5)\,u(p_1) \nonumber\\
\nonumber\\ 
&=& \Gamma_Z^\mu - \biggl[ \frac{g^\prime b_{02}}{(g (1-x_1) b_{22} + 
  \tilde{g} x_1 b_{12} - g^\prime
  b_{02})} + \frac{g^\prime b_{01}}{(g (1-x_1) b_{21} +
  \tilde{g} x_1 b_{11} - g^\prime
  b_{01})} \biggr]\,\Gamma_\gamma^\mu\,,
\label{eq:Gam-Zp}
\end{eqnarray} 
where $\Gamma_Z^\mu$ and $\Gamma_\gamma^\mu$ are given respectively by
Eqs.~(\ref{eq:Gamma_Z}) and (\ref{eq:Gamma_pho}).  Then, the corrections
to the $\gamma\ell\ell$ ($Z\ell\ell$) vertex from $\gamma (Z)-Z^\prime$ 
mixing are given by:
\begin{eqnarray} 
\Delta {\cal{A}}_{V,\gamma(Z)}^\mu |_{\gamma(Z)-Z^\prime} &=&
  -i g_{Z^\prime\ell\ell} \Gamma_{Z^\prime}^\mu 
  \frac{- i}{q^2 - M_{Z^\prime}^2}\,\Delta \Pi_{\gamma (Z)-Z^\prime} 
\nonumber\\
\nonumber\\
&=& \frac{g_{Z^\prime\ell\ell}}{M_{Z^\prime}^2} \biggl\{
\Gamma_Z^\mu 
 - \biggl[ \frac{g^\prime b_{02}}{(g (1-x_1) b_{22} +
  \tilde{g} x_1 b_{12} - g^\prime
  b_{02})} \nonumber\\
\nonumber\\
&& \,\,\,\,\,\,\,\,\,\,\,\,\,\,\,
  + \frac{g^\prime b_{01}}{(g (1-x_1) b_{21} +
  \tilde{g} x_1 b_{11} - g^\prime
  b_{01})} \biggr]\,\Gamma_\gamma^\mu \biggr\}
  \Delta \Pi_{\gamma (Z)-Z^\prime}  \,,
\label{eq:amp-phoZ-Zp-Mix}
\end{eqnarray}
where $\Delta \Pi_{\gamma (Z)-Z^\prime}$ can be calculated using the 
results of Section \ref{subsec:neutral-two-pt} with one of the external 
light gauge bosons replaced by a $Z^\prime$ and we have assumed that
$q^2 \approx M_{Z}^2 \ll M_{Z^\prime}^2$ in order to expand the denominator
of the $Z^\prime$ propagator.

The corrections from $W-W^\prime$ mixing prove to be much simpler given the
fact that the $W^\prime$ current is identical to the current associated with
the SM-like $W$.  In fact, we find that the contribution from $W-W^\prime$
mixing takes the compact form:
\begin{eqnarray}
\Delta {\cal{A}}_{V,W}^\mu |_{W-W^\prime} = \frac{g_{\ell\nu W^\prime}}
  {M_{W^\prime}^2}\,\Gamma_W^\mu \,\Delta \Pi_{W-W^\prime} \,,
\label{eq:amp-W-Wp-Mix}
\end{eqnarray}
where $\Delta \Pi_{W-W^\prime}$ can be calculated using the results of 
Section \ref{subsec:charged-two-pt} and, again, we have expanded the 
denominator of the $W^\prime$ propagator assuming $q^2 \approx M_W^2 \ll
M_{W^\prime}^2$.

Finally, Eqs.~(\ref{eq:amp-phoZ-Zp-Mix}) and (\ref{eq:amp-W-Wp-Mix}) can
be combined with the corrections from the standard vertex corrections
(Eqs.~(\ref{eq:delta-Av}) and (\ref{eq:delta-Avw}) respectively) in order
to extract the total pinch contributions $\{V_{\gamma(Z)}\}$ and 
$\{V_{W}\}$ needed to construct the PT self-energies.

\bibliography{STpinch}

\begin{thebibliography}{86}
\expandafter\ifx\csname natexlab\endcsname\relax\def\natexlab#1{#1}\fi
\expandafter\ifx\csname bibnamefont\endcsname\relax
  \def\bibnamefont#1{#1}\fi
\expandafter\ifx\csname bibfnamefont\endcsname\relax
  \def\bibfnamefont#1{#1}\fi
\expandafter\ifx\csname citenamefont\endcsname\relax
  \def\citenamefont#1{#1}\fi
\expandafter\ifx\csname url\endcsname\relax
  \def\url#1{\texttt{#1}}\fi
\expandafter\ifx\csname urlprefix\endcsname\relax\def\urlprefix{URL }\fi
\providecommand{\bibinfo}[2]{#2}
\providecommand{\eprint}[2][]{\url{#2}}

\bibitem[{\citenamefont{Arkani-Hamed et~al.}(1998)\citenamefont{Arkani-Hamed,
  Dimopoulos, and Dvali}}]{Arkani-Hamed:1998rs}
\bibinfo{author}{\bibfnamefont{N.}~\bibnamefont{Arkani-Hamed}},
  \bibinfo{author}{\bibfnamefont{S.}~\bibnamefont{Dimopoulos}},
  \bibnamefont{and} \bibinfo{author}{\bibfnamefont{G.~R.} \bibnamefont{Dvali}},
  \bibinfo{journal}{Phys. Lett.} \textbf{\bibinfo{volume}{B429}},
  \bibinfo{pages}{263} (\bibinfo{year}{1998}), \eprint{hep-ph/9803315}.

\bibitem[{\citenamefont{Randall and Sundrum}(1999)}]{Randall:1999ee}
\bibinfo{author}{\bibfnamefont{L.}~\bibnamefont{Randall}} \bibnamefont{and}
  \bibinfo{author}{\bibfnamefont{R.}~\bibnamefont{Sundrum}},
  \bibinfo{journal}{Phys. Rev. Lett.} \textbf{\bibinfo{volume}{83}},
  \bibinfo{pages}{3370} (\bibinfo{year}{1999}), \eprint{hep-ph/9905221}.

\bibitem[{\citenamefont{Csaki et~al.}(2004{\natexlab{a}})\citenamefont{Csaki,
  Grojean, Murayama, Pilo, and Terning}}]{Csaki:2003dt}
\bibinfo{author}{\bibfnamefont{C.}~\bibnamefont{Csaki}},
  \bibinfo{author}{\bibfnamefont{C.}~\bibnamefont{Grojean}},
  \bibinfo{author}{\bibfnamefont{H.}~\bibnamefont{Murayama}},
  \bibinfo{author}{\bibfnamefont{L.}~\bibnamefont{Pilo}}, \bibnamefont{and}
  \bibinfo{author}{\bibfnamefont{J.}~\bibnamefont{Terning}},
  \bibinfo{journal}{Phys. Rev.} \textbf{\bibinfo{volume}{D69}},
  \bibinfo{pages}{055006} (\bibinfo{year}{2004}{\natexlab{a}}),
  \eprint{hep-ph/0305237}.

\bibitem[{\citenamefont{Cacciapaglia
  et~al.}(2005{\natexlab{a}})\citenamefont{Cacciapaglia, Csaki, Grojean, and
  Terning}}]{Cacciapaglia:2004rb}
\bibinfo{author}{\bibfnamefont{G.}~\bibnamefont{Cacciapaglia}},
  \bibinfo{author}{\bibfnamefont{C.}~\bibnamefont{Csaki}},
  \bibinfo{author}{\bibfnamefont{C.}~\bibnamefont{Grojean}}, \bibnamefont{and}
  \bibinfo{author}{\bibfnamefont{J.}~\bibnamefont{Terning}},
  \bibinfo{journal}{Phys. Rev.} \textbf{\bibinfo{volume}{D71}},
  \bibinfo{pages}{035015} (\bibinfo{year}{2005}{\natexlab{a}}),
  \eprint{hep-ph/0409126}.

\bibitem[{\citenamefont{Nomura}(2003)}]{Nomura:2003du}
\bibinfo{author}{\bibfnamefont{Y.}~\bibnamefont{Nomura}},
  \bibinfo{journal}{JHEP} \textbf{\bibinfo{volume}{11}}, \bibinfo{pages}{050}
  (\bibinfo{year}{2003}), \eprint{hep-ph/0309189}.

\bibitem[{\citenamefont{Csaki et~al.}(2004{\natexlab{b}})\citenamefont{Csaki,
  Grojean, Pilo, and Terning}}]{Csaki:2003zu}
\bibinfo{author}{\bibfnamefont{C.}~\bibnamefont{Csaki}},
  \bibinfo{author}{\bibfnamefont{C.}~\bibnamefont{Grojean}},
  \bibinfo{author}{\bibfnamefont{L.}~\bibnamefont{Pilo}}, \bibnamefont{and}
  \bibinfo{author}{\bibfnamefont{J.}~\bibnamefont{Terning}},
  \bibinfo{journal}{Phys. Rev. Lett.} \textbf{\bibinfo{volume}{92}},
  \bibinfo{pages}{101802} (\bibinfo{year}{2004}{\natexlab{b}}),
  \eprint{hep-ph/0308038}.

\bibitem[{\citenamefont{Lee et~al.}(1977)\citenamefont{Lee, Quigg, and
  Thacker}}]{Lee:1977eg}
\bibinfo{author}{\bibfnamefont{B.~W.} \bibnamefont{Lee}},
  \bibinfo{author}{\bibfnamefont{C.}~\bibnamefont{Quigg}}, \bibnamefont{and}
  \bibinfo{author}{\bibfnamefont{H.~B.} \bibnamefont{Thacker}},
  \bibinfo{journal}{Phys. Rev.} \textbf{\bibinfo{volume}{D16}},
  \bibinfo{pages}{1519} (\bibinfo{year}{1977}).

\bibitem[{\citenamefont{Arkani-Hamed et~al.}(2001)\citenamefont{Arkani-Hamed,
  Cohen, and Georgi}}]{Arkani-Hamed:2001ca}
\bibinfo{author}{\bibfnamefont{N.}~\bibnamefont{Arkani-Hamed}},
  \bibinfo{author}{\bibfnamefont{A.~G.} \bibnamefont{Cohen}}, \bibnamefont{and}
  \bibinfo{author}{\bibfnamefont{H.}~\bibnamefont{Georgi}},
  \bibinfo{journal}{Phys. Rev. Lett.} \textbf{\bibinfo{volume}{86}},
  \bibinfo{pages}{4757} (\bibinfo{year}{2001}), \eprint{hep-th/0104005}.

\bibitem[{\citenamefont{Hill et~al.}(2001)\citenamefont{Hill, Pokorski, and
  Wang}}]{Hill:2000mu}
\bibinfo{author}{\bibfnamefont{C.~T.} \bibnamefont{Hill}},
  \bibinfo{author}{\bibfnamefont{S.}~\bibnamefont{Pokorski}}, \bibnamefont{and}
  \bibinfo{author}{\bibfnamefont{J.}~\bibnamefont{Wang}},
  \bibinfo{journal}{Phys. Rev.} \textbf{\bibinfo{volume}{D64}},
  \bibinfo{pages}{105005} (\bibinfo{year}{2001}), \eprint{hep-th/0104035}.

\bibitem[{\citenamefont{Appelquist and Bernard}(1980)}]{Appelquist:1980vg}
\bibinfo{author}{\bibfnamefont{T.}~\bibnamefont{Appelquist}} \bibnamefont{and}
  \bibinfo{author}{\bibfnamefont{C.~W.} \bibnamefont{Bernard}},
  \bibinfo{journal}{Phys. Rev.} \textbf{\bibinfo{volume}{D22}},
  \bibinfo{pages}{200} (\bibinfo{year}{1980}).

\bibitem[{\citenamefont{Longhitano}(1980)}]{Longhitano:1980iz}
\bibinfo{author}{\bibfnamefont{A.~C.} \bibnamefont{Longhitano}},
  \bibinfo{journal}{Phys. Rev.} \textbf{\bibinfo{volume}{D22}},
  \bibinfo{pages}{1166} (\bibinfo{year}{1980}).

\bibitem[{\citenamefont{Longhitano}(1981)}]{Longhitano:1980tm}
\bibinfo{author}{\bibfnamefont{A.~C.} \bibnamefont{Longhitano}},
  \bibinfo{journal}{Nucl. Phys.} \textbf{\bibinfo{volume}{B188}},
  \bibinfo{pages}{118} (\bibinfo{year}{1981}).

\bibitem[{\citenamefont{Bagger et~al.}(1993)\citenamefont{Bagger, Dawson, and
  Valencia}}]{Bagger:1992vu}
\bibinfo{author}{\bibfnamefont{J.}~\bibnamefont{Bagger}},
  \bibinfo{author}{\bibfnamefont{S.}~\bibnamefont{Dawson}}, \bibnamefont{and}
  \bibinfo{author}{\bibfnamefont{G.}~\bibnamefont{Valencia}},
  \bibinfo{journal}{Nucl. Phys.} \textbf{\bibinfo{volume}{B399}},
  \bibinfo{pages}{364} (\bibinfo{year}{1993}), \eprint{hep-ph/9204211}.

\bibitem[{\citenamefont{Perelstein}(2004)}]{Perelstein:2004sc}
\bibinfo{author}{\bibfnamefont{M.}~\bibnamefont{Perelstein}},
  \bibinfo{journal}{JHEP} \textbf{\bibinfo{volume}{10}}, \bibinfo{pages}{010}
  (\bibinfo{year}{2004}), \eprint{hep-ph/0408072}.

\bibitem[{\citenamefont{Chivukula et~al.}(2007)\citenamefont{Chivukula,
  Matsuzaki, Simmons, and Tanabashi}}]{Chivukula:2007ic}
\bibinfo{author}{\bibfnamefont{R.~S.} \bibnamefont{Chivukula}},
  \bibinfo{author}{\bibfnamefont{S.}~\bibnamefont{Matsuzaki}},
  \bibinfo{author}{\bibfnamefont{E.~H.} \bibnamefont{Simmons}},
  \bibnamefont{and} \bibinfo{author}{\bibfnamefont{M.}~\bibnamefont{Tanabashi}}
  (\bibinfo{year}{2007}), \eprint{hep-ph/0702218}.

\bibitem[{\citenamefont{Foadi et~al.}(2004)\citenamefont{Foadi, Gopalakrishna,
  and Schmidt}}]{Foadi:2003xa}
\bibinfo{author}{\bibfnamefont{R.}~\bibnamefont{Foadi}},
  \bibinfo{author}{\bibfnamefont{S.}~\bibnamefont{Gopalakrishna}},
  \bibnamefont{and} \bibinfo{author}{\bibfnamefont{C.}~\bibnamefont{Schmidt}},
  \bibinfo{journal}{JHEP} \textbf{\bibinfo{volume}{03}}, \bibinfo{pages}{042}
  (\bibinfo{year}{2004}), \eprint{hep-ph/0312324}.

\bibitem[{\citenamefont{Hirn and Stern}(2004)}]{Hirn:2004ze}
\bibinfo{author}{\bibfnamefont{J.}~\bibnamefont{Hirn}} \bibnamefont{and}
  \bibinfo{author}{\bibfnamefont{J.}~\bibnamefont{Stern}},
  \bibinfo{journal}{Eur. Phys. J.} \textbf{\bibinfo{volume}{C34}},
  \bibinfo{pages}{447} (\bibinfo{year}{2004}), \eprint{hep-ph/0401032}.

\bibitem[{\citenamefont{Casalbuoni et~al.}(2004)\citenamefont{Casalbuoni,
  De~Curtis, and Dominici}}]{Casalbuoni:2004id}
\bibinfo{author}{\bibfnamefont{R.}~\bibnamefont{Casalbuoni}},
  \bibinfo{author}{\bibfnamefont{S.}~\bibnamefont{De~Curtis}},
  \bibnamefont{and} \bibinfo{author}{\bibfnamefont{D.}~\bibnamefont{Dominici}},
  \bibinfo{journal}{Phys. Rev.} \textbf{\bibinfo{volume}{D70}},
  \bibinfo{pages}{055010} (\bibinfo{year}{2004}), \eprint{hep-ph/0405188}.

\bibitem[{\citenamefont{Chivukula
  et~al.}(2004{\natexlab{a}})\citenamefont{Chivukula, Simmons, He, Kurachi, and
  Tanabashi}}]{Chivukula:2004pk}
\bibinfo{author}{\bibfnamefont{R.~S.} \bibnamefont{Chivukula}},
  \bibinfo{author}{\bibfnamefont{E.~H.} \bibnamefont{Simmons}},
  \bibinfo{author}{\bibfnamefont{H.-J.} \bibnamefont{He}},
  \bibinfo{author}{\bibfnamefont{M.}~\bibnamefont{Kurachi}}, \bibnamefont{and}
  \bibinfo{author}{\bibfnamefont{M.}~\bibnamefont{Tanabashi}},
  \bibinfo{journal}{Phys. Rev.} \textbf{\bibinfo{volume}{D70}},
  \bibinfo{pages}{075008} (\bibinfo{year}{2004}{\natexlab{a}}),
  \eprint{hep-ph/0406077}.

\bibitem[{\citenamefont{Georgi}(2005)}]{Georgi:2004iy}
\bibinfo{author}{\bibfnamefont{H.}~\bibnamefont{Georgi}},
  \bibinfo{journal}{Phys. Rev.} \textbf{\bibinfo{volume}{D71}},
  \bibinfo{pages}{015016} (\bibinfo{year}{2005}), \eprint{hep-ph/0408067}.

\bibitem[{\citenamefont{Sekhar~Chivukula
  et~al.}(2005{\natexlab{a}})\citenamefont{Sekhar~Chivukula, Simmons, He,
  Kurachi, and Tanabashi}}]{SekharChivukula:2004mu}
\bibinfo{author}{\bibfnamefont{R.}~\bibnamefont{Sekhar~Chivukula}},
  \bibinfo{author}{\bibfnamefont{E.~H.} \bibnamefont{Simmons}},
  \bibinfo{author}{\bibfnamefont{H.-J.} \bibnamefont{He}},
  \bibinfo{author}{\bibfnamefont{M.}~\bibnamefont{Kurachi}}, \bibnamefont{and}
  \bibinfo{author}{\bibfnamefont{M.}~\bibnamefont{Tanabashi}},
  \bibinfo{journal}{Phys. Rev.} \textbf{\bibinfo{volume}{D71}},
  \bibinfo{pages}{035007} (\bibinfo{year}{2005}{\natexlab{a}}),
  \eprint{hep-ph/0410154}.

\bibitem[{\citenamefont{Sekhar~Chivukula
  et~al.}(2006)}]{SekharChivukula:2006cg}
\bibinfo{author}{\bibfnamefont{R.}~\bibnamefont{Sekhar~Chivukula}}
  \bibnamefont{et~al.} (\bibinfo{year}{2006}), \eprint{hep-ph/0607124}.

\bibitem[{\citenamefont{Cacciapaglia
  et~al.}(2005{\natexlab{b}})\citenamefont{Cacciapaglia, Csaki, Grojean, Reece,
  and Terning}}]{Cacciapaglia:2005pa}
\bibinfo{author}{\bibfnamefont{G.}~\bibnamefont{Cacciapaglia}},
  \bibinfo{author}{\bibfnamefont{C.}~\bibnamefont{Csaki}},
  \bibinfo{author}{\bibfnamefont{C.}~\bibnamefont{Grojean}},
  \bibinfo{author}{\bibfnamefont{M.}~\bibnamefont{Reece}}, \bibnamefont{and}
  \bibinfo{author}{\bibfnamefont{J.}~\bibnamefont{Terning}},
  \bibinfo{journal}{Phys. Rev.} \textbf{\bibinfo{volume}{D72}},
  \bibinfo{pages}{095018} (\bibinfo{year}{2005}{\natexlab{b}}),
  \eprint{hep-ph/0505001}.

\bibitem[{\citenamefont{Foadi et~al.}(2005)\citenamefont{Foadi, Gopalakrishna,
  and Schmidt}}]{Foadi:2004ps}
\bibinfo{author}{\bibfnamefont{R.}~\bibnamefont{Foadi}},
  \bibinfo{author}{\bibfnamefont{S.}~\bibnamefont{Gopalakrishna}},
  \bibnamefont{and} \bibinfo{author}{\bibfnamefont{C.}~\bibnamefont{Schmidt}},
  \bibinfo{journal}{Phys. Lett.} \textbf{\bibinfo{volume}{B606}},
  \bibinfo{pages}{157} (\bibinfo{year}{2005}), \eprint{hep-ph/0409266}.

\bibitem[{\citenamefont{Foadi and Schmidt}(2006)}]{Foadi:2005hz}
\bibinfo{author}{\bibfnamefont{R.}~\bibnamefont{Foadi}} \bibnamefont{and}
  \bibinfo{author}{\bibfnamefont{C.}~\bibnamefont{Schmidt}},
  \bibinfo{journal}{Phys. Rev.} \textbf{\bibinfo{volume}{D73}},
  \bibinfo{pages}{075011} (\bibinfo{year}{2006}), \eprint{hep-ph/0509071}.

\bibitem[{\citenamefont{Chivukula
  et~al.}(2005{\natexlab{a}})\citenamefont{Chivukula, Simmons, He, Kurachi, and
  Tanabashi}}]{Chivukula:2005bn}
\bibinfo{author}{\bibfnamefont{R.~S.} \bibnamefont{Chivukula}},
  \bibinfo{author}{\bibfnamefont{E.~H.} \bibnamefont{Simmons}},
  \bibinfo{author}{\bibfnamefont{H.-J.} \bibnamefont{He}},
  \bibinfo{author}{\bibfnamefont{M.}~\bibnamefont{Kurachi}}, \bibnamefont{and}
  \bibinfo{author}{\bibfnamefont{M.}~\bibnamefont{Tanabashi}},
  \bibinfo{journal}{Phys. Rev.} \textbf{\bibinfo{volume}{D71}},
  \bibinfo{pages}{115001} (\bibinfo{year}{2005}{\natexlab{a}}),
  \eprint{hep-ph/0502162}.

\bibitem[{\citenamefont{Casalbuoni et~al.}(2005)\citenamefont{Casalbuoni,
  De~Curtis, Dolce, and Dominici}}]{Casalbuoni:2005rs}
\bibinfo{author}{\bibfnamefont{R.}~\bibnamefont{Casalbuoni}},
  \bibinfo{author}{\bibfnamefont{S.}~\bibnamefont{De~Curtis}},
  \bibinfo{author}{\bibfnamefont{D.}~\bibnamefont{Dolce}}, \bibnamefont{and}
  \bibinfo{author}{\bibfnamefont{D.}~\bibnamefont{Dominici}},
  \bibinfo{journal}{Phys. Rev.} \textbf{\bibinfo{volume}{D71}},
  \bibinfo{pages}{075015} (\bibinfo{year}{2005}), \eprint{hep-ph/0502209}.

\bibitem[{\citenamefont{Sekhar~Chivukula
  et~al.}(2005{\natexlab{b}})\citenamefont{Sekhar~Chivukula, Simmons, He,
  Kurachi, and Tanabashi}}]{SekharChivukula:2005xm}
\bibinfo{author}{\bibfnamefont{R.}~\bibnamefont{Sekhar~Chivukula}},
  \bibinfo{author}{\bibfnamefont{E.~H.} \bibnamefont{Simmons}},
  \bibinfo{author}{\bibfnamefont{H.-J.} \bibnamefont{He}},
  \bibinfo{author}{\bibfnamefont{M.}~\bibnamefont{Kurachi}}, \bibnamefont{and}
  \bibinfo{author}{\bibfnamefont{M.}~\bibnamefont{Tanabashi}},
  \bibinfo{journal}{Phys. Rev.} \textbf{\bibinfo{volume}{D72}},
  \bibinfo{pages}{015008} (\bibinfo{year}{2005}{\natexlab{b}}),
  \eprint{hep-ph/0504114}.

\bibitem[{\citenamefont{Casalbuoni et~al.}(1985)\citenamefont{Casalbuoni,
  De~Curtis, Dominici, and Gatto}}]{Casalbuoni:1985kq}
\bibinfo{author}{\bibfnamefont{R.}~\bibnamefont{Casalbuoni}},
  \bibinfo{author}{\bibfnamefont{S.}~\bibnamefont{De~Curtis}},
  \bibinfo{author}{\bibfnamefont{D.}~\bibnamefont{Dominici}}, \bibnamefont{and}
  \bibinfo{author}{\bibfnamefont{R.}~\bibnamefont{Gatto}},
  \bibinfo{journal}{Phys. Lett.} \textbf{\bibinfo{volume}{B155}},
  \bibinfo{pages}{95} (\bibinfo{year}{1985}).

\bibitem[{\citenamefont{Casalbuoni et~al.}(1987)\citenamefont{Casalbuoni,
  De~Curtis, Dominici, and Gatto}}]{Casalbuoni:1986vq}
\bibinfo{author}{\bibfnamefont{R.}~\bibnamefont{Casalbuoni}},
  \bibinfo{author}{\bibfnamefont{S.}~\bibnamefont{De~Curtis}},
  \bibinfo{author}{\bibfnamefont{D.}~\bibnamefont{Dominici}}, \bibnamefont{and}
  \bibinfo{author}{\bibfnamefont{R.}~\bibnamefont{Gatto}},
  \bibinfo{journal}{Nucl. Phys.} \textbf{\bibinfo{volume}{B282}},
  \bibinfo{pages}{235} (\bibinfo{year}{1987}).

\bibitem[{\citenamefont{Peskin and Takeuchi}(1992)}]{Peskin:1991sw}
\bibinfo{author}{\bibfnamefont{M.~E.} \bibnamefont{Peskin}} \bibnamefont{and}
  \bibinfo{author}{\bibfnamefont{T.}~\bibnamefont{Takeuchi}},
  \bibinfo{journal}{Phys. Rev.} \textbf{\bibinfo{volume}{D46}},
  \bibinfo{pages}{381} (\bibinfo{year}{1992}).

\bibitem[{\citenamefont{Degrassi and
  Sirlin}(1992{\natexlab{a}})}]{Degrassi:1992ue}
\bibinfo{author}{\bibfnamefont{G.}~\bibnamefont{Degrassi}} \bibnamefont{and}
  \bibinfo{author}{\bibfnamefont{A.}~\bibnamefont{Sirlin}},
  \bibinfo{journal}{Phys. Rev.} \textbf{\bibinfo{volume}{D46}},
  \bibinfo{pages}{3104} (\bibinfo{year}{1992}{\natexlab{a}}).

\bibitem[{\citenamefont{Degrassi et~al.}(1993)\citenamefont{Degrassi, Kniehl,
  and Sirlin}}]{Degrassi:1993kn}
\bibinfo{author}{\bibfnamefont{G.}~\bibnamefont{Degrassi}},
  \bibinfo{author}{\bibfnamefont{B.~A.} \bibnamefont{Kniehl}},
  \bibnamefont{and} \bibinfo{author}{\bibfnamefont{A.}~\bibnamefont{Sirlin}},
  \bibinfo{journal}{Phys. Rev.} \textbf{\bibinfo{volume}{D48}},
  \bibinfo{pages}{3963} (\bibinfo{year}{1993}).

\bibitem[{\citenamefont{Degrassi and
  Sirlin}(1992{\natexlab{b}})}]{Degrassi:1992ff}
\bibinfo{author}{\bibfnamefont{G.}~\bibnamefont{Degrassi}} \bibnamefont{and}
  \bibinfo{author}{\bibfnamefont{A.}~\bibnamefont{Sirlin}},
  \bibinfo{journal}{Nucl. Phys.} \textbf{\bibinfo{volume}{B383}},
  \bibinfo{pages}{73} (\bibinfo{year}{1992}{\natexlab{b}}).

\bibitem[{\citenamefont{Papavassiliou and Sirlin}(1994)}]{Papavassiliou:1994fp}
\bibinfo{author}{\bibfnamefont{J.}~\bibnamefont{Papavassiliou}}
  \bibnamefont{and} \bibinfo{author}{\bibfnamefont{A.}~\bibnamefont{Sirlin}},
  \bibinfo{journal}{Phys. Rev.} \textbf{\bibinfo{volume}{D50}},
  \bibinfo{pages}{5951} (\bibinfo{year}{1994}), \eprint{hep-ph/9403378}.

\bibitem[{\citenamefont{Cornwall}(1981)}]{Cornwall:1981ru}
\bibinfo{author}{\bibfnamefont{J.~M.} \bibnamefont{Cornwall}}
  (\bibinfo{year}{1981}), \eprint{UCLA/81/TEP/12}.

\bibitem[{\citenamefont{Cornwall}(1982)}]{Cornwall:1981zr}
\bibinfo{author}{\bibfnamefont{J.~M.} \bibnamefont{Cornwall}},
  \bibinfo{journal}{Phys. Rev.} \textbf{\bibinfo{volume}{D26}},
  \bibinfo{pages}{1453} (\bibinfo{year}{1982}).

\bibitem[{\citenamefont{Cornwall and Papavassiliou}(1989)}]{Cornwall:1989gv}
\bibinfo{author}{\bibfnamefont{J.~M.} \bibnamefont{Cornwall}} \bibnamefont{and}
  \bibinfo{author}{\bibfnamefont{J.}~\bibnamefont{Papavassiliou}},
  \bibinfo{journal}{Phys. Rev.} \textbf{\bibinfo{volume}{D40}},
  \bibinfo{pages}{3474} (\bibinfo{year}{1989}).

\bibitem[{\citenamefont{Papavassiliou}(1990)}]{Papavassiliou:1989zd}
\bibinfo{author}{\bibfnamefont{J.}~\bibnamefont{Papavassiliou}},
  \bibinfo{journal}{Phys. Rev.} \textbf{\bibinfo{volume}{D41}},
  \bibinfo{pages}{3179} (\bibinfo{year}{1990}).

\bibitem[{\citenamefont{Matsuzaki et~al.}(2006)\citenamefont{Matsuzaki,
  Chivukula, Tanabashi, and Simmons}}]{Matsuzaki:2006wn}
\bibinfo{author}{\bibfnamefont{S.}~\bibnamefont{Matsuzaki}},
  \bibinfo{author}{\bibfnamefont{R.~S.} \bibnamefont{Chivukula}},
  \bibinfo{author}{\bibfnamefont{M.}~\bibnamefont{Tanabashi}},
  \bibnamefont{and} \bibinfo{author}{\bibfnamefont{E.~H.}
  \bibnamefont{Simmons}} (\bibinfo{year}{2006}), \eprint{hep-ph/0607191}.

\bibitem[{\citenamefont{Agashe et~al.}(2003)\citenamefont{Agashe, Delgado, May,
  and Sundrum}}]{Agashe:2003zs}
\bibinfo{author}{\bibfnamefont{K.}~\bibnamefont{Agashe}},
  \bibinfo{author}{\bibfnamefont{A.}~\bibnamefont{Delgado}},
  \bibinfo{author}{\bibfnamefont{M.~J.} \bibnamefont{May}}, \bibnamefont{and}
  \bibinfo{author}{\bibfnamefont{R.}~\bibnamefont{Sundrum}},
  \bibinfo{journal}{JHEP} \textbf{\bibinfo{volume}{08}}, \bibinfo{pages}{050}
  (\bibinfo{year}{2003}), \eprint{hep-ph/0308036}.

\bibitem[{\citenamefont{Barbieri
  et~al.}(2004{\natexlab{a}})\citenamefont{Barbieri, Pomarol, and
  Rattazzi}}]{Barbieri:2003pr}
\bibinfo{author}{\bibfnamefont{R.}~\bibnamefont{Barbieri}},
  \bibinfo{author}{\bibfnamefont{A.}~\bibnamefont{Pomarol}}, \bibnamefont{and}
  \bibinfo{author}{\bibfnamefont{R.}~\bibnamefont{Rattazzi}},
  \bibinfo{journal}{Phys. Lett.} \textbf{\bibinfo{volume}{B591}},
  \bibinfo{pages}{141} (\bibinfo{year}{2004}{\natexlab{a}}),
  \eprint{hep-ph/0310285}.

\bibitem[{\citenamefont{Csaki et~al.}(2004{\natexlab{c}})\citenamefont{Csaki,
  Grojean, Hubisz, Shirman, and Terning}}]{Csaki:2003sh}
\bibinfo{author}{\bibfnamefont{C.}~\bibnamefont{Csaki}},
  \bibinfo{author}{\bibfnamefont{C.}~\bibnamefont{Grojean}},
  \bibinfo{author}{\bibfnamefont{J.}~\bibnamefont{Hubisz}},
  \bibinfo{author}{\bibfnamefont{Y.}~\bibnamefont{Shirman}}, \bibnamefont{and}
  \bibinfo{author}{\bibfnamefont{J.}~\bibnamefont{Terning}},
  \bibinfo{journal}{Phys. Rev.} \textbf{\bibinfo{volume}{D70}},
  \bibinfo{pages}{015012} (\bibinfo{year}{2004}{\natexlab{c}}),
  \eprint{hep-ph/0310355}.

\bibitem[{\citenamefont{Davoudiasl
  et~al.}(2004{\natexlab{a}})\citenamefont{Davoudiasl, Hewett, Lillie, and
  Rizzo}}]{Davoudiasl:2003me}
\bibinfo{author}{\bibfnamefont{H.}~\bibnamefont{Davoudiasl}},
  \bibinfo{author}{\bibfnamefont{J.~L.} \bibnamefont{Hewett}},
  \bibinfo{author}{\bibfnamefont{B.}~\bibnamefont{Lillie}}, \bibnamefont{and}
  \bibinfo{author}{\bibfnamefont{T.~G.} \bibnamefont{Rizzo}},
  \bibinfo{journal}{Phys. Rev.} \textbf{\bibinfo{volume}{D70}},
  \bibinfo{pages}{015006} (\bibinfo{year}{2004}{\natexlab{a}}),
  \eprint{hep-ph/0312193}.

\bibitem[{\citenamefont{Burdman and Nomura}(2004)}]{Burdman:2003ya}
\bibinfo{author}{\bibfnamefont{G.}~\bibnamefont{Burdman}} \bibnamefont{and}
  \bibinfo{author}{\bibfnamefont{Y.}~\bibnamefont{Nomura}},
  \bibinfo{journal}{Phys. Rev.} \textbf{\bibinfo{volume}{D69}},
  \bibinfo{pages}{115013} (\bibinfo{year}{2004}), \eprint{hep-ph/0312247}.

\bibitem[{\citenamefont{Cacciapaglia et~al.}(2004)\citenamefont{Cacciapaglia,
  Csaki, Grojean, and Terning}}]{Cacciapaglia:2004jz}
\bibinfo{author}{\bibfnamefont{G.}~\bibnamefont{Cacciapaglia}},
  \bibinfo{author}{\bibfnamefont{C.}~\bibnamefont{Csaki}},
  \bibinfo{author}{\bibfnamefont{C.}~\bibnamefont{Grojean}}, \bibnamefont{and}
  \bibinfo{author}{\bibfnamefont{J.}~\bibnamefont{Terning}},
  \bibinfo{journal}{Phys. Rev.} \textbf{\bibinfo{volume}{D70}},
  \bibinfo{pages}{075014} (\bibinfo{year}{2004}), \eprint{hep-ph/0401160}.

\bibitem[{\citenamefont{Davoudiasl
  et~al.}(2004{\natexlab{b}})\citenamefont{Davoudiasl, Hewett, Lillie, and
  Rizzo}}]{Davoudiasl:2004pwWY}
\bibinfo{author}{\bibfnamefont{H.}~\bibnamefont{Davoudiasl}},
  \bibinfo{author}{\bibfnamefont{J.~L.} \bibnamefont{Hewett}},
  \bibinfo{author}{\bibfnamefont{B.}~\bibnamefont{Lillie}}, \bibnamefont{and}
  \bibinfo{author}{\bibfnamefont{T.~G.} \bibnamefont{Rizzo}},
  \bibinfo{journal}{JHEP} \textbf{\bibinfo{volume}{05}}, \bibinfo{pages}{015}
  (\bibinfo{year}{2004}{\natexlab{b}}), \eprint{hep-ph/0403300}.

\bibitem[{\citenamefont{Barbieri
  et~al.}(2004{\natexlab{b}})\citenamefont{Barbieri, Pomarol, Rattazzi, and
  Strumia}}]{Barbieri:2004qk}
\bibinfo{author}{\bibfnamefont{R.}~\bibnamefont{Barbieri}},
  \bibinfo{author}{\bibfnamefont{A.}~\bibnamefont{Pomarol}},
  \bibinfo{author}{\bibfnamefont{R.}~\bibnamefont{Rattazzi}}, \bibnamefont{and}
  \bibinfo{author}{\bibfnamefont{A.}~\bibnamefont{Strumia}},
  \bibinfo{journal}{Nucl. Phys.} \textbf{\bibinfo{volume}{B703}},
  \bibinfo{pages}{127} (\bibinfo{year}{2004}{\natexlab{b}}),
  \eprint{hep-ph/0405040}.

\bibitem[{\citenamefont{Agashe et~al.}(2004)\citenamefont{Agashe, Perez, and
  Soni}}]{Agashe:2004ay}
\bibinfo{author}{\bibfnamefont{K.}~\bibnamefont{Agashe}},
  \bibinfo{author}{\bibfnamefont{G.}~\bibnamefont{Perez}}, \bibnamefont{and}
  \bibinfo{author}{\bibfnamefont{A.}~\bibnamefont{Soni}},
  \bibinfo{journal}{Phys. Rev. Lett.} \textbf{\bibinfo{volume}{93}},
  \bibinfo{pages}{201804} (\bibinfo{year}{2004}), \eprint{hep-ph/0406101}.

\bibitem[{\citenamefont{Hewett et~al.}(2004)\citenamefont{Hewett, Lillie, and
  Rizzo}}]{Hewett:2004dv}
\bibinfo{author}{\bibfnamefont{J.~L.} \bibnamefont{Hewett}},
  \bibinfo{author}{\bibfnamefont{B.}~\bibnamefont{Lillie}}, \bibnamefont{and}
  \bibinfo{author}{\bibfnamefont{T.~G.} \bibnamefont{Rizzo}},
  \bibinfo{journal}{JHEP} \textbf{\bibinfo{volume}{10}}, \bibinfo{pages}{014}
  (\bibinfo{year}{2004}), \eprint{hep-ph/0407059}.

\bibitem[{\citenamefont{Agashe et~al.}(2005)\citenamefont{Agashe, Perez, and
  Soni}}]{Agashe:2004cp}
\bibinfo{author}{\bibfnamefont{K.}~\bibnamefont{Agashe}},
  \bibinfo{author}{\bibfnamefont{G.}~\bibnamefont{Perez}}, \bibnamefont{and}
  \bibinfo{author}{\bibfnamefont{A.}~\bibnamefont{Soni}},
  \bibinfo{journal}{Phys. Rev.} \textbf{\bibinfo{volume}{D71}},
  \bibinfo{pages}{016002} (\bibinfo{year}{2005}), \eprint{hep-ph/0408134}.

\bibitem[{\citenamefont{Lillie}(2006)}]{Lillie:2005pt}
\bibinfo{author}{\bibfnamefont{B.}~\bibnamefont{Lillie}},
  \bibinfo{journal}{JHEP} \textbf{\bibinfo{volume}{02}}, \bibinfo{pages}{019}
  (\bibinfo{year}{2006}), \eprint{hep-ph/0505074}.

\bibitem[{\citenamefont{Agashe et~al.}(2007{\natexlab{a}})\citenamefont{Agashe,
  Perez, and Soni}}]{Agashe:2006wa}
\bibinfo{author}{\bibfnamefont{K.}~\bibnamefont{Agashe}},
  \bibinfo{author}{\bibfnamefont{G.}~\bibnamefont{Perez}}, \bibnamefont{and}
  \bibinfo{author}{\bibfnamefont{A.}~\bibnamefont{Soni}},
  \bibinfo{journal}{Phys. Rev.} \textbf{\bibinfo{volume}{D75}},
  \bibinfo{pages}{015002} (\bibinfo{year}{2007}{\natexlab{a}}),
  \eprint{hep-ph/0606293}.

\bibitem[{\citenamefont{Carena et~al.}(2006)\citenamefont{Carena, Ponton,
  Santiago, and Wagner}}]{Carena:2006bn}
\bibinfo{author}{\bibfnamefont{M.}~\bibnamefont{Carena}},
  \bibinfo{author}{\bibfnamefont{E.}~\bibnamefont{Ponton}},
  \bibinfo{author}{\bibfnamefont{J.}~\bibnamefont{Santiago}}, \bibnamefont{and}
  \bibinfo{author}{\bibfnamefont{C.~E.~M.} \bibnamefont{Wagner}},
  \bibinfo{journal}{Nucl. Phys.} \textbf{\bibinfo{volume}{B759}},
  \bibinfo{pages}{202} (\bibinfo{year}{2006}), \eprint{hep-ph/0607106}.

\bibitem[{\citenamefont{Cacciapaglia
  et~al.}(2007{\natexlab{a}})\citenamefont{Cacciapaglia, Csaki, Marandella, and
  Terning}}]{Cacciapaglia:2006gp}
\bibinfo{author}{\bibfnamefont{G.}~\bibnamefont{Cacciapaglia}},
  \bibinfo{author}{\bibfnamefont{C.}~\bibnamefont{Csaki}},
  \bibinfo{author}{\bibfnamefont{G.}~\bibnamefont{Marandella}},
  \bibnamefont{and} \bibinfo{author}{\bibfnamefont{J.}~\bibnamefont{Terning}},
  \bibinfo{journal}{Phys. Rev.} \textbf{\bibinfo{volume}{D75}},
  \bibinfo{pages}{015003} (\bibinfo{year}{2007}{\natexlab{a}}),
  \eprint{hep-ph/0607146}.

\bibitem[{\citenamefont{Djouadi et~al.}(2006)\citenamefont{Djouadi, Moreau, and
  Richard}}]{Djouadi:2006rk}
\bibinfo{author}{\bibfnamefont{A.}~\bibnamefont{Djouadi}},
  \bibinfo{author}{\bibfnamefont{G.}~\bibnamefont{Moreau}}, \bibnamefont{and}
  \bibinfo{author}{\bibfnamefont{F.}~\bibnamefont{Richard}}
  (\bibinfo{year}{2006}), \eprint{hep-ph/0610173}.

\bibitem[{\citenamefont{Cacciapaglia
  et~al.}(2007{\natexlab{b}})\citenamefont{Cacciapaglia, Csaki, Marandella, and
  Terning}}]{Cacciapaglia:2006mz}
\bibinfo{author}{\bibfnamefont{G.}~\bibnamefont{Cacciapaglia}},
  \bibinfo{author}{\bibfnamefont{C.}~\bibnamefont{Csaki}},
  \bibinfo{author}{\bibfnamefont{G.}~\bibnamefont{Marandella}},
  \bibnamefont{and} \bibinfo{author}{\bibfnamefont{J.}~\bibnamefont{Terning}},
  \bibinfo{journal}{JHEP} \textbf{\bibinfo{volume}{02}}, \bibinfo{pages}{036}
  (\bibinfo{year}{2007}{\natexlab{b}}), \eprint{hep-ph/0611358}.

\bibitem[{\citenamefont{Contino et~al.}(2006)\citenamefont{Contino, Kramer,
  Son, and Sundrum}}]{Contino:2006nn}
\bibinfo{author}{\bibfnamefont{R.}~\bibnamefont{Contino}},
  \bibinfo{author}{\bibfnamefont{T.}~\bibnamefont{Kramer}},
  \bibinfo{author}{\bibfnamefont{M.}~\bibnamefont{Son}}, \bibnamefont{and}
  \bibinfo{author}{\bibfnamefont{R.}~\bibnamefont{Sundrum}}
  (\bibinfo{year}{2006}), \eprint{hep-ph/0612180}.

\bibitem[{\citenamefont{Carena et~al.}(2007)\citenamefont{Carena, Ponton,
  Santiago, and Wagner}}]{Carena:2007ua}
\bibinfo{author}{\bibfnamefont{M.}~\bibnamefont{Carena}},
  \bibinfo{author}{\bibfnamefont{E.}~\bibnamefont{Ponton}},
  \bibinfo{author}{\bibfnamefont{J.}~\bibnamefont{Santiago}}, \bibnamefont{and}
  \bibinfo{author}{\bibfnamefont{C.~E.~M.} \bibnamefont{Wagner}}
  (\bibinfo{year}{2007}), \eprint{hep-ph/0701055}.

\bibitem[{\citenamefont{Lillie et~al.}(2007)\citenamefont{Lillie, Randall, and
  Wang}}]{Lillie:2007yh}
\bibinfo{author}{\bibfnamefont{B.}~\bibnamefont{Lillie}},
  \bibinfo{author}{\bibfnamefont{L.}~\bibnamefont{Randall}}, \bibnamefont{and}
  \bibinfo{author}{\bibfnamefont{L.-T.} \bibnamefont{Wang}}
  (\bibinfo{year}{2007}), \eprint{hep-ph/0701166}.

\bibitem[{\citenamefont{Agashe et~al.}(2007{\natexlab{b}})\citenamefont{Agashe,
  Davoudiasl, Perez, and Soni}}]{Agashe:2007zd}
\bibinfo{author}{\bibfnamefont{K.}~\bibnamefont{Agashe}},
  \bibinfo{author}{\bibfnamefont{H.}~\bibnamefont{Davoudiasl}},
  \bibinfo{author}{\bibfnamefont{G.}~\bibnamefont{Perez}}, \bibnamefont{and}
  \bibinfo{author}{\bibfnamefont{A.}~\bibnamefont{Soni}}
  (\bibinfo{year}{2007}{\natexlab{b}}), \eprint{hep-ph/0701186}.

\bibitem[{\citenamefont{Chivukula
  et~al.}(2004{\natexlab{b}})\citenamefont{Chivukula, He, Howard, and
  Simmons}}]{Chivukula:2003wj}
\bibinfo{author}{\bibfnamefont{R.~S.} \bibnamefont{Chivukula}},
  \bibinfo{author}{\bibfnamefont{H.-J.} \bibnamefont{He}},
  \bibinfo{author}{\bibfnamefont{J.}~\bibnamefont{Howard}}, \bibnamefont{and}
  \bibinfo{author}{\bibfnamefont{E.~H.} \bibnamefont{Simmons}},
  \bibinfo{journal}{Phys. Rev.} \textbf{\bibinfo{volume}{D69}},
  \bibinfo{pages}{015009} (\bibinfo{year}{2004}{\natexlab{b}}),
  \eprint{hep-ph/0307209}.

\bibitem[{\citenamefont{Chivukula
  et~al.}(2004{\natexlab{c}})\citenamefont{Chivukula, Simmons, He, Kurachi, and
  Tanabashi}}]{Chivukula:2004af}
\bibinfo{author}{\bibfnamefont{R.~S.} \bibnamefont{Chivukula}},
  \bibinfo{author}{\bibfnamefont{E.~H.} \bibnamefont{Simmons}},
  \bibinfo{author}{\bibfnamefont{H.-J.} \bibnamefont{He}},
  \bibinfo{author}{\bibfnamefont{M.}~\bibnamefont{Kurachi}}, \bibnamefont{and}
  \bibinfo{author}{\bibfnamefont{M.}~\bibnamefont{Tanabashi}},
  \bibinfo{journal}{Phys. Lett.} \textbf{\bibinfo{volume}{B603}},
  \bibinfo{pages}{210} (\bibinfo{year}{2004}{\natexlab{c}}),
  \eprint{hep-ph/0408262}.

\bibitem[{\citenamefont{Chivukula
  et~al.}(2005{\natexlab{b}})\citenamefont{Chivukula, Simmons, He, Kurachi, and
  Tanabashi}}]{Chivukula:2005ji}
\bibinfo{author}{\bibfnamefont{R.~S.} \bibnamefont{Chivukula}},
  \bibinfo{author}{\bibfnamefont{E.~H.} \bibnamefont{Simmons}},
  \bibinfo{author}{\bibfnamefont{H.-J.} \bibnamefont{He}},
  \bibinfo{author}{\bibfnamefont{M.}~\bibnamefont{Kurachi}}, \bibnamefont{and}
  \bibinfo{author}{\bibfnamefont{M.}~\bibnamefont{Tanabashi}},
  \bibinfo{journal}{Phys. Rev.} \textbf{\bibinfo{volume}{D72}},
  \bibinfo{pages}{075012} (\bibinfo{year}{2005}{\natexlab{b}}),
  \eprint{hep-ph/0508147}.

\bibitem[{\citenamefont{Sekhar~Chivukula
  et~al.}(2005{\natexlab{c}})\citenamefont{Sekhar~Chivukula, Simmons, He,
  Kurachi, and Tanabashi}}]{SekharChivukula:2005cc}
\bibinfo{author}{\bibfnamefont{R.}~\bibnamefont{Sekhar~Chivukula}},
  \bibinfo{author}{\bibfnamefont{E.~H.} \bibnamefont{Simmons}},
  \bibinfo{author}{\bibfnamefont{H.-J.} \bibnamefont{He}},
  \bibinfo{author}{\bibfnamefont{M.}~\bibnamefont{Kurachi}}, \bibnamefont{and}
  \bibinfo{author}{\bibfnamefont{M.}~\bibnamefont{Tanabashi}},
  \bibinfo{journal}{Phys. Rev.} \textbf{\bibinfo{volume}{D72}},
  \bibinfo{pages}{095013} (\bibinfo{year}{2005}{\natexlab{c}}),
  \eprint{hep-ph/0509110}.

\bibitem[{\citenamefont{Sekhar~Chivukula
  et~al.}(2007)\citenamefont{Sekhar~Chivukula, Simmons, He, Kurachi, and
  Tanabashi}}]{SekharChivukula:2006we}
\bibinfo{author}{\bibfnamefont{R.}~\bibnamefont{Sekhar~Chivukula}},
  \bibinfo{author}{\bibfnamefont{E.~H.} \bibnamefont{Simmons}},
  \bibinfo{author}{\bibfnamefont{H.-J.} \bibnamefont{He}},
  \bibinfo{author}{\bibfnamefont{M.}~\bibnamefont{Kurachi}}, \bibnamefont{and}
  \bibinfo{author}{\bibfnamefont{M.}~\bibnamefont{Tanabashi}},
  \bibinfo{journal}{Phys. Rev.} \textbf{\bibinfo{volume}{D75}},
  \bibinfo{pages}{035005} (\bibinfo{year}{2007}), \eprint{hep-ph/0612070}.

\bibitem[{\citenamefont{Birkedal et~al.}(2005)\citenamefont{Birkedal, Matchev,
  and Perelstein}}]{Birkedal:2004au}
\bibinfo{author}{\bibfnamefont{A.}~\bibnamefont{Birkedal}},
  \bibinfo{author}{\bibfnamefont{K.}~\bibnamefont{Matchev}}, \bibnamefont{and}
  \bibinfo{author}{\bibfnamefont{M.}~\bibnamefont{Perelstein}},
  \bibinfo{journal}{Phys. Rev. Lett.} \textbf{\bibinfo{volume}{94}},
  \bibinfo{pages}{191803} (\bibinfo{year}{2005}), \eprint{hep-ph/0412278}.

\bibitem[{\citenamefont{Passarino and Veltman}(1979)}]{Passarino:1978jh}
\bibinfo{author}{\bibfnamefont{G.}~\bibnamefont{Passarino}} \bibnamefont{and}
  \bibinfo{author}{\bibfnamefont{M.~J.~G.} \bibnamefont{Veltman}},
  \bibinfo{journal}{Nucl. Phys.} \textbf{\bibinfo{volume}{B160}},
  \bibinfo{pages}{151} (\bibinfo{year}{1979}).

\bibitem[{\citenamefont{Yao et~al.}(2006)}]{Yao:2006px}
\bibinfo{author}{\bibfnamefont{W.~M.} \bibnamefont{Yao}} \bibnamefont{et~al.}
  (\bibinfo{collaboration}{Particle Data Group}), \bibinfo{journal}{J. Phys.}
  \textbf{\bibinfo{volume}{G33}}, \bibinfo{pages}{1} (\bibinfo{year}{2006}).

\bibitem[{\citenamefont{Li and Pagels}(1971)}]{Li:1971vr}
\bibinfo{author}{\bibfnamefont{L.-F.} \bibnamefont{Li}} \bibnamefont{and}
  \bibinfo{author}{\bibfnamefont{H.}~\bibnamefont{Pagels}},
  \bibinfo{journal}{Phys. Rev. Lett.} \textbf{\bibinfo{volume}{26}},
  \bibinfo{pages}{1204} (\bibinfo{year}{1971}).

\bibitem[{\citenamefont{Ling-Fong and Pagels}(1972)}]{Ling-Fong:1972hu}
\bibinfo{author}{\bibfnamefont{L.}~\bibnamefont{Ling-Fong}} \bibnamefont{and}
  \bibinfo{author}{\bibfnamefont{H.}~\bibnamefont{Pagels}},
  \bibinfo{journal}{Phys. Rev.} \textbf{\bibinfo{volume}{D5}},
  \bibinfo{pages}{1509} (\bibinfo{year}{1972}).

\bibitem[{\citenamefont{Langacker and Pagels}(1973)}]{Langacker:1973hh}
\bibinfo{author}{\bibfnamefont{P.}~\bibnamefont{Langacker}} \bibnamefont{and}
  \bibinfo{author}{\bibfnamefont{H.}~\bibnamefont{Pagels}},
  \bibinfo{journal}{Phys. Rev.} \textbf{\bibinfo{volume}{D8}},
  \bibinfo{pages}{4595} (\bibinfo{year}{1973}).

\bibitem[{\citenamefont{Anichini et~al.}(1995)\citenamefont{Anichini,
  Casalbuoni, and De~Curtis}}]{Anichini:1994xx}
\bibinfo{author}{\bibfnamefont{L.}~\bibnamefont{Anichini}},
  \bibinfo{author}{\bibfnamefont{R.}~\bibnamefont{Casalbuoni}},
  \bibnamefont{and}
  \bibinfo{author}{\bibfnamefont{S.}~\bibnamefont{De~Curtis}},
  \bibinfo{journal}{Phys. Lett.} \textbf{\bibinfo{volume}{B348}},
  \bibinfo{pages}{521} (\bibinfo{year}{1995}), \eprint{hep-ph/9410377}.

\bibitem[{\citenamefont{Csaki et~al.}(2005)\citenamefont{Csaki, Hubisz, and
  Meade}}]{Csaki:2005vy}
\bibinfo{author}{\bibfnamefont{C.}~\bibnamefont{Csaki}},
  \bibinfo{author}{\bibfnamefont{J.}~\bibnamefont{Hubisz}}, \bibnamefont{and}
  \bibinfo{author}{\bibfnamefont{P.}~\bibnamefont{Meade}}
  (\bibinfo{year}{2005}), \eprint{hep-ph/0510275}.

\bibitem[{\citenamefont{Bando et~al.}(1985{\natexlab{a}})\citenamefont{Bando,
  Kugo, Uehara, Yamawaki, and Yanagida}}]{Bando:1984ej}
\bibinfo{author}{\bibfnamefont{M.}~\bibnamefont{Bando}},
  \bibinfo{author}{\bibfnamefont{T.}~\bibnamefont{Kugo}},
  \bibinfo{author}{\bibfnamefont{S.}~\bibnamefont{Uehara}},
  \bibinfo{author}{\bibfnamefont{K.}~\bibnamefont{Yamawaki}}, \bibnamefont{and}
  \bibinfo{author}{\bibfnamefont{T.}~\bibnamefont{Yanagida}},
  \bibinfo{journal}{Phys. Rev. Lett.} \textbf{\bibinfo{volume}{54}},
  \bibinfo{pages}{1215} (\bibinfo{year}{1985}{\natexlab{a}}).

\bibitem[{\citenamefont{Bando et~al.}(1985{\natexlab{b}})\citenamefont{Bando,
  Kugo, and Yamawaki}}]{Bando:1985rf}
\bibinfo{author}{\bibfnamefont{M.}~\bibnamefont{Bando}},
  \bibinfo{author}{\bibfnamefont{T.}~\bibnamefont{Kugo}}, \bibnamefont{and}
  \bibinfo{author}{\bibfnamefont{K.}~\bibnamefont{Yamawaki}},
  \bibinfo{journal}{Nucl. Phys.} \textbf{\bibinfo{volume}{B259}},
  \bibinfo{pages}{493} (\bibinfo{year}{1985}{\natexlab{b}}).

\bibitem[{\citenamefont{Bando et~al.}(1988{\natexlab{a}})\citenamefont{Bando,
  Fujiwara, and Yamawaki}}]{Bando:1987ym}
\bibinfo{author}{\bibfnamefont{M.}~\bibnamefont{Bando}},
  \bibinfo{author}{\bibfnamefont{T.}~\bibnamefont{Fujiwara}}, \bibnamefont{and}
  \bibinfo{author}{\bibfnamefont{K.}~\bibnamefont{Yamawaki}},
  \bibinfo{journal}{Prog. Theor. Phys.} \textbf{\bibinfo{volume}{79}},
  \bibinfo{pages}{1140} (\bibinfo{year}{1988}{\natexlab{a}}).

\bibitem[{\citenamefont{Bando et~al.}(1988{\natexlab{b}})\citenamefont{Bando,
  Kugo, and Yamawaki}}]{Bando:1987br}
\bibinfo{author}{\bibfnamefont{M.}~\bibnamefont{Bando}},
  \bibinfo{author}{\bibfnamefont{T.}~\bibnamefont{Kugo}}, \bibnamefont{and}
  \bibinfo{author}{\bibfnamefont{K.}~\bibnamefont{Yamawaki}},
  \bibinfo{journal}{Phys. Rept.} \textbf{\bibinfo{volume}{164}},
  \bibinfo{pages}{217} (\bibinfo{year}{1988}{\natexlab{b}}).

\bibitem[{\citenamefont{Herrero and Ruiz~Morales}(1994)}]{Herrero:1993nc}
\bibinfo{author}{\bibfnamefont{M.~J.} \bibnamefont{Herrero}} \bibnamefont{and}
  \bibinfo{author}{\bibfnamefont{E.}~\bibnamefont{Ruiz~Morales}},
  \bibinfo{journal}{Nucl. Phys.} \textbf{\bibinfo{volume}{B418}},
  \bibinfo{pages}{431} (\bibinfo{year}{1994}), \eprint{hep-ph/9308276}.

\bibitem[{\citenamefont{Herrero and Ruiz~Morales}(1995)}]{Herrero:1994iu}
\bibinfo{author}{\bibfnamefont{M.~J.} \bibnamefont{Herrero}} \bibnamefont{and}
  \bibinfo{author}{\bibfnamefont{E.}~\bibnamefont{Ruiz~Morales}},
  \bibinfo{journal}{Nucl. Phys.} \textbf{\bibinfo{volume}{B437}},
  \bibinfo{pages}{319} (\bibinfo{year}{1995}), \eprint{hep-ph/9411207}.

\bibitem[{\citenamefont{Dittmaier and Grosse-Knetter}(1996)}]{Dittmaier:1995ee}
\bibinfo{author}{\bibfnamefont{S.}~\bibnamefont{Dittmaier}} \bibnamefont{and}
  \bibinfo{author}{\bibfnamefont{C.}~\bibnamefont{Grosse-Knetter}},
  \bibinfo{journal}{Nucl. Phys.} \textbf{\bibinfo{volume}{B459}},
  \bibinfo{pages}{497} (\bibinfo{year}{1996}), \eprint{hep-ph/9505266}.

\bibitem[{\citenamefont{Matias}(1996)}]{Matias:1996hf}
\bibinfo{author}{\bibfnamefont{J.}~\bibnamefont{Matias}},
  \bibinfo{journal}{Nucl. Phys.} \textbf{\bibinfo{volume}{B478}},
  \bibinfo{pages}{90} (\bibinfo{year}{1996}), \eprint{hep-ph/9604390}.

\bibitem[{\citenamefont{Alam et~al.}(1998)\citenamefont{Alam, Dawson, and
  Szalapski}}]{Alam:1997nk}
\bibinfo{author}{\bibfnamefont{S.}~\bibnamefont{Alam}},
  \bibinfo{author}{\bibfnamefont{S.}~\bibnamefont{Dawson}}, \bibnamefont{and}
  \bibinfo{author}{\bibfnamefont{R.}~\bibnamefont{Szalapski}},
  \bibinfo{journal}{Phys. Rev.} \textbf{\bibinfo{volume}{D57}},
  \bibinfo{pages}{1577} (\bibinfo{year}{1998}), \eprint{hep-ph/9706542}.

\bibitem[{\citenamefont{Bagger et~al.}(2000)\citenamefont{Bagger, Falk, and
  Swartz}}]{Bagger:1999te}
\bibinfo{author}{\bibfnamefont{J.~A.} \bibnamefont{Bagger}},
  \bibinfo{author}{\bibfnamefont{A.~F.} \bibnamefont{Falk}}, \bibnamefont{and}
  \bibinfo{author}{\bibfnamefont{M.}~\bibnamefont{Swartz}},
  \bibinfo{journal}{Phys. Rev. Lett.} \textbf{\bibinfo{volume}{84}},
  \bibinfo{pages}{1385} (\bibinfo{year}{2000}), \eprint{hep-ph/9908327}.

\bibitem[{\citenamefont{Chivukula et~al.}(2000)\citenamefont{Chivukula,
  Hoelbling, and Evans}}]{Chivukula:2000px}
\bibinfo{author}{\bibfnamefont{R.~S.} \bibnamefont{Chivukula}},
  \bibinfo{author}{\bibfnamefont{C.}~\bibnamefont{Hoelbling}},
  \bibnamefont{and} \bibinfo{author}{\bibfnamefont{N.~J.} \bibnamefont{Evans}},
  \bibinfo{journal}{Phys. Rev. Lett.} \textbf{\bibinfo{volume}{85}},
  \bibinfo{pages}{511} (\bibinfo{year}{2000}), \eprint{hep-ph/0002022}.

\bibitem[{\citenamefont{Chivukula and Evans}(1999)}]{Chivukula:1999az}
\bibinfo{author}{\bibfnamefont{R.~S.} \bibnamefont{Chivukula}}
  \bibnamefont{and} \bibinfo{author}{\bibfnamefont{N.~J.} \bibnamefont{Evans}},
  \bibinfo{journal}{Phys. Lett.} \textbf{\bibinfo{volume}{B464}},
  \bibinfo{pages}{244} (\bibinfo{year}{1999}), \eprint{hep-ph/9907414}.

\end{thebibliography}

\end{document}